%% file: measles_paper.tex
\documentclass[11pt]{article}
\usepackage[margin=2cm]{geometry}

\usepackage{verbatim}

\usepackage{dcolumn}
\usepackage{comment}
\usepackage{fancyhdr}

\usepackage[T1]{fontenc}% Must be loaded for proper fontencoding when using pdfLaTeX
\usepackage[utf8]{inputenx}% For proper input encoding

\usepackage{booktabs}% Pretty tables
\usepackage{threeparttable}% For Notes below table
\usepackage[skip=5pt, justification=centering]{caption}
\usepackage{longtable}
\usepackage{pdflscape}
\usepackage{amsmath}
\usepackage{morefloats}
\usepackage{xcolor}
\usepackage[bookmarksnumbered=true, bookmarksdepth=1,citecolor=blue, colorlinks=true, filecolor=blue,hyperfootnotes=false,linkcolor=blue, pdffitwindow=true,urlcolor=blue]{hyperref}
\usepackage{graphicx}
\usepackage{subcaption}

\usepackage{placeins}

\usepackage[title]{appendix}

\usepackage{natbib}
\usepackage[onehalfspacing]{setspace}

\usepackage{mweights}
\usepackage{lmodern}

      \makeatletter
       \edef\originalbmathcode{%
           \noexpand\mathchardef\noexpand\@tempa\the\mathcode`\(\relax}
       \def\resetMathstrut@{%
         \setbox\z@\hbox{%
           \originalbmathcode
          \def\@tempb##1"##2##3{\the\textfont"##3\char"}%
           \expandafter\@tempb\meaning\@tempa \relax
         }%
         \ht\Mathstrutbox@\ht\z@ \dp\Mathstrutbox@\dp\z@
       }
       \makeatother

 \makeatletter\let\expandableinput\@@input\makeatother

\usepackage{siunitx}
\begin{document}

\begin{titlepage}

\title{\textbf{Early life exposure to measles and later-life outcomes: Evidence from the introduction of a vaccine\thanks{We gratefully acknowledge funding of this project from the European Research Council (ERC) under the European Union's Horizon 2020 research and innovation programme, grant agreement no. 851725. We thank Samuel Baker for his extensive work on digitizing disease notifications from the Registrar General Weekly Returns. We thank Pietro Biroli, Hans van Kippersluis, Fleur Meddens, Dilnoza Muslimova, Niels Rietveld and Denny V{\aa}ger\"o for helpful discussions, and Dilnoza Muslimova for constructing the polygenic scores. We also thank seminar participants at the European Social Science Genomics Network Conference, the GEIGHEI meetings and the University of Bristol for their helpful comments and suggestions. We gratefully acknowledge employees and participants of the 23andMe, Inc. cohort for sharing GWAS summary statistics for educational attainment.}}}
\author{Gerard J. van den Berg\thanks{University of Groningen, University Medical Center Groningen, IFAU, IZA, ZEW, CEPR and CESIfo.} 
\and
Stephanie von Hinke\thanks{Corresponding author; University of Bristol, Institute for Fiscal Studies.}
\and 
Nicolai Vitt\thanks{University of Bristol.}}

\maketitle
\thispagestyle{empty}

\bigskip

\onehalfspacing

\begin{abstract}
\noindent Until the mid 1960s, the UK experienced regular measles epidemics, with the vast majority of children being infected in early childhood. The introduction of a measles vaccine substantially reduced its incidence. The first part of this paper examines the long-term human capital and health effects of this change in the early childhood disease environment. The second part investigates interactions between the vaccination campaign and individuals' endowments as captured using molecular genetic data, shedding light on complementarities between public health investments and individual endowments. We use two identification approaches, based on the nationwide introduction of the vaccine in 1968 and local vaccination trials in 1966. Our results show that exposure to the vaccination in early childhood positively affects adult height, but only among those with high genetic endowments for height. We find no effects on years of education; neither a direct effect, nor evidence of complementarities.  
\end{abstract}

\strut


\strut


\pagebreak

\end{titlepage}

\doublespacing

\section{Introduction} \label{intro}
Before the discovery of antibiotics, infections such as pneumonia and tuberculosis had high mortality rates, with up to 70 annual deaths per 100,000 population in the UK \citep{registrar_gen_reviews}. The symptoms associated with such infections were severe, and could last anywhere from one month to three years \citep[see e.g.,][]{Kipple1993, tiemersma2011natural, asthma2022}. In light of the evidence on developmental origins of later life outcomes \citep[see e.g.,][]{almond2018childhood}, it is therefore perhaps not surprising that exposure to these infections has been shown to impact individuals' health and human capital in older age \citep[see e.g.,][]{Bhalotra2015, Butikofer2020}. The potential long-term effects of less severe infections, however, are less well known. On the one hand, infections such as measles may have fewer long-term consequences due to the lower severity of the disease as compared to pneumonia and tuberculosis. On the other hand, the inflammatory response that is associated with such infections may still divert substantial nutritional resources from child development to survival and it may program the bodily infrastructure in a way that is disadvantageous at high ages. With the recent rise in measles cases globally, it is crucial to better understand the potential long-term effects of such infections as well as the long-term impact of the vaccination campaigns against it. Similarly, in the aftermath of the Covid-19 pandemic, there is an increasing awareness of the importance of long-run health distortions due to viral diseases. 

In the first part of this paper, we investigate the extent to which exposure to measles in early childhood affects individuals in older age. We focus on two key outcomes capturing individuals' accumulation of human capital and health: years of education and height in adulthood. To account for the endogeneity of measles exposure, we use two identification approaches. Both exploit a quasi-experiment in the form of a vaccination campaign. They were characterized by different implementation strategies and were introduced at different times, and hence the identification approaches rely on somewhat different assumptions.

First, we exploit the nationwide introduction of the measles vaccine in the UK in 1968, which, as we show, led to a sharp drop in measles cases. The identification approach relies on regional variation in exposure to measles prior to the vaccine introduction and cohort differences in vaccine exposure. This approach is similar to that used in, e.g., \citet{Bleakley2007} and \citet{Atwood2022}. 
Second, we exploit a relatively small-scale `blanketing trial' in 1966/67, during which the UK Medical Research Council (MRC) offered vaccination to all susceptible and eligible children in a number of local geographic regions in England and Wales. We compare outcomes of eligible and ineligible children (where eligibility is based on their year-month of birth) born in `treated' districts (i.e., those that were included in the trial) to those in similar control districts using a difference-in-difference design. 

The second part of this paper builds on the literature highlighting that individuals' later life human capital and health outcomes are not only shaped by their early life circumstances (e.g., their disease environment; `nurture'), but also by their genetic endowment to the outcomes of interest (i.e., `nature'), as well as the `nature-nurture' interplay. Some recent evidence suggests that genetic endowments may moderate the effect of environmental circumstances \citep[see e.g.,][]{muslimova2020dynamic, pereira2020interplay, von2022long, berg2022rationing}. Hence, we explore the extent to which individuals' genetic endowments for education and height may exacerbate or alleviate the effects of early life disease exposure on these outcomes.

In addition to improving our understanding of how nature and nurture interact to shape individuals' later life outcomes, estimating such gene-environment ($G \times E$) interplay is informative about key features of the human capital production function. The literature on human capital production emphasizes the importance of \textit{complementarities} between investments and endowments: investments are more productive for individuals with higher endowments \citep[see e.g.,][]{becker1986human, cunha2007technology}. Modelling the nature-nurture interplay speaks directly to this literature \citep[see e.g.][]{muslimova2020dynamic, muslimova2022diffusion}. We interpret the introduction of the measles vaccination campaign as an exogenous public investment that improved the childhood health environment. In addition to estimating the direct effects of this investment on individuals' later life human capital and health outcomes, any complementarities between investments and endowments would then be reflected by a positive $G \times E$ interaction effect; by larger effects of the vaccination campaign among those with higher genetic endowments.

We use the UK Biobank, a prospective, population-based cohort that contains detailed information on the health and well-being of approximately 500,000 individuals in the United Kingdom. Using information on individuals' location and year-month of birth, we merge these data with monthly local area-level measles infection rates. We obtain the latter by collecting and digitizing (weekly) local area-level ($n=1472$) disease notifications for the years 1941--1974 from the Registrar General reports for England and Wales \citep{RG, Baker2021}, combined with annual local area-level population estimates \citep{registrar_gen_reviews} from the Great Britain Historical Database \citep{GBHD}, allowing us to construct monthly measles infection rates between 1941 and 1974. Measles has been a notifiable disease in England and Wales since 1940, meaning that medical practitioners have been required to report any cases to their local authority within three days of occurrence. In sum, the data of measles cases and outbreaks have a high frequency and are highly reliable. 

Not surprisingly, both quasi-experiments -- the 1968 nationwide measles vaccination campaign and the 1966/67 blanketing trials -- reveal that the introduction of measles vaccination dramatically reduced measles infections. We find no evidence though, that this affected individuals' human capital or health accumulation, as proxied by their years of education and height in adulthood. However, the average effects for height conceal underlying heterogeneity with respect to individuals' genetic endowments. Indeed, we present evidence of complementarities between public health investments and endowments for height, with measles vaccinations leading to increased height for those with higher genetic endowments. We find no such evidence for education.  

Our paper speaks to at least three literatures, alluded to above. We now discuss connections to studies in those literatures in more detail. First, consider the literature on the long-term consequences of early-life exposure to infectious disease. This generally shows negative long-term effects, including losses in education, earnings, longevity, and health.\footnote{See e.g., \cite{bengtsson2000childhood, bengtsson2003airborne, almond20061918, kelly2011scourge, quaranta2013scarred, Bhalotra2015, Butikofer2020, daysal2021germs}. Similarly, \citet{kelly2011scourge} and \citet{schwandt2018lasting} examine the effect of influenza exposure, and \citet{mosca2022long} investigate the effect of exposure to rubella, though their focus is on the prenatal period, and hence on \textit{maternal} exposure to disease.} A set of papers concurrent to ours has focused specifically on the long-term effects of exposure to measles. These studies exploit cross-state differences in pre-vaccination measles rates and the differential vaccine exposure driven by year of birth in a difference-in-differences design to estimate the long-term effects of the measles vaccine in the United States \citep{Atwood2022, barteska2022mass, chuard2022economic} and Mexico \citep{AtwoodPearlman2022}.\footnote{We are aware of four further studies that explore the longer-term effects of measles infections. \citet{barker1991relation} show no relationship between contracting measles in infancy and adult lung function, whereas \citet{barker1986childhood} find a positive correlation using regional data between infant mortality rates from measles and mortality from lung disease 70 years later. \citet{nandi2019anthropometric} conclude that measles-vaccinated children are taller, have better test scores and more years of education, though their focus is on late childhood rather than adulthood. Finally \citet{driessen2015effect} use the staggered roll-out of the vaccination campaign in Bangladesh and find that vaccination increased boys' school enrolment. Other than the latter, however, these studies do not account for the possible endogeneity of measles exposure.} They find that lifetime exposure to the vaccine increases education, employment as well as earnings and improves health outcomes. 
In contrast, we find no evidence of effects on individuals' human capital and health outcomes. There are multiple potential reasons for these conflicting findings. First, the setting of our paper differs substantially from the setting in the concurrent measles literature. For example, the take-up of the measles vaccine was near universal in the United States and Mexico, with measles rates dropping to almost zero in the years after the vaccine introduction. In contrast, measles rates in England and Wales did not decrease to such low levels. This suggests that the studies examining the US and Mexico estimate the effect of \textit{no} versus \textit{full} vaccine compliance, whereas we estimate the effect of \textit{no} versus \textit{partial} compliance. 
Similarly, as illustrated by the widely different effect sizes found for Mexico compared to the US \citep{AtwoodPearlman2022}, the economic and policy setting matters for the long-term effect estimates of measles vaccination. Hence, the distinctive features of the UK's labour market, welfare, education and health systems may further help explain the differences between our results and those observed for the United States and Mexico.
Finally, the analysis in our setting suggests that our null results can be explained by differential area-specific trends in the outcomes of interest in England and Wales. We illustrate this in two ways. First, we show that taking into account the differential effects of government policy\footnote{Specifically, the increase in minimum school leaving age from 15 to 16 years for those born from September 1957 onwards.} on areas with high and low pre-vaccination measles rates reduces the human capital estimates and renders them insignificantly different from zero. Second, when we control for local area-specific time trends, the treatment effect attenuates to zero and turns insignificant. 

A second literature that we speak to is that on the importance of $G \times E$ interplay. Although $G \times E$ interaction studies are not new, they historically did not take into account the endogeneity of the environment, nor the endogeneity of one's genotype. The geographical environment may be endogenous as the disease exposure in local areas may be correlated to the area's socio-economic conditions, and individuals may select into their environments based on unobserved characteristics. Individual genetic variation may be endogenous since it is inherited from one's parents, whose genetic endowments can shape the offspring environment \citep[so-called `genetic nurture', see e.g.][]{kong2018nature}. 

To deal with endogeneity of the disease environment, we exploit exogenous variation in disease exposure from the national vaccination campaign. Additionally including family fixed effects (or controlling for the parental genotypes) would account for the endogeneity of genetic variation \citep[for a detailed discussion of $G \times E$ estimation, including the endogeneity of the environment/genetic variation, see][]{biroli2022}. However, observing genetic data for multiple family members in combination with exogenous variation in environments for a large sample of individuals is currently rare. Our baseline specifications do not deal with the potential endogeneity of the genotype. The implication of this is that we may over-estimate the genetic effect as its coefficient may partially reflect the environments that are shaped by the parents. In an additional analysis, however, we use the smaller subsample of siblings from the UK Biobank to identify the causal \textit{genetic}, \textit{environmental}, as well as $G \times E$ effect, estimating causal complementarities between endowments and investments.

A third strand of literature that our paper speaks to is on health and human capital formation \cite[see e.g.][]{becker1986human, cunha2007technology}. In this realm, we mention \citet{muslimova2020dynamic} who explore the complementarity between genetic endowments and birth order within families, the latter proxying for parental time investments. Rather than a measure of \textit{parental} investments, we use the introduction of vaccinations as \textit{public} investments in the health of children. This allows us to explore whether improvements in children's health environment interact with individual genetic endowments, providing direct estimates of the complementarity between endowments and public investments.

The rest of this paper is structured as follows. The next section provides the background on measles and the introduction of the measles vaccine in the UK. \autoref{sec:data} describes the data and provides the descriptive statistics, and \autoref{sec:methods} outlines the empirical approach. We discuss the results in \autoref{sec:results}, and report the robustness analysis in \autoref{sec:robustness}. \autoref{sec:conclusion} concludes.

\section{Measles and the vaccine introduction in the UK}
\label{sec:background}

\subsection{Measles}
Measles is a highly infectious viral disease that spreads through water droplets in the air and direct contact. Symptoms include fever, a cough, inflamed eyes, cold-like symptoms and a typical skin rash. It can cause potentially fatal complications, including infections of the lung (pneumonia), brain (encephalitis) and severe diarrhoea. 
While measles still has a high mortality rate in the developing world, parents and physicians in Western Europe in the 1960s generally considered it an unpleasant but inevitable stage of a child's development. Most cases were thought to be uncomplicated, with children recovering within ten days of the appearance of the rash, followed by lifelong immunity. In an average epidemic year, over half a million cases were reported in England and Wales alone, with 7\% ($\sim$35,000) having serious complications and 1.2\% ($\sim$6,000) being hospitalized \citep{hendriks2013measles}. Going further back to the late 1800s however, the measles mortality rate could be as high as 5-10\% among malnourished children \citep{mackenbach2020history}, mainly among those in infancy \citep{woods2000demography}. Its demise as a perceived major health hazard started in the first decades of the 20th century. Although incidence rates remained at similar levels, case fatality declined due to reduced crowding and improved nutrition. The incidence started to decline only with the introduction of mass vaccination \citep{mackenbach2020history}. 
Although there has historically not been any systematic data collection on age-specific measles infection rates in the UK, the evidence suggests that -- prior to the introduction of the measles vaccine in the UK -- approximately 50-60\% of measles cases occurred in children aged 1 to 4 years \citep{registrar_gen_reviews, woods2000demography}. More detailed analysis of the age profile of UK measles cases at the end of the 19th century shows almost 90\% of cases to occur in children aged 1 to 6 years \citep{woods2000demography}.

Recent research has highlighted extended detrimental effects on the immune system of the inflammatory response to a measles infection \citep[see e.g.,][]{Gadroen2018, Mina2019, Petrova2019}. The immunosuppression associated with a measles infection diminishes the immune memory for other pathogens\footnote{This is called `immune amnesia', referring to the loss of previously acquired immune memory cells due to a measles infection \citep{Mina2015, Mina2019}.} and this can increase the vulnerability to other diseases for up to three years after a measles infection \citep{Mina2015}. 
Some of those diseases may have their own long-run effects on the outcomes we consider. Moreover, children may lose time at school in the first few years after measles because of sickness spells and this may be a mediator for our outcomes.  
The mirror image of these pathways is that measles vaccinations lead to reductions in diseases in the years after vaccination and reductions in the ensuing other adversities.\footnote{There is evidence that infections with the measles virus can kill a range of cancer cell types, e.g. in leukaemia. In line with this, the measles vaccine has been used in therapies against ovarian cancer, myeloma and cutaneous non-Hodgkin lymphoma \citep[see][for a review]{Aref2016}. However these are rare diseases and their occurrence should not quantitatively affect our findings.}

\subsection{Measles in England and Wales}
\begin{figure}[h]
   \caption{Measles notification in England and Wales, 1941-1974}
   \label{figure_weekly_measles} 
     {\centering \includegraphics[scale=0.55, trim=5 15 5 5, clip]{"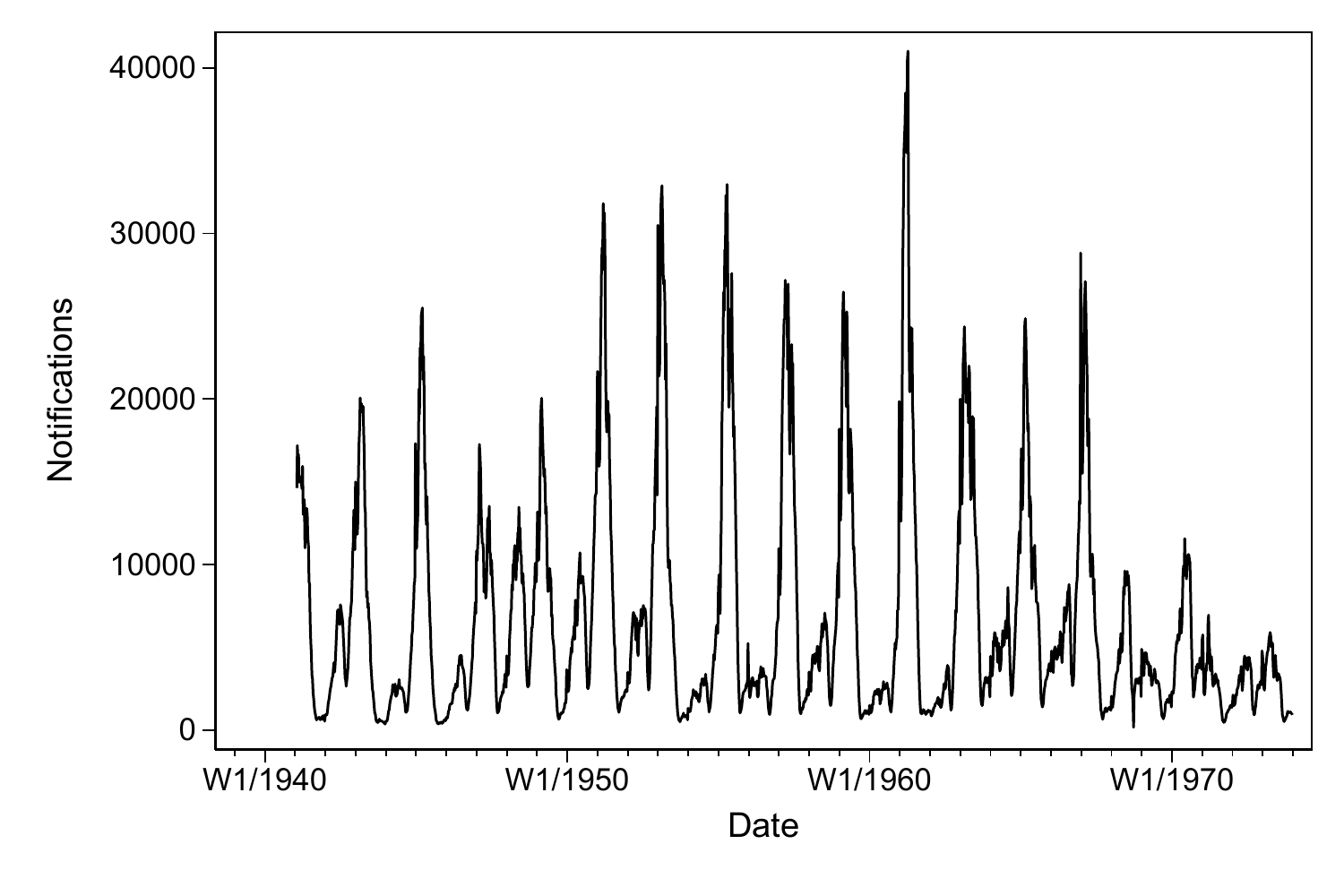"}
     \includegraphics[scale=0.55, trim=5 15 5 5, clip]{"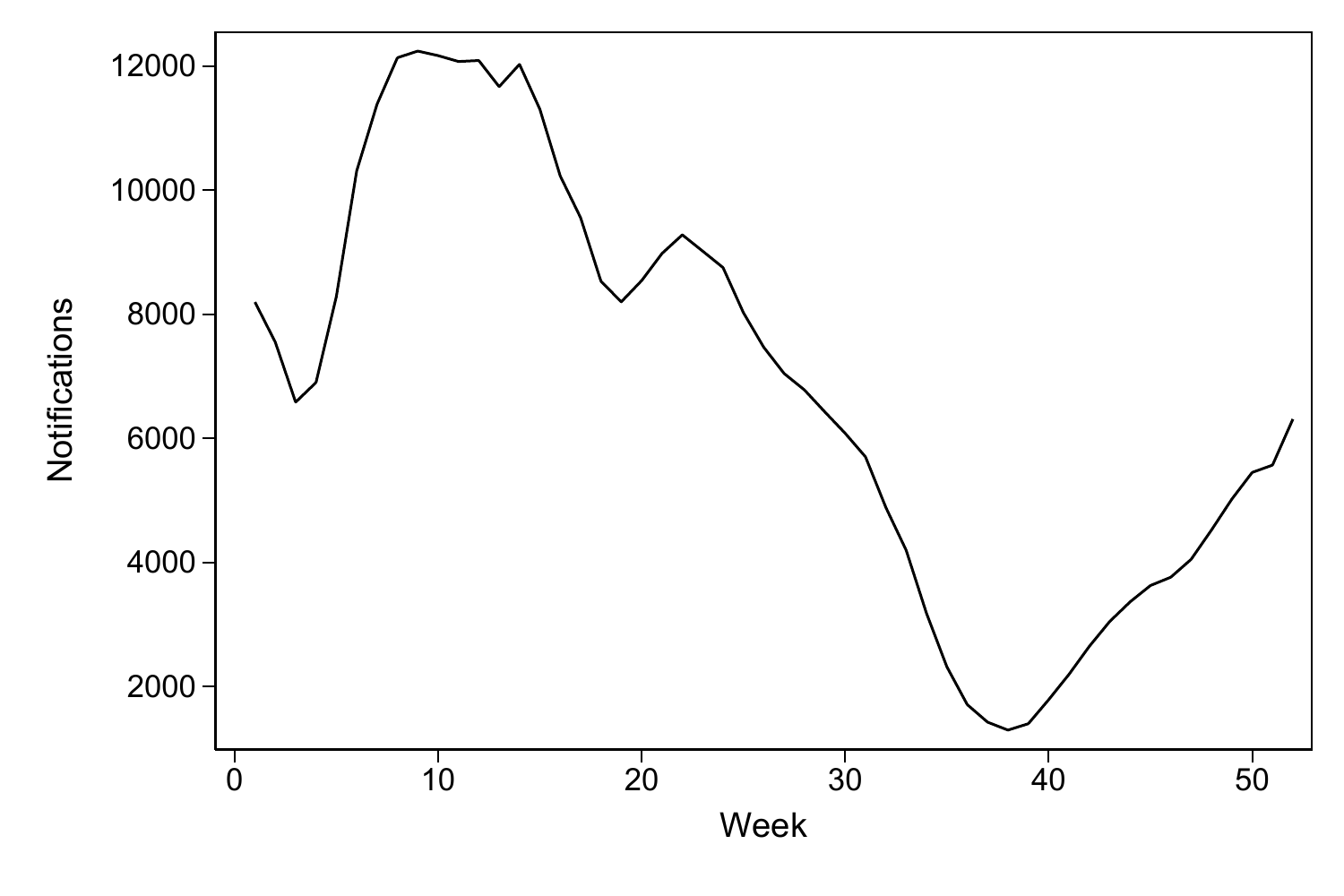"} \par}
     {\scriptsize \singlespacing Note: The left panel shows the weekly measles notifications between 1941 and 1974 in England and Wales. The right panel presents average measles notifications between 1941 and 1974 for each calendar week of the year. The data are obtained from the Registrar General weekly reports on regional disease notifications \citep{RG}.\par}
\end{figure}

Prior to vaccination, measles incidence in England and Wales followed regular cycles. To secure a chain of infection, measles needs a sufficiently large susceptible population. This population is depleted through measles infections, but is continuously replenished through new births. In England and Wales, this translated into highly regular two-year cycles, with a high-incidence year followed by a low-incidence year \citep{anderson1984oscillatory}. The left panel of \autoref{figure_weekly_measles} shows this regularity by plotting weekly cases in England and Wales between 1941 and 1974. This bi-annual pattern was only interrupted in 1946--1950, when an increased birth rate led to a faster replenishment of the susceptible population, causing major annual outbreaks. 

Another important driver of measles is the mixing of children. As such, measles incidence shows a clear seasonal pattern, starting at the beginning of the schoolyear and peaking after the Christmas and Easter breaks \citep{fine1982measles}. This is illustrated in the right panel of \autoref{figure_weekly_measles}, showing relatively low measles incidence during school holidays and during the warmer summer months when less indoor mixing of children takes place, and an increase from September onwards, with the start of the school year.

Population density and family size are other important drivers of measles infections. Densely populated areas, cramped housing conditions and increased family sizes encourage the spread of measles \citep{mackenbach2020history}. For example, older siblings are likely to bring viruses home, infecting their younger siblings at earlier ages than they would have become infected themselves, at which age the younger siblings are more vulnerable \citep{daysal2021germs}. Consistent with this, the reduction in family size in the 1900s led to children being infected at higher ages, when they were less vulnerable, contributing to the reduction in the measles fatality rate \citep{mackenbach2020history}. 

\subsection{The vaccine introduction}
During the early 1960s, measles vaccines were developed and successfully trialled worldwide. Although the first two vaccines in the US were approved in 1963 \citep{hendriks2013measles}, the process was slower in the UK. The UK MRC started a series of small trials in 1964-65 that explored different vaccination protocols and their effectiveness relative to a control group \citep{MRCtrialreport1966,MRCtrialreport1968}. Subsequently, so-called `blanketing trials' in 1966-67 aimed to measure the effects of large-scale vaccination in eight selected areas \citep{Black2017, JCVI1965Jul, JCVI1966Oct, Warin1967, Warin1968}.\footnote{Details of the trial implementation can be found in the annual reports of the Medical Officers of Health \citep{MOHreportBedford1966, MOHreportBedford1967,MOHreportBristol1966, MOHreportBristol1967,MOHreportCardiff1966, MOHreportCardiff1967,MOHreportHull1966, MOHreportHull1967,MOHreportLeicestershire1966, MOHreportLeicestershire1967,MOHreportNewcastle1966, MOHreportNewcastle1967,MOHreportOxford1966, MOHreportOxford1967,MOHreportSouthampton1966, MOHreportSouthampton1967}.} 
Vaccination was offered to all susceptible children (i.e., children without prior immunity from previously contracting measles) aged between 10/18 months and 10/12 years in Bedford, Kingston Upon Hull, Newcastle upon Tyne, and Oxford.\footnote{The age windows differed between areas, with e.g. some areas vaccinating children between 10 months and 12 years while others vaccinated children between 18 months and 10 years.} In Bristol, Cardiff, Southampton and Leicestershire (excl. the city of Leicester), vaccination was offered to all susceptible children aged between 10/12 months and 2 years.
 
Following these trials, the Joint Committee on Vaccination and Immunisation (JCVI) recommended in November 1967 to offer the measles vaccine to all susceptible children over the age of one \citep{JCVI1967Nov}. Vaccinations began in May 1968 \citep{JCVI1968Jul}. Due to limited supplies, however, the vaccine was initially only offered to susceptible children aged 4 to 6 years and to younger children attending day nurseries, nursery schools or living in residential establishments. During the late summer, supply of the vaccine was sufficient to offer vaccination to other susceptible children over the age of 1 year.\footnote{The vaccines are known to have side effects. The most common are fever and a rash, about one to two weeks after the vaccination. These are directly attributable to the measles component in the vaccine; see e.g. \cite{Gastanaduy2021}. Other well-documented side effects are extremely rare and most do not seem to have long-run consequences \citep{Patja2000}.}

While precise vaccination statistics at the regional or even national level are not available, the JCVI estimated that over 700,000 children were vaccinated in England and Wales by the end of 1968 \citep{JCVI1969May}.\footnote{\autoref{figure_vaccination_rate_timeseries}, \autoref{appendix_tables_figures}, shows measles vaccination rates for two-year-olds from 1971 onwards. This shows that, throughout the 1970s, take-up of the routine vaccination within the recommended schedule was around 50\%.}

\section{Data and descriptive statistics}\label{sec:data}

\subsection{Data}
We use the UK Biobank \citep{Sudlow2015}, a population-based cohort study with data on the health and well-being of approximately 500,000 individuals across the UK. Individuals aged between 40 and 69 were invited to join the data collection between 2006 and 2010. At the interviews, data was collected on participants' demographics, physical and mental health, health behaviours, well-being, cognition and personality. Furthermore, the data has been linked to hospital records, as well as the national death registry, providing detailed information on participant's medical history. Finally, trained nurses took anthropometric measures and collected blood, urine and saliva samples, allowing for the genotyping of all individuals.

A limitation of the data is that it contains little information on individuals' early life circumstances. However, the residential location at birth is recorded. We exploit this (i.e., the eastings and northings of birth) to merge-in external contextual information on the early life disease environment. Each individual is assigned to one of the 1472 Local Government Districts of birth for England and Wales. We collect and digitize weekly district-level disease notifications for the years 1941-1974 from the Registrar General's Weekly Reports \citep{RG, Baker2021}. We use the district-level annual population data from The Registrar General's Statistical Review of England and Wales \citep{registrar_gen_reviews} and digitized in the Great Britain Historical Database \citep{GBHD} to calculate measles exposure \textit{rates} as the number of cases per 100 population in a district.\footnote{To account for changes in district boundaries during the sample period, we take the 1951 boundaries as the base year and use weightGIS \citep{Baker2021weightGIS} to convert the data (using a population-weighting mechanism) to time-invariant district units.} Following the fact that 90\% of measles cases occur between age 1--6, we focus on exposure to the vaccine during this age range. The average ratio of measles notifications to lagged birth counts during the pre-vaccination period is 55-60\%, substantially below the share of children that contracted measles according to other sources \citep[see e.g.,][]{Gastanaduy2021}. As mild cases were more likely to be unnoticed or misdiagnosed, the measles notification data can be viewed as capturing variation in both the number and the severity of measles cases in a district.

For our purposes, the fact that the data only includes individuals born between 1934 and 1971 (with smaller samples in the early (1930s) and later (1970s) cohorts) constitutes a second limitation of the data for studying the introduction of the vaccine in 1968. Due to different recruitment dates at the assessment centres where the initial interviews were conducted, some areas are over- or under-represented in the early and late cohorts (compared to the middle cohorts). For this reason we restrict our sample to only include individuals born up to August 1969, ensuring similar coverage of individuals across England and Wales. In addition, high birth rates in the late 1940s lead to somewhat unusual measles dynamics and may have affected our outcomes of interest (e.g. due to larger class sizes). Therefore we also restrict the sample to cohorts born from September 1949 onwards.

We focus on two outcomes of interest, broadly capturing the accumulation of human capital and health. First, we explore the effects of early life disease exposure on one's years of education. In the absence of direct data on years of education, we follow the literature \citep[see e.g.,][]{Rietveld2013,Okbay2016,lee2018gene,okbay2022polygenic} and define this based on individuals' qualifications.\footnote{See \autoref{table_education_mapping}, \autoref{appendix_tables_figures}, for the mapping we use to derive years of education from individuals' highest qualifications.} To allow us to investigate potential non-linear effects, we also explore binary outcomes indicating whether individuals have any qualification, upper secondary qualifications, or a degree. Second, we explore the effects on adult height. This is often used as a proxy for health outcomes later in life
insofar as the latter are affected by conditions before adulthood \citep[see][for an overview]{vandenberg2014}). During the UK Biobank interviews, participants' standing height in centimetres was measured following a standardized protocol.

To study the complementarity between investments and endowments, we explore interactions between the vaccination programme and individuals' genetic endowment for the relevant outcome of interest. For the latter, we use molecular genetic data to construct so-called polygenic indices (PGIs, also known as polygenic scores) for education and height. The PGIs combine information on genetic variation across $>1$ million locations of the genome using weights derived in genome-wide association studies (GWAS) of educational attainment \citep{lee2018gene} and height \citep{Wood2014GIANT}.\footnote{For more information on the genetic terms used here, see \autoref{appendix_genetics}.} In our sensitivity analysis, we explore the robustness of our findings to the use of differently constructed PGIs.

The full UK Biobank includes over 500,000 participants. Our sample selection is as follows. First, we drop individuals for whom we cannot identify their district of birth (mainly due to missing eastings and northings), or who were born in Scotland or Northern Ireland, since we do not observe their disease notifications. Second, we restrict the sample to those born between September 1949 and August 1969, leaving us with 215,538 participants. Third, we follow the genetics literature and drop those who are of non-European ancestry to ensure a more genetically homogeneous sample, resulting in a sample of 187,078 individuals. For the analysis of the nationwide introduction of the measles vaccine in 1968, we additionally exclude individuals born in districts that participated in the 1966 blanketing trial. This leaves a final sample size of 171,685 for these estimations.

For the analysis of the blanketing trial of the measles vaccine in 1966, we focus on those trial districts which targeted susceptible children up to age 10 or 12. Individuals born in trial districts targeting children up to age 2 are excluded from the sample. To ensure a comparison of the trial districts with similar control districts, we furthermore restrict our sample to individuals born in a district with a population density within one standard deviation of the mean population density among the trial districts. The final sample size for the analysis of the 1966 blanketing trial is 96,315. This includes 5328 individuals in treated districts, of whom 1739 are potentially exposed to at least one month of the trial (based on their year-month of birth), with only 891 individuals potentially exposed for the full two year period. Note that the actual number of vaccine recipients will be lower, since take-up was not universal and individuals with prior immunity would not have received the vaccine. The small number of treated individuals implies that the blanketing trial analyses have relatively low power.

\subsection{Descriptive statistics}

The summary statistics of the variables used in our analyses are presented in \autoref{table_descriptives}. This shows that, on average, there were 0.86 measles notifications at age 1--6 per 100 population.\footnote{Using participants' year-month and district of birth, we construct, for each individual and each year of childhood, their annual measles exposure rate, assuming births occur mid-month. For example, for an individual born in February 1962, their exposure rate in the first year of life is defined as $\frac{\sum_{w=1}^{52} m_{wd}}{population_{dt}}$, where the numerator is the sum of the 52 weekly ($w$) notifications in district $d$ starting in mid February 1962, and the denominator is the population of district $d$ in year $t$. We then report the average rate across ages 1--6.} Participants in our sample have 13.4 years of education on average, with 91\% having any qualification, 72\% an upper secondary qualification, and 35\% a degree. The average height is 170cm, and 55\% of our sample is female.

%==================BEGIN TABLE=================%
\begin{table}[tb!]\centering\small
  \begin{threeparttable}
    \caption{\label{table_descriptives}Descriptive statistics} %%TABLE TITLE
    
                       \begin{tabular*}{\linewidth}{@{\hskip\tabcolsep\extracolsep\fill}l*{5}{S[table-format=4.2  ] S[table-format=1.2  ] S[table-format=4.2  ] S[table-format=4.2  ]  S[table-format=3.3  ]}}
                       \toprule
                       \input{"tables/descriptives_tablefragment.tex"}
                       \bottomrule
                       \end{tabular*}
   
  \end{threeparttable}
\end{table}
%==================END TABLE=================

The PGIs for educational attainment and height explain a substantial share of the variation in these outcomes. After accounting for year of birth and gender, the PGI for education explains 8.56\% of the remaining variation in years of education.\footnote{We follow the approach in \citet{Becker2021} in the calculations of the incremental $R^2$. We residualize each outcome on a third-order polynomial of year of birth, sex and their interactions. The residuals are then regressed on the first 20 genetic principal components, genotyping array fixed effects and the polygenic index to obtain the incremental $R^2$.} The PGI for height explains 21.21\% of the remaining variation in height.

\begin{figure}[tb!]
	\captionsetup[subfigure]{aboveskip=0pt,belowskip=-1.5pt}
   	\caption{Average years of education and height, by genetic propensity}
   	\label{figure_educ_height_histograms_by_pgigroup} 
     {\centering 
		\begin{subfigure}{0.475\linewidth}
		\caption{Years of education}
     	\includegraphics[width=\linewidth, trim=10 10 10 10, clip]{"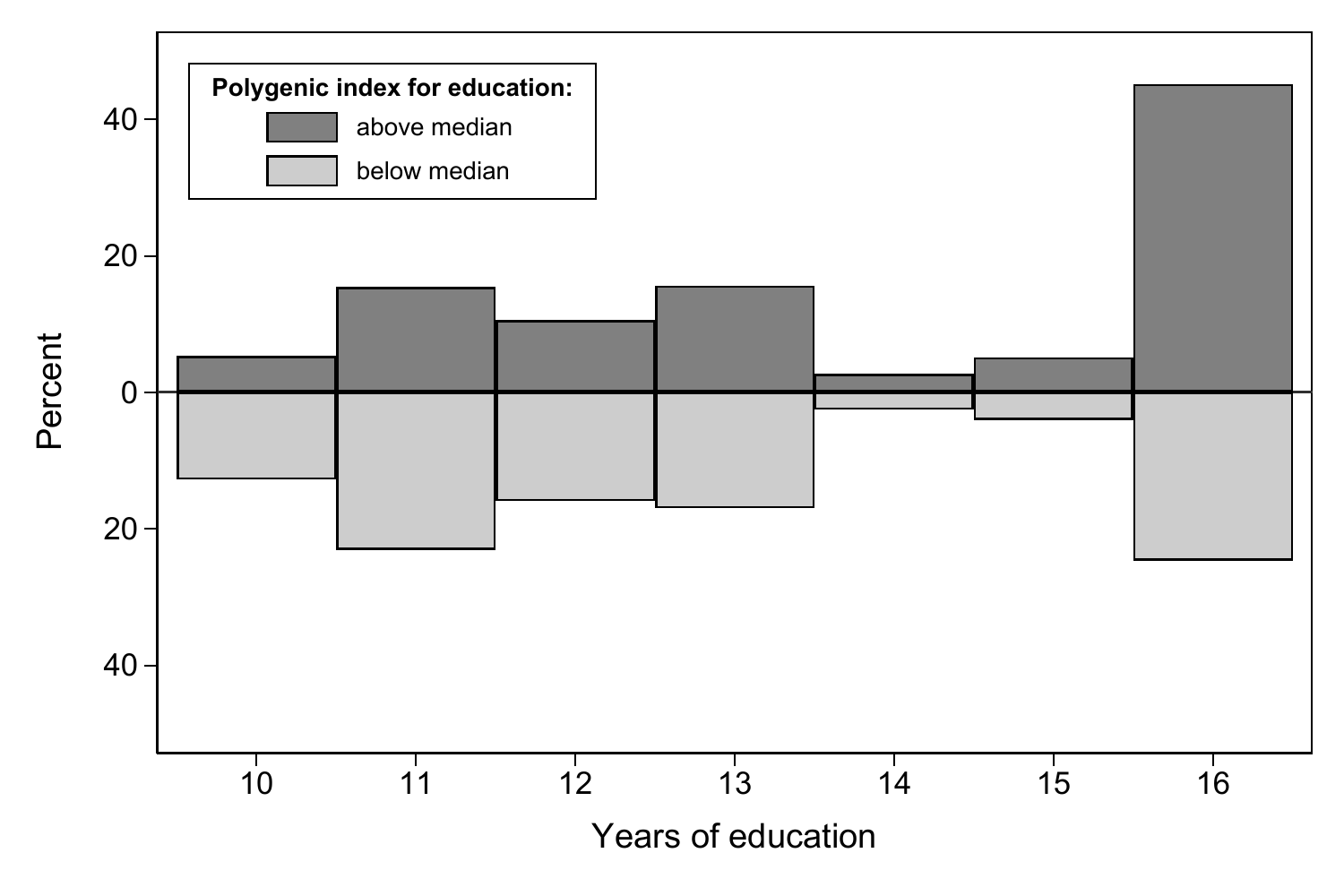"}
		\end{subfigure}   
		\hfill
		\begin{subfigure}{0.475\linewidth}
		\caption{Height}  
		\includegraphics[width=\linewidth, trim=10 10 10 10, clip]{"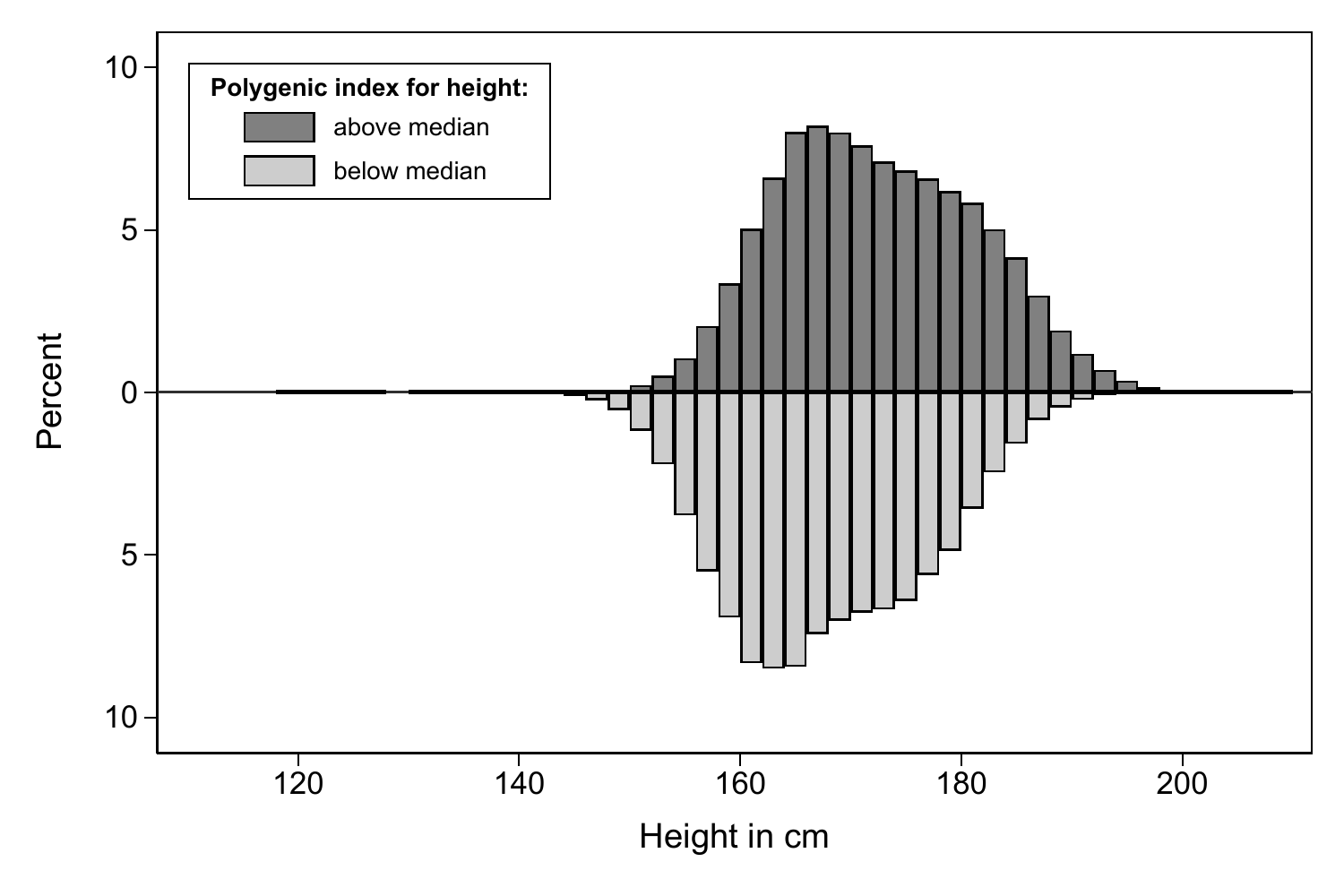"}
		\end{subfigure} \par}
     {\scriptsize \singlespacing Note: The figures show the distributions of years of education and height in centimetres in the sub-samples with above- / below-median genetic propensity for the respective trait. \par}
\end{figure}

\autoref{figure_educ_height_histograms_by_pgigroup} shows a comparison of the education and height distributions across the sub-samples with above-median and below-median polygenic indices. The sample with an above-median PGI for education are substantially more likely to have completed 16 years of education (45.1\% compared to 24.7\% in the below-median sample) and less likely to have left education after 10 or 11 years (20.8\% compared to 35.8\% in the below-median sample). On average, those with an above-median polygenic index obtained 1.07 additional years of education. The average adult height in the sub-sample with an above-median PGI for height is 171.95 cm compared to 167.28 cm in the below-median sample. To put this into perspective: this difference of 4.67 cm is equivalent to approximately 35\% of the gender gap in height.

\section{Empirical strategy}\label{sec:methods}
Our empirical strategy exploits two approaches with different identifying assumptions. Both exploit a quasi-experiment in the form of a vaccination campaign, though these were characterized by different implementation strategies and introduced at different times. This, in turn, allows us to explore them separately and examine the robustness of our estimates of the long-term effects of exposure to measles in early life. Our first identification strategy exploits the nationwide introduction of the measles vaccine in 1968. Our second identification strategy uses the smaller blanketing trials of 1966/67 that were introduced in some areas, but not others. We discuss our empirical approach, as well as the implicit assumptions of each strategy in turn below. 

\subsection{Vaccine introduction in 1968}\label{sec:method_vacc_introduction}
We follow an identification strategy similar to, e.g., \cite{Bleakley2007}, \cite{Butikofer2020} and \cite{Atwood2022}. More specifically, we consider whether the benefits of the vaccine introduction were stronger for districts with previously high rates of measles infections as they had a larger potential for measles reductions compared to districts with low rates. We estimate the following specification: 

\begin{align}\label{eq:vaccine2}
Y_{idc} &= \alpha + \sum_{a} \beta_a Post_{age_i=a} \times PreRate_{d} + \mathbf{X_{i}'} \boldsymbol{\psi} + \gamma_d + \delta_{county_{d}} \times c + \lambda_c + u_{idc},
\end{align}
where $Y_{idc}$ denotes either the education (in years) or the height (in cm) for individual $i$ born in district $d$ in cohort $c$. The variable $Post_{age_i=a} \times PreRate_{d}$ denotes the interaction between the share of the relevant age range in which individual $i$ was exposed to vaccination ($Post_{age_i=a}$) and the pre-vaccination measles infection rate in district $d$ ($PreRate_{d}$). These are the key covariates. We now define their components more precisely. $PreRate_{d}$ is the infection rate (per 100 population) in district $d$ between September 1950 and August 1960,\footnote{In a sensitivity analysis, we show that the results are robust to using other time windows for the pre-vaccination infection rate. This is important because the number of peaks in the local infection rate may vary across different time windows.} 
and we call this 
the ``treatment intensity''. As we have seen, variation in $PreRate_{d}$ around the average national value may reflect variation in the \textit{number} as well as the \textit{severity} of measles cases in the district. However, it is well known that before vaccinations, infection with measles was nearly universal during childhood \citep[see e.g.,][]{Gastanaduy2021}. This suggests that the actual prevalence of measles is not a prime driver of variation in $PreRate_{d}$ across districts, or at least that this is dominated by variation in the severity or the reporting of severe cases. The severity may be affected by contextual indicators such as local health care provision, weather conditions or the general socio-economic environment in the district. Later in this subsection we provide some empirical evidence of this and in Subsection \ref{ss:6.3} we return to this. We emphasize that the above equation contains district-specific fixed effects $\gamma_d$, controlling for different conditions across districts throughout the full observation window.

The other key covariate, $Post_{age_i=a}$, denotes the degree to which individual $i$ was exposed at age $a$ to the post-vaccination period.\footnote{With generally low infection rates during the summer months and the annual cycle of measles following the schoolyear, we define September 1968 as the start of the post-vaccination period. This corresponds to the start of the vaccinations campaign with sufficient vaccine supply.} Since the majority of measles cases occurred in children aged between 1--6 \citep{woods2000demography}, we focus on exposure to measles in this age range. In our main analysis, it is defined as the share of the relevant age period during which the individual was exposed to the vaccination program. \autoref{figure_postvacc_share}, \autoref{appendix_tables_figures}, illustrates how the post-vaccination share for ages 1 to 6 is computed based on individuals' date of birth.\footnote{Babies born to mothers with measles antibodies are protected in their first year of life, hence we do not consider exposure at age zero. Anyone born prior to September 1961 turned 7 before the vaccination was introduced and therefore is assigned a share of 0. Anyone born after September 1967 was exposed to the post-vaccination period throughout the age range of 1--6 years and is therefore assigned a share of 1. For those born between September 1961 and September 1967, the share increases linearly in the date of birth, reflecting the increasing exposure to the post-vaccination period. In our sensitivity checks, we show our results are robust to the use of a binary indicator for any exposure to the post-vaccination period (i.e., for at least one month).}  
If contracting measles negatively affects individuals' human capital and health outcomes, we would expect to find improvements in these outcomes for individuals in districts with higher pre-vaccination incidence rates and among cohorts with increased exposure to the vaccine.

$\mathbf{X_{i}}$ is a vector of individual-level controls, including gender and month of birth fixed effects. The variables $\delta_{county_{d}} \times c$ are county-of-birth-specific linear cohort trends, $\gamma_d$ are district of birth fixed effects, $\lambda_c$ are schoolyear of birth fixed effects,\footnote{Since the cyclical pattern of measles follows the schoolyear, we use schoolyears (i.e., the year from September of year $t$ to August of year $t+1$) rather than calendar years when accounting for birth cohort effects.} and $u_{idc}$ is an idiosyncratic error term. Note that our results are unchanged if we control for linear or quadratic trends instead of schoolyear of birth fixed effects. We report heteroskedasticity-robust standard errors, clustered by district of birth throughout, and we explore the robustness of our findings to clustering by schoolyear of birth and to two-way clustering by district and schoolyear of birth in \autoref{sec:robustness}. 

The coefficients of interest are the $\beta_a$'s, capturing the later life effect of a reduction in measles exposure \textit{at age $a$}. This captures the direct effect of a reduction in disease through two potential channels. First, a biological channel, since the reduction in infections may have mitigated potential barriers to child development, such as the diversion of nutritional resources. Furthermore, due to measles' immunosuppresive effects, a reduction in measles exposure may have reduced children's susceptibility to other infectious diseases, which in turn could have affected child development. The second channel is more `behavioural', with the decrease in infections potentially reducing school absences. Furthermore, there may have been parental responses to measles exposure, which in turn may have affected child development. 

\begin{figure}[tb!]
   \caption{Pre-vaccination measles rates across England and Wales}
   \label{figure_measles_prevacc_map} 
     {\centering \includegraphics[scale=0.5, trim=2 2 2 2, clip]{"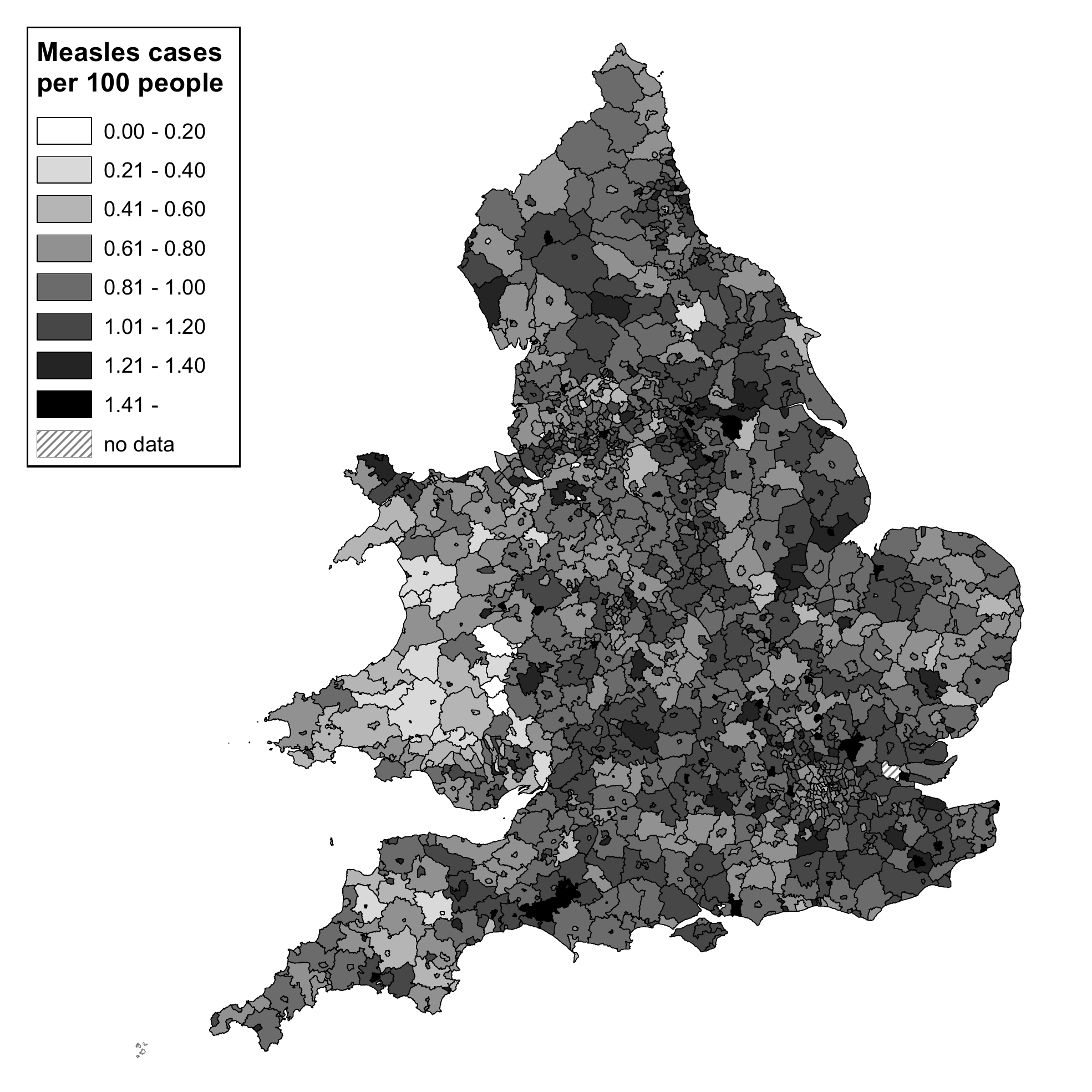"} \par}
     {\scriptsize \singlespacing Note: The maps shows the average district-level annual measles rate between September 1950 and August 1960, in cases per 100 people. \par}
\end{figure}

\begin{figure}[tb!]
   \caption{Average annual measles rates in districts with high and low pre-vaccination rates}
   \label{figure_measles_tseries_by_prevacc} 
   {\centering \includegraphics[scale=0.7, trim=15 15 0 10, clip]{"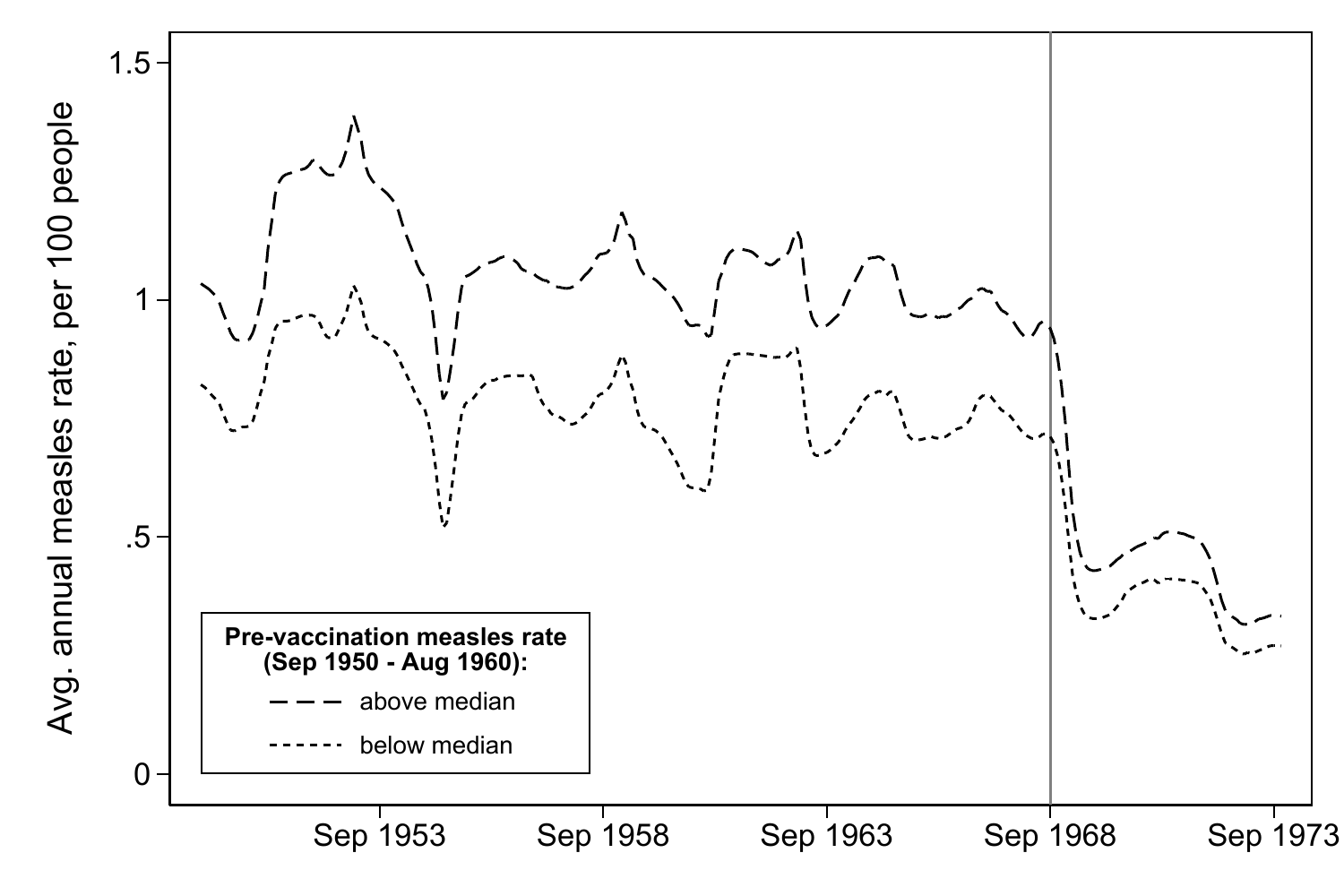"} \par}
     {\scriptsize \singlespacing Note: The grey vertical line represents the beginning of the vaccine roll-out in September 1968. Each monthly observation corresponds to the average annual measles rate (per 100 people) over the preceding 24 months. 11 out of 1472 districts were excluded due to (partially) missing data on measles cases or population size. Districts participating in the 1966 trial are excluded from the figure.\par}
\end{figure}

Identification of the coefficients of interest benefits from regional variation in measles infection rates. We show this in  \autoref{figure_measles_prevacc_map}, mapping regional average infection rates over the pre-vaccination period $1950-1960$: $PreRate_d$. This reveals large regional variation in measles infection rates across England and Wales.\footnote{\autoref{figure_measles_maps}, \autoref{appendix_tables_figures}, separately shows the annual infection rates for the years $1950-1952$ and $1960-1962$. This shows that areas with low rates in one year tend to have high rates in the subsequent year, and that this biennial pattern is not identical across the country, with some areas showing high rates of measles in years where the rest of the country is characterised by low infection rates.} \autoref{figure_measles_tseries_by_prevacc} shows that districts with high rates of measles infections benefited \textit{more} from the introduction of the vaccination campaign than districts with lower rates of infections and, with that, experienced a larger decline in cases. The vertical axis presents the two-year annual measles rates to account for the sawtooth pattern of measles infections. We see a slight downward trend in the measles rate prior to the introduction of the vaccine in 1968 for districts with both above- and below-median rates, with a sharp drop immediately after the vaccine introduction. Districts with above-median and below-median measles rates show similar trends prior to the vaccine-introduction, but following the vaccine introduction there is a larger decline for districts with previously high rates of measles.\footnote{\autoref{figure_measles_tseries_by_prevacc_decile}, \autoref{appendix_tables_figures}, shows the development of measles rates for the deciles of the pre-vaccination measles rate distribution. The larger drop in infections for districts with higher rates of measles prior to vaccination is also shown in \autoref{figure_measles_prevacc_vs_change}, \autoref{appendix_tables_figures}, plotting the reduction in measles infection rates on the vertical axis, with the horizontal axis displaying the average annual pre-vaccination measles rate. The left panel plots the values for each individual district, showing that districts with high rates of infection prior to vaccination (i.e., moving to the right on the horizontal axis) benefited more from the vaccination campaign (i.e., had larger reductions on the vertical axis) compared to districts with lower rates of infection. The panel on the right shows the distribution of pre-vaccination measles rates across districts, with the solid line representing the conditional mean of the post-vaccination change in measles rates based on a local linear regression against pre-vaccination rates. Both figures show that infection rates converged across districts after the introduction of the vaccination campaign. To better understand how districts with above- and below-median levels of measles infections compare, \autoref{table_descriptives_lowhighdistricts}, \autoref{appendix_tables_figures}, presents some descriptive statistics of the two groups, with the relevant district-level variables obtained from the 1951 Census. This shows that those with above-median infection levels are on average of slightly lower social class, living in more crowded conditions, and leaving education at earlier ages.}

\autoref{figure_measles_tseries_by_prevacc} shows that the vaccination campaign was effective in reducing measles infections. Indeed, in the absence of the vaccination campaign, we would have expected an epidemic measles year in 1968, but the introduction of measles vaccinations ensured this never materialised (see also the left hand panel of \autoref{figure_weekly_measles} and \autoref{figure_monthly_measles}, presenting weekly and monthly measles rates in England and Wales, respectively). 

We need to assume that the vaccination campaign is exogenous with respect to the prevalence of measles prior to vaccination. Historical documents highlight that the introduction of the vaccine was nationwide, rather than targeting e.g. high-prevalence areas. The campaign was funded by the government, with vaccines supplied free of charge to local authorities. GP practices and health clinics were offering vaccinations in all local authorities; schools were additionally involved in some.

\begin{figure}[tb!]
	\captionsetup[subfigure]{aboveskip=0pt,belowskip=-1.5pt}
   	\caption{Average years of education and height in districts with high and low pre-vaccination measles rates}
   	\label{figure_educ_height_tseries_by_prevacc} 
     {\centering 
		\begin{subfigure}{0.475\linewidth}
		\caption{Years of education}
     	\includegraphics[width=\linewidth, trim=15 15 10 10, clip]{"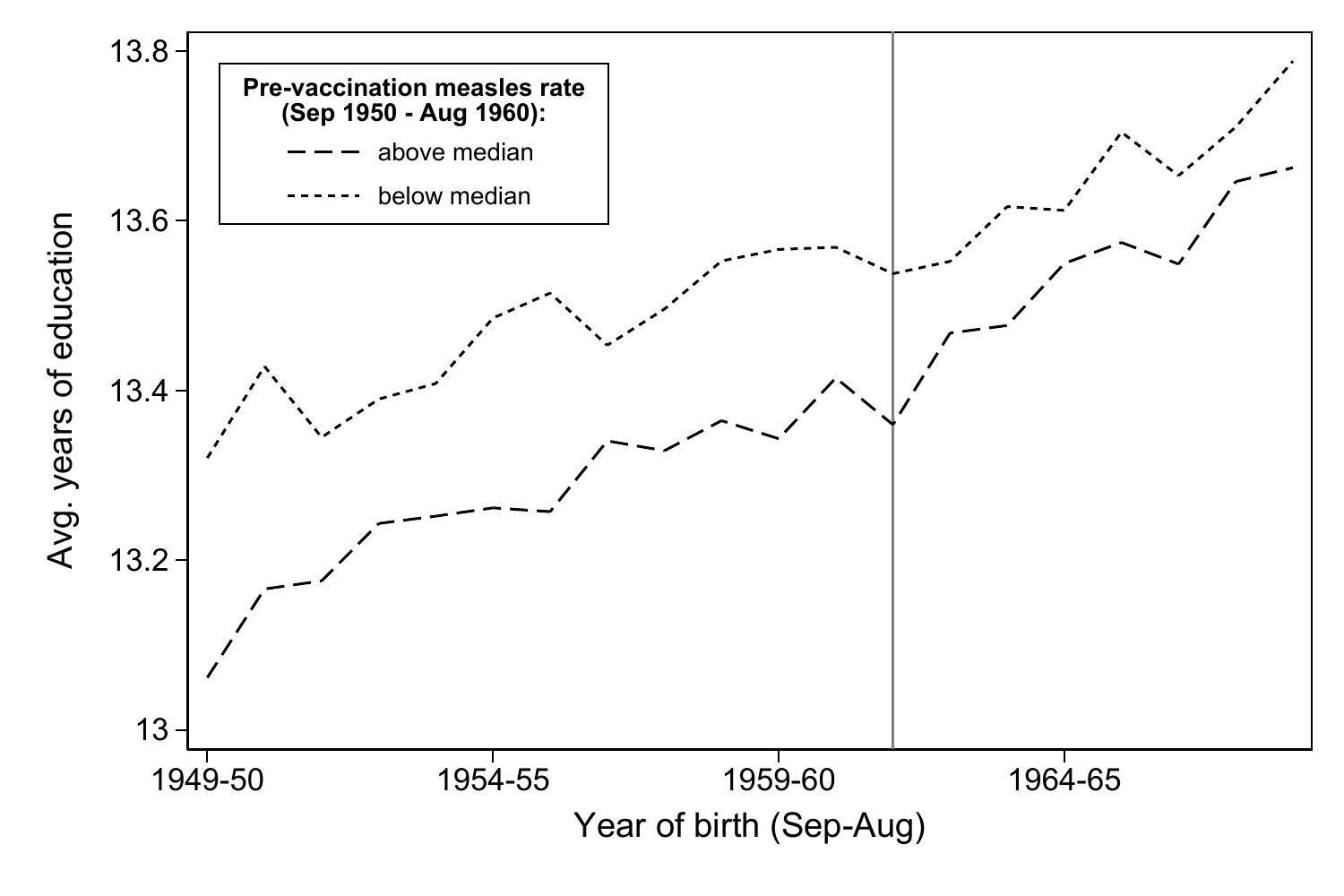"}
		\end{subfigure}   
		\hfill
		\begin{subfigure}{0.475\linewidth}
		\caption{Height}  
		\includegraphics[width=\linewidth, trim=15 15 10 10, clip]{"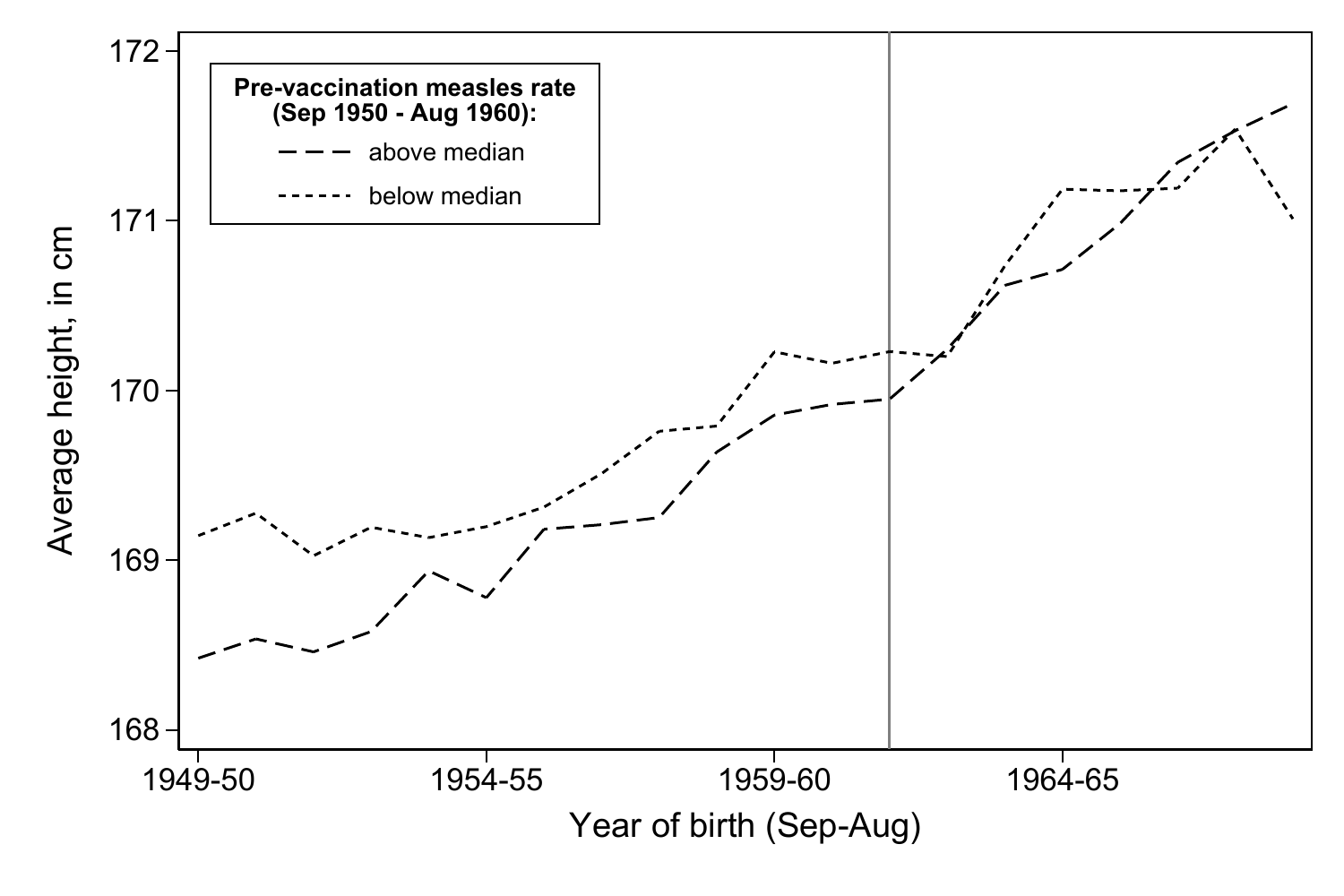"}
		\end{subfigure} \par}
     {\scriptsize \singlespacing Note: Years of education and height in centimetres are averaged across schoolyears of birth (September to August). E.g. the 1949-1950 period covers the cohorts born between September 1949 and August 1950. The grey vertical line represents the last cohort likely unaffected by the vaccine introduction, namely those born between September 1961 and August 1962. Districts participating in the 1966 trial are excluded from the figures. \par}
\end{figure}

A potential issue in the above empirical approach is that the outcomes of interest may have followed different trends in districts with high and low measles infection rates prior to the introduction of vaccination. If districts with high rates of infections show stronger increases in years of education before the introduction of the vaccine compared to districts with lower rates of infections, this suggests that improvements in education may have occurred even in the absence of the vaccination campaign. We explore this empirically, plotting the average years of education and height in districts with high and low levels of measles infections before the introduction of the vaccination campaign. \autoref{figure_educ_height_tseries_by_prevacc} shows that individuals who are born in districts with relatively low infection rates, on average, have higher levels of education and are taller. However, the upward trends in education and height are similar for individuals born in areas characterized by above- and below-median levels of infection.

\begin{figure}[tb!]
	\captionsetup[subfigure]{aboveskip=0pt,belowskip=-1.5pt}
   	\caption{Qualification completion in districts with high and low pre-vaccination measles rates}
   	\label{figure_quals_tseries_by_prevacc} 
     {\centering 
		\begin{subfigure}{0.475\linewidth}
		\caption{Any qualification}
		\includegraphics[width=\linewidth, trim=5 15 5 5, clip]{"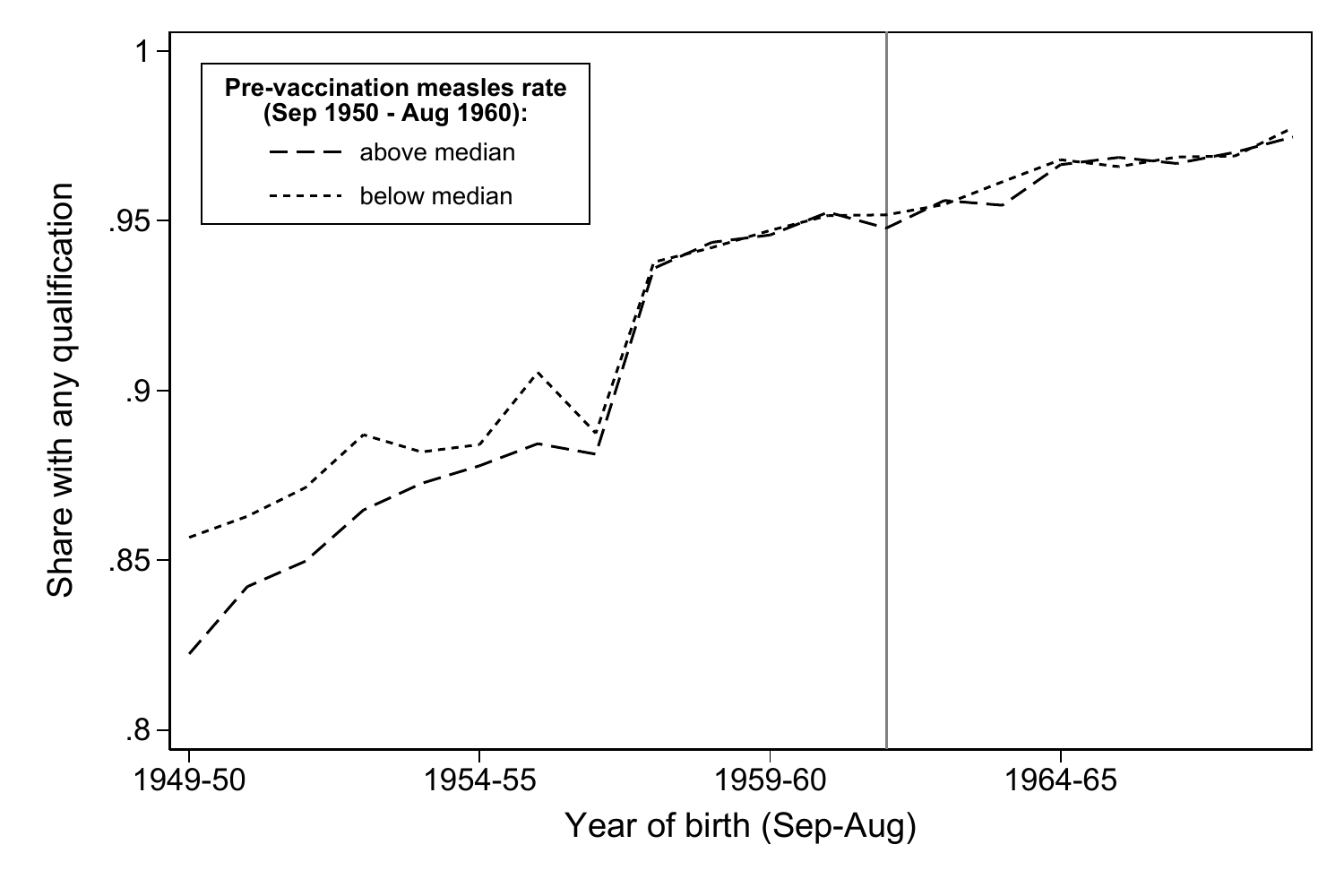"}
		\end{subfigure}
		\hfill
		\begin{subfigure}{0.475\linewidth}
		\caption{Upper secondary qualification}
     	\includegraphics[width=\linewidth, trim=5 15 5 5, clip]{"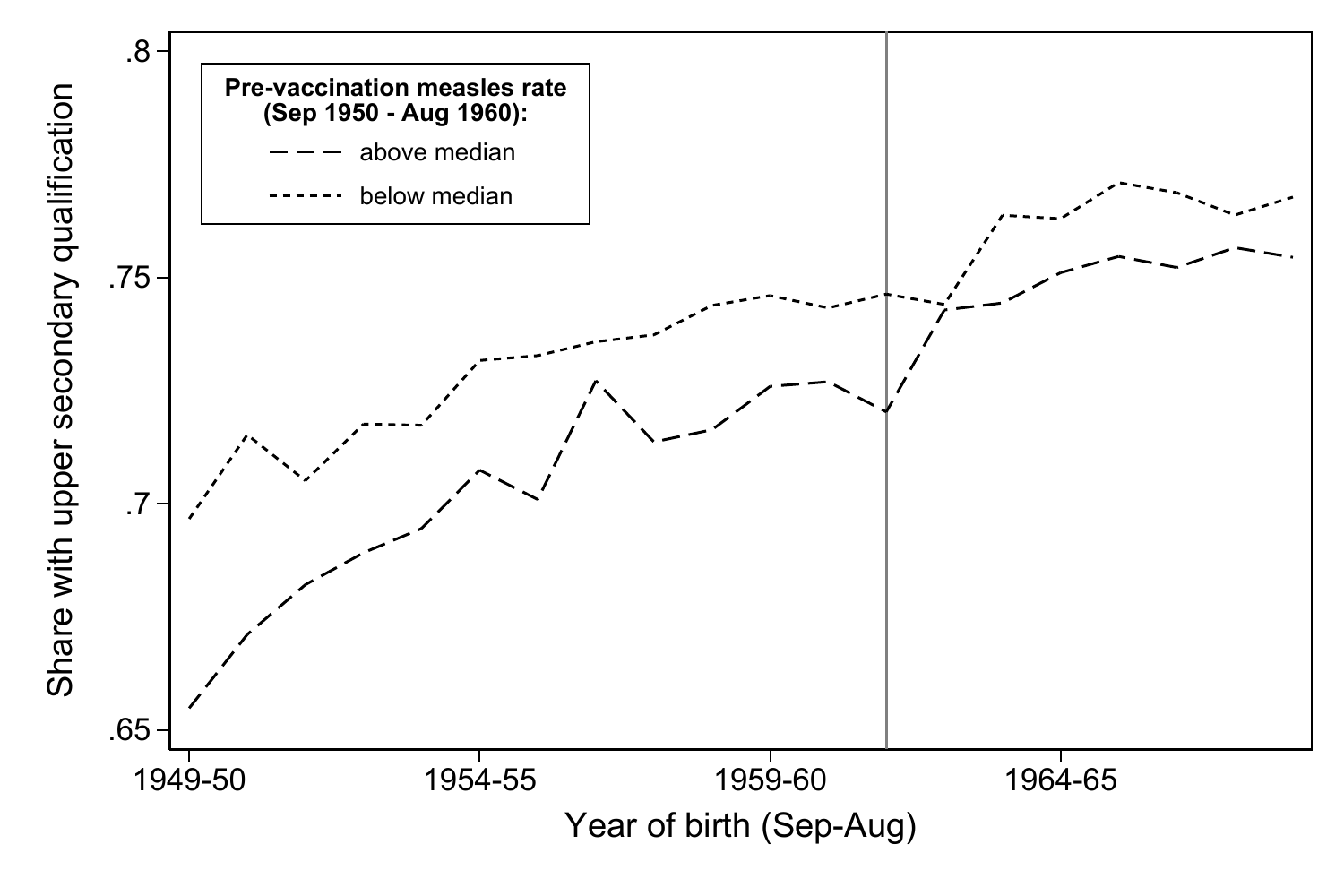"}
		\end{subfigure}
		\begin{subfigure}{0.475\linewidth}    
		\caption{Degree qualification}				
		\includegraphics[width=\linewidth, trim=5 15 5 5, clip]{"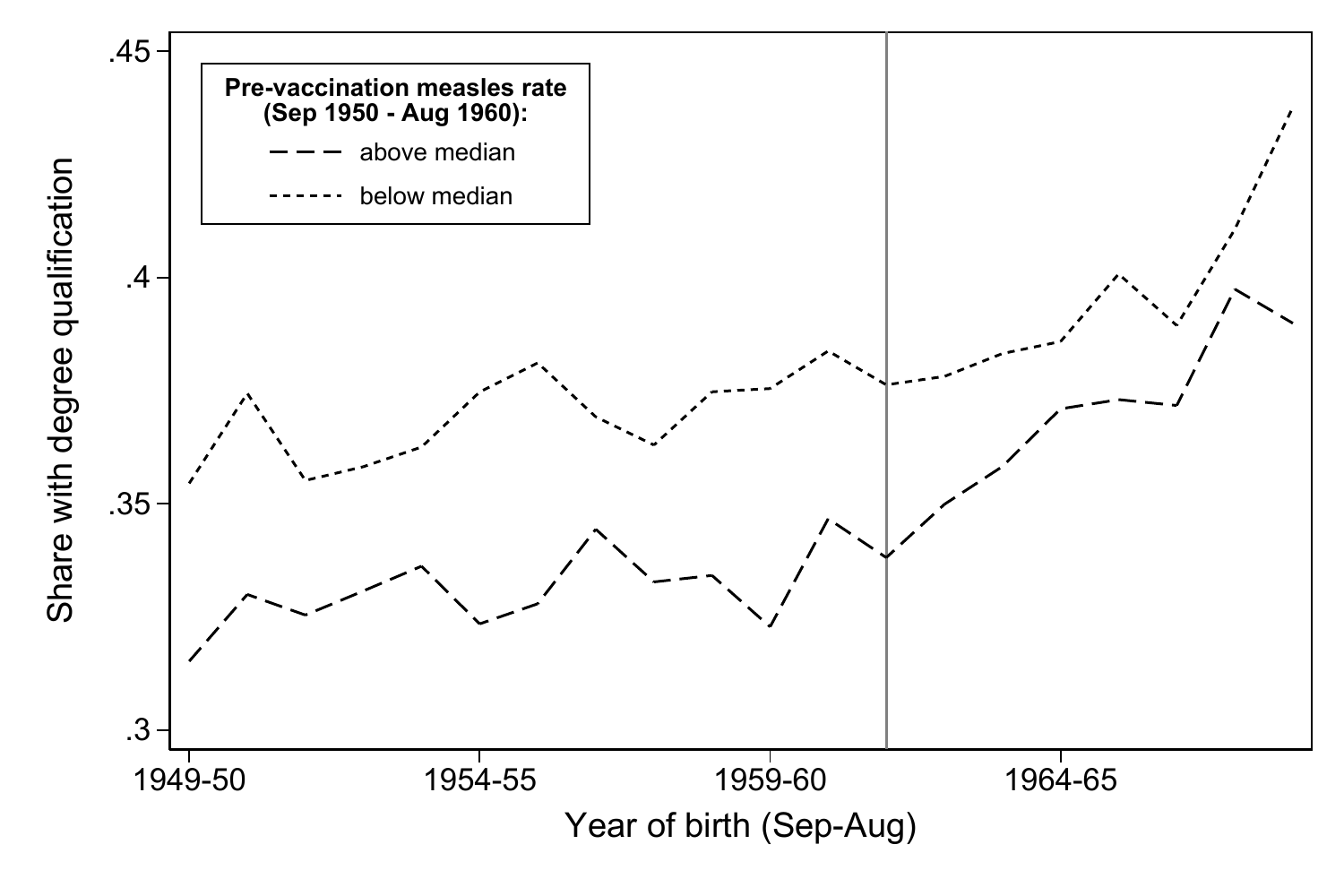"}
		\end{subfigure} \par}
     {\scriptsize \singlespacing Note: Qualification shares are averaged across schoolyears of birth (September to August). E.g. the 1949-1950 period covers the cohorts born between September 1949 and August 1950. The grey vertical lines represents the last cohort likely unaffected by the vaccine introduction, namely those born between September 1961 and August 1962. Districts participating in the 1966 trial are excluded from the figures. \par}
\end{figure}

There is one important issue that deserves additional examination: the education reform that increased the school leaving age in the UK from 15 to 16 years. This affected cohorts born from September 1957 and caused a substantial increase in the average school leaving age, in particular among those who would have left education at age 15 in the absence of the reform.  Given that districts with a lower educated population are more likely to experience higher rates of infections (see \autoref{table_descriptives_lowhighdistricts}), this reform may have increased the level of education disproportionally for these higher-measles districts, even in the absence of the vaccination campaign. Although \autoref{figure_educ_height_tseries_by_prevacc}, showing the \textit{average years of education} on the vertical axis, does not suggest such differential trends, we plot the probability of having any qualifications, an upper secondary and a degree qualification in \autoref{figure_quals_tseries_by_prevacc}. This shows that the raising of the school leaving age eliminated the differential population share with any qualifications between districts with above- and below-median pre-vaccination measles rates. To empirically account for these differential trends, we therefore include a dummy that equals one for cohorts affected by the reform (i.e., born after September 1957), as well as an interaction between this and $PreRate_d$. This captures the differential trends in education that are driven by the education reform for districts with different pre-vaccination measles rates.\footnote{We present the results without these additional covariates in \autoref{appendix_no_rosla}, where we show that these additional controls are important in capturing the differential trends in education across districts with high and low infection rates.}

For our purposes it is important to address insights from the recent literature on difference-in-difference (DiD) estimation regarding the use of two-way fixed effects (TWFE) estimators in DiD applications that go beyond the canonical setting of a binary treatment at a single point in time. The empirical approach described here is such an extension of the simple DiD case: treatment is continuous, with the interaction of (temporal) exposure to the post-vaccination period ($Post_{age=a}$) and the district's pre-vaccination measles rate ($PreRate_d$) representing the ``treatment dose''. \autoref{appendix_did} provides a detailed discussion of the issues that may arise in this DiD setting with continuous treatment. In short, the main concern is a potential ``selection bias'' in TWFE estimates due to differences in treatment effects for a given dose between groups actually receiving different doses. 
We show that non-linearities in the relationship between rates of measles infections and the outcome of interest would cause such a selection bias, though we do not have strong priors to suggest a specific direction of the potential bias. Furthermore, note that our analysis of the 1966/67 blanketing trial, discussed in the following section, is not subject to these concerns.

\subsection{Vaccine trial in 1966}\label{sec:method_vacc_trial}

We next exploit the ``blanketing trial'' of 1966/67 when vaccination was offered to all eligible susceptible children in some districts, but not others. While the trials took place in eight areas of England and Wales, our analysis focuses on four of these that targeted a larger age range than the others: Bedford, Kingston upon Hull, Newcastle upon Tyne and Oxford. These areas offered vaccination to all susceptible children (those without prior immunity from a measles infection) from age 10 or 18 months up to age 10 or 12.

To compare the long-term outcomes of individuals exposed to the trial at a given age to those not exposed, we estimate the following difference-in-difference specification:
\begin{align}\label{eq:trial}
\begin{split}
Y_{idc} = \eta + \sum_{a} \pi_a TrialYears_{age_i=a} + \sum_{a} \theta_a TrialYears_{age_i=a} \times TrialDistrict_{d} \\ 
+ \mathbf{X_{i}'} \boldsymbol{\psi} + \gamma_d + \lambda_c + u_{idc}
\end{split}
\end{align}
where $Y_{idc}$ denotes the outcome (e.g., years of education or height) for individual $i$, born in district $d$ in cohort $c$. $TrialYears_{age_i=a}$ denotes the years of exposure to the vaccine trial that ran from September 1966 to August 1968 during the year(s) in which individual $i$ was of age(s) $a$.\footnote{In line with our previous analysis, we define September 1966 as the start of the trial period. This corresponds to the beginning of the schoolyear following the start of the trials. We define August 1968 as the end of the trial period corresponding to the general introduction of the vaccine.} It takes values between 0 (no exposure to the trial period) and 2 (full exposure to the 2-year trial period). $TrialDistrict_d$ is a dummy indicating whether district $d$ participated in the blanketing trial. Since districts participating in the trial were predominantly urban areas, we restrict the control districts to those with a similar population density to the trial districts. More specifically, using population data for 1961--1964 we include districts with a population density within one standard deviation of the mean among trial districts. As above, we control for a vector of individual-level controls $\mathbf{X_{i}}$, including gender and month of birth fixed effects, as well as district of birth fixed effects $\gamma_d$ and school-year of birth fixed effects $\lambda_c$. The parameters of interest in \autoref{eq:trial} are the $\theta_a$'s, capturing the extent to which exposure to the 1966 trial affected the education or height of individuals in the trial districts, compared to the control districts.

\begin{figure}[b!]
   \caption{Average annual measles rates in districts participating in the 1966/67 trial and control districts}
   \label{figure_measles_tseries_by_trial_mov_avg} 
     {\centering \includegraphics[scale=0.7, trim=15 15 0 10, clip]{"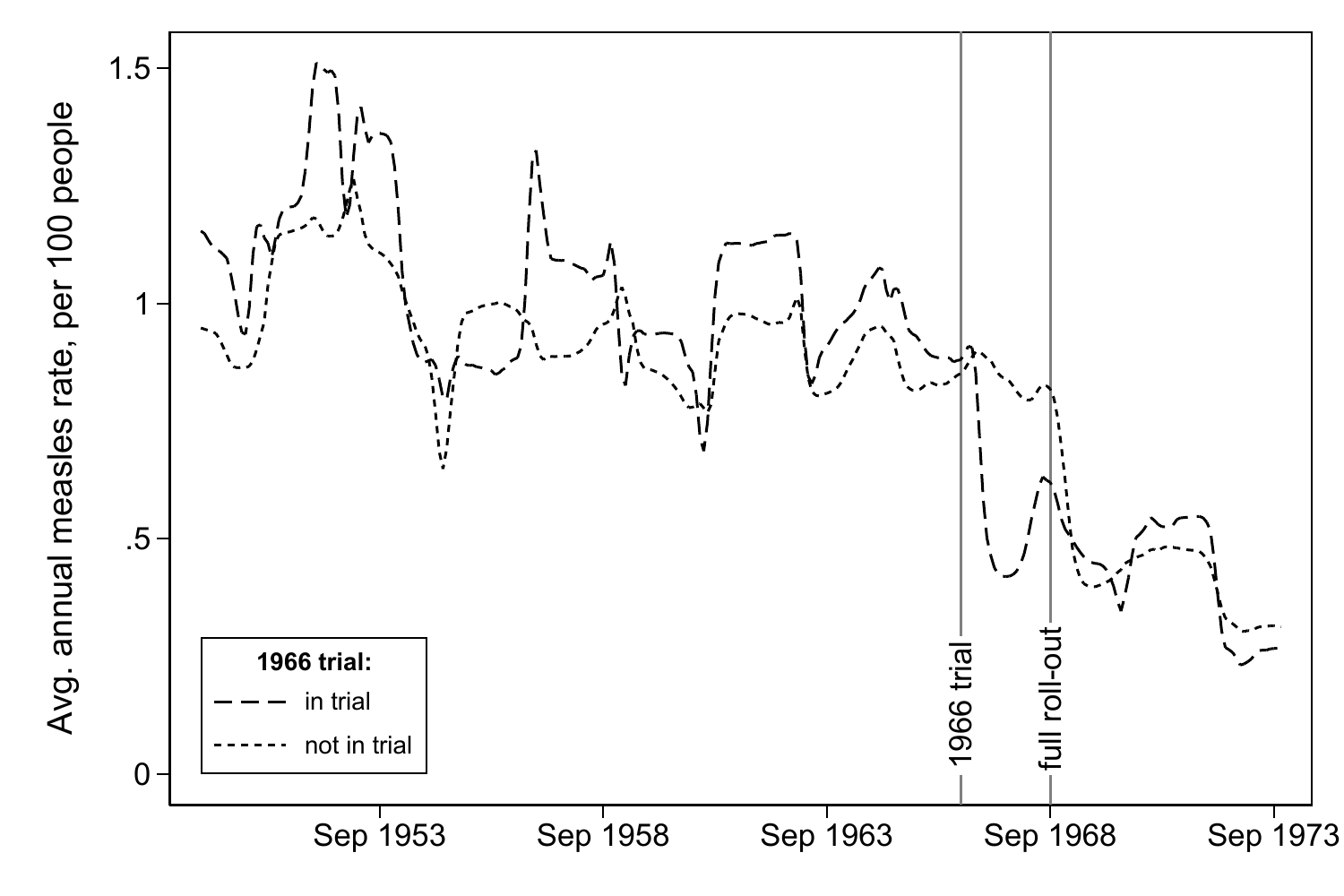"} \par}
     {\scriptsize \singlespacing Note: We focus here on the trial targeting susceptible children up to age 10 or 12. Districts with trials targeting children up to age 2 are excluded from the graph. The sample was furthermore restricted to control districts with a population density within one standard deviation of the mean among the trial districts. Each monthly observation corresponds to the average annual measles rate (per 100 people) over the preceding 24 months. 11 out of 1472 districts were excluded due to (partially) missing data on measles cases or population size. \par}
\end{figure}

The difference-in-difference specification assumes that the trial led to a reduction in measles rates for treated districts compared to control districts. We confirm this in  \autoref{figure_measles_tseries_by_trial_mov_avg}, showing the average annual measles rates for treated and control districts, represented by the dashed and dotted line respectively. This shows a large reduction in measles rates for the treated districts immediately after the start of the blanketing trial, which is followed by a reduction in the control districts immediately after the full roll-out of the vaccination campaign. Furthermore, the figure shows that both drops are of similar magnitude: prior to the 1966 trial (1968 roll-out for control districts), districts reported approximately 0.9 cases per 100 population, reducing to 0.5 per 100 population after the 1966 trial (1968 roll-out); equivalent to a 45\% reduction.

\begin{figure}[tb!]
	\captionsetup[subfigure]{aboveskip=0pt,belowskip=-1.5pt}
   	\caption{Average years of education and height in districts participating in the 1966/67 trial and control districts}
  	\label{figure_educ_height_tseries_by_trial} 
     {\centering 
		\begin{subfigure}{0.475\linewidth}
		\caption{Years of education}
		\includegraphics[width=\linewidth, trim=15 15 10 10, clip]{"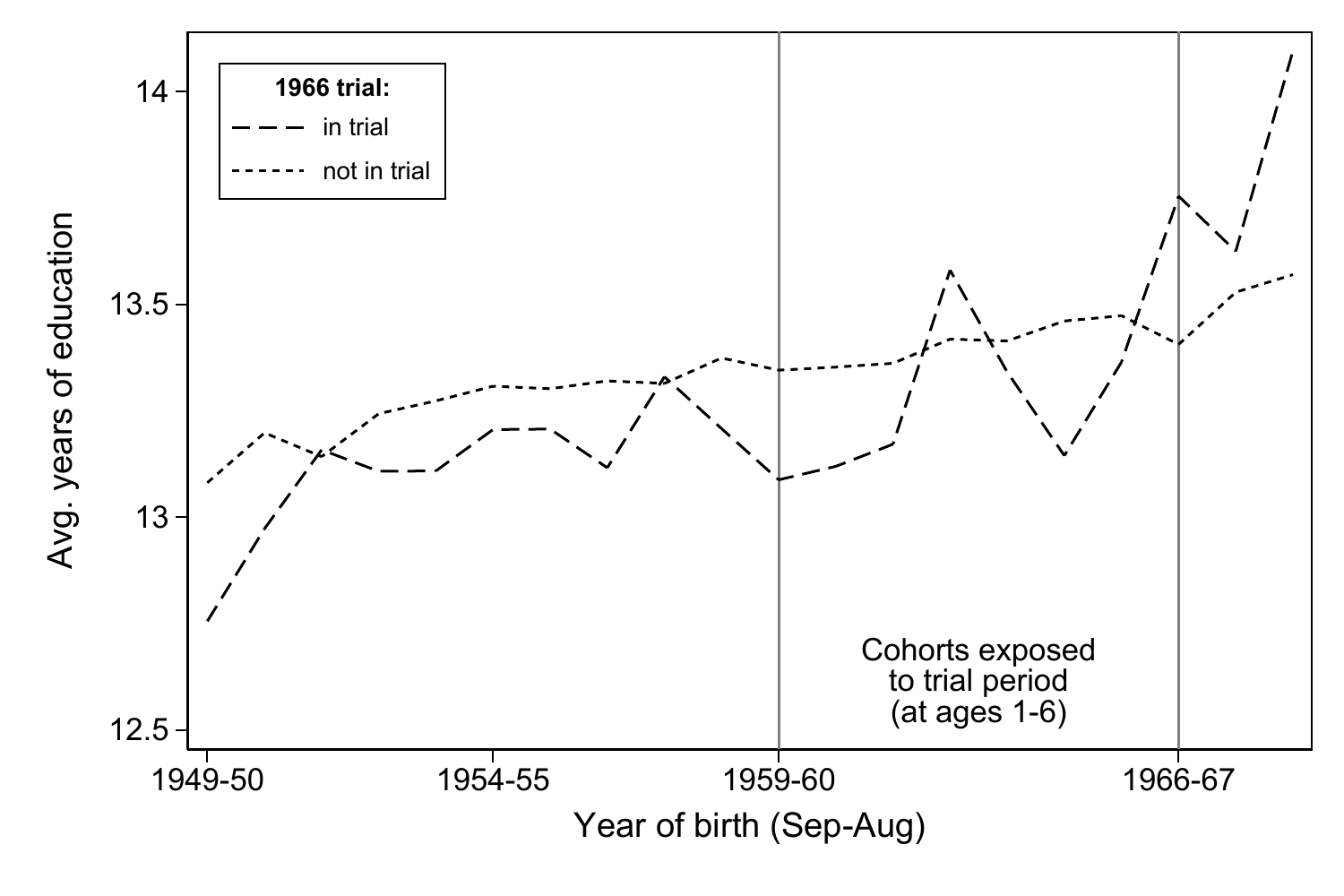"}
		\end{subfigure}   
		\hfill
		\begin{subfigure}{0.475\linewidth}
		\caption{Height}  
      	\includegraphics[width=\linewidth, trim=15 15 10 10, clip]{"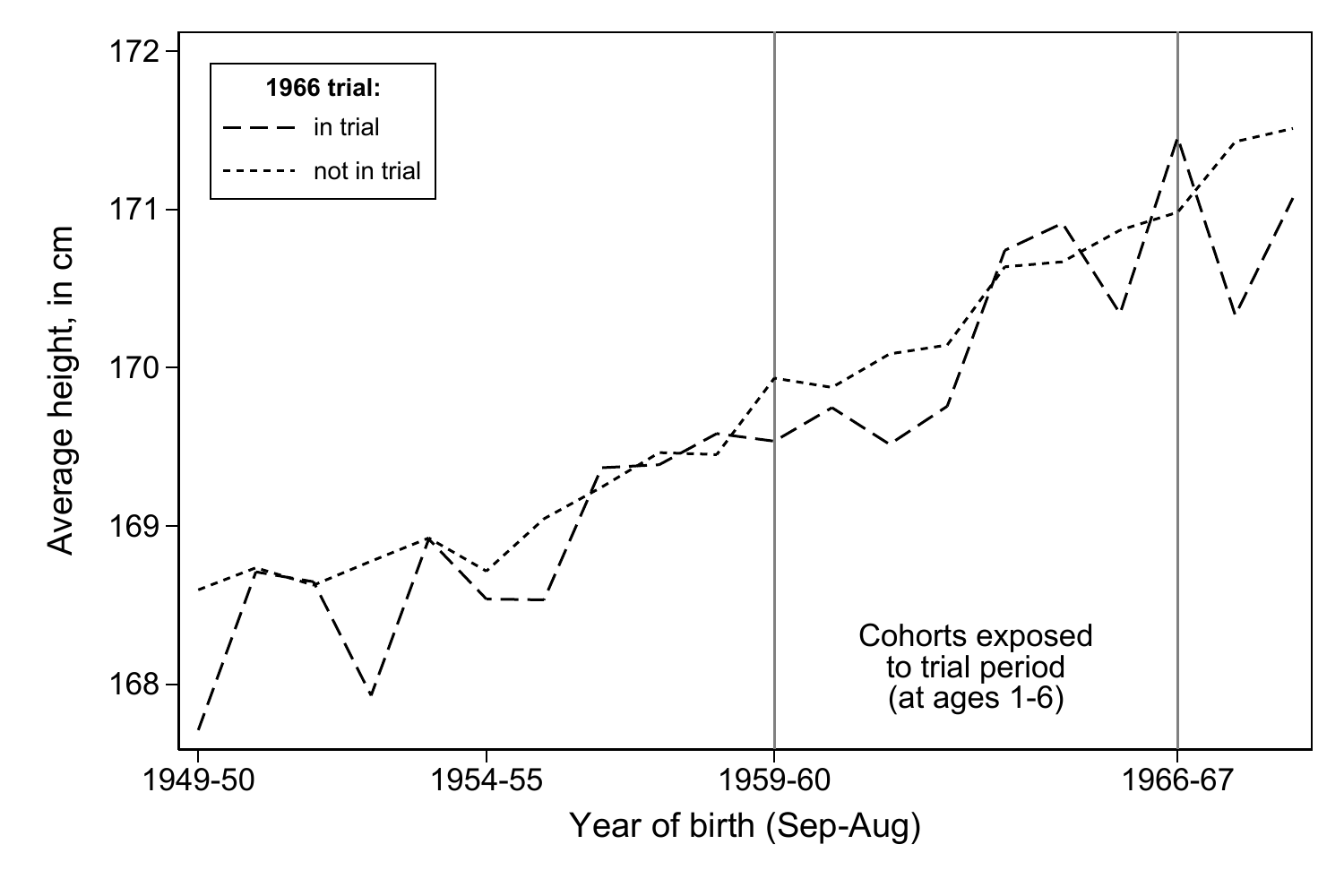"} 
		\end{subfigure} \par}
     {\scriptsize \singlespacing Note: We focus here on the trial  targeting susceptible children up to age 10 or 12. Districts with trials targeting children up to age 2 are excluded from the graph. The sample was furthermore restricted to control districts with a population density within one standard deviation of the mean among the trial districts. Years of education and height in centimetres are averaged across schoolyears of birth (September to August). E.g. the 1949-1950 period covers the cohorts born between September 1949 and August 1950. The grey vertical lines represent the first and last cohorts (partially) exposed to the period of the vaccine trial at age 1 to 6. \par}
\end{figure}

Another important assumption in the difference-in-difference specification is that the average years of education and average height for individuals in treated districts would have followed the same trends as that in the control districts in the absence of the trial. We explore this graphically in \autoref{figure_educ_height_tseries_by_trial}, showing the average years of education (Panel a) and height (Panel b) for treated and control districts by year-of-birth. This shows increased variability in outcomes in treated compared to control districts, driven by their relatively small sample size. We find increasing trends in years of schooling and height before the start of the trial, but with no strong evidence of differential trends across treated and control districts.

\subsection{Complementarity between investments and endowments}

To estimate the complementarity between public investments and endowments, we allow for our treatment effect -- the effect of the introduction of the measles vaccine in 1968  
-- to differ depending on individuals' genetic endowment for the outcomes of interest. More specifically, for the analysis on the 1968 vaccine introduction, we additionally include as regressors in \autoref{eq:vaccine2} the PGI for the respective outcome (i.e., education or height) and its interaction with our explanatory variable of interest $Post_{age_i=a} \times PreRate_{d}$.\footnote{In alternative specifications reported in our sensitivity analyses, we estimate \autoref{eq:vaccine2} separately for those with PGIs above and below the median. We also show the robustness of our estimates to the use of alternative PGIs, constructed using a different GWAS discovery sample \citep{muslimova2022rank}.} Due to the very small number of individuals who are treated in the blanketing trial (see \autoref{sec:data}), we focus on the national roll-out when exploring complementarities.
The theory of human capital production predicts that public health investments (i.e., the measles vaccination campaign) are more productive for individuals with higher genetic endowments. This would be reflected by a positive coefficient on the interaction term.

Although polygenic indices are constructed using molecular genetic data that is fixed at conception, this does not imply that the association with the outcome of interest is immutable or even biological. In fact, in addition to potential biological effects, the associations may capture gene-environment \textit{correlation} (so-called $rGE$), either via genetic variation invoking certain environmental responses (\textit{evocative} $rGE$), or via individuals selecting into certain environments based on their genetic variation (\textit{active} $rGE$). Furthermore it may capture `genetic nurture' effects: the fact that parental genetic variation can shape the environment that the child is exposed to, which in turn can influence their outcome (an example of \textit{passive} $rGE$). In short, this implies that the predictive power of the polygenic index may capture both genetic as well as environmental components. By exploiting exogenous changes in investments, we account for any evocative as well as active $rGE$. Hence we can interpret the coefficient on the interaction as reflecting the complementarity through a direct (causal) genetic effect as well as passive $rGE$.

To account for passive $rGE$, we can exploit the random inheritance of genes using a within-family analysis (i.e., by including family fixed effects or alternatively controlling for parental genotypes using parent-child trio-data). Experiments that combine both exogenous environments and exogenous genetic variation, however, are rare due to very limited data availability.

The UK Biobank includes a relatively small sample of siblings identified based on the genetic relatedness matrix. In a supplementary analysis, we use the subset of 10,832 siblings (5296 families, excluding multiple births) from the sample used in our main analysis to estimate causal gene-environment interplay. Specifically, we exploit random within-sibling variation in genetic endowments by using the deviation of the PGI from the mean PGI among the sampled siblings in the family, as in \cite{howe2021within}. This ``PGI sibling mean deviation'' (PGI-SMD) captures random genetic differences among siblings who share the same biological parents. A negative PGI-SMD captures lower genetic endowments than the sibling average, but also exposure to an environment of siblings with a higher genetic endowments than the sibling average. Hence, estimates from these analyses cannot distinguish between direct genetic effects and indirect effects via siblings' genetic endowments. Nevertheless, it does allow us to account for genetic nurture effects stemming from the parents, which have been shown to be a large contributor to the genetic effect, especially for human capital outcomes such as education.

\section{Results}\label{sec:results}
We discuss our results in three subsections. First, we present the estimates that exploit the introduction of the measles vaccine in 1968 to identify the effects of early life disease exposure on years of education and height in adulthood. Second, we report the difference-in-difference estimates that exploit the blanketing trials that took place in treated, but not control, districts in 1966/67. Finally, we investigate the extent to which public health investments and individuals' endowments are complementarities in the production of human capital and health.

\subsection{Vaccine introduction in 1968}\label{sec:vacc_introduction}
We estimate reduced form models, exploring whether the vaccine introduction differentially affected adult outcomes for those born in districts with previously high measles rates compared to those with previously low rates, controlling for district, year and month of birth fixed effects as well as county-specific linear trends for year-month of birth. 

\autoref{table_vaccine_intensity_education_height} shows the estimates of \autoref{eq:vaccine2} for years of education (columns 1-4) and adult height (columns 5-8). Column (1) of Panel A shows that full exposure to the post-vaccine period at ages 1--6 is associated with 0.23 more years of education for each additional annual measles case per 100 population at the district-level prior to the introduction of the vaccination.\footnote{If fully exposed to the post-vaccine period, this translates to a one standard deviation increase in the districts' pre-vaccination measles rate being associated with a 0.043 standard deviation increase in years of schooling.} Controlling for district fixed effects (column 2) does not affect this estimate much. However, since education shows strong upward trends with year of birth, this estimate is likely to be partially driven by cohort differences. Indeed, once we account for school-year of birth fixed effects in column (3), the estimate attenuates to zero. Additionally controlling for county-specific trends in column (4) does not substantially alter the results. We find the same pattern of results when we model the probability of obtaining any qualification, an upper secondary qualification, and a degree qualification, as shown in Panels A.1, B.1 and C.1 of \autoref{table_vaccine_intensity_quals}, \autoref{appendix_tables_figures}.\footnote{\autoref{appendix_no_rosla} presents estimates that do \textit{not} account for differential effects of the education reform that increased the minimum school leaving age from 15 to 16 years for those born in or after September 1957. The significant positive impact on the probability of obtaining any qualification found in these estimations (prior to including county-specific trends) highlights the importance of accounting for such differential effects in our analyses.}

%==================BEGIN TABLE=================%
\begin{table}[tb!]\centering\scriptsize
  \begin{threeparttable}
    \caption{\label{table_vaccine_intensity_education_height}Long-term effects of the measles vaccine introduction} %%TABLE TITLE
    \setlength\tabcolsep{5.5pt}
                       \begin{tabular*}{\linewidth}{@{\hskip\tabcolsep\extracolsep\fill}l*{8}{S[table-format=1.3]}}
                       \toprule
                       \input{"tables/height_education_measlesvacc_tablefragment.tex"}
                       \bottomrule
                       \end{tabular*}
                     
\begin{tablenotes}[para,flushleft]
{\tiny  Note: The explanatory variables of interest are the share of the given age periods during which the individual was exposed to the vaccination program, interacted with the measles cases per 100 people prior to the vaccination program. Individuals born in districts that participated in the 1966 trial are excluded from the sample. Standard errors clustered at the district of birth level are shown in parentheses. Significance levels are indicated as follows: * p$<$0.1, ** p$<$0.05, *** p$<$0.01}
\end{tablenotes}

  \end{threeparttable}
\end{table}
%==================END TABLE=================

These estimates suggest that the vaccination campaign did not affect individuals' years of education throughout the education distribution. It may be, however, that there are particular ages at which vaccination is more important. We therefore next explore whether we can identify a sensitive age at which exposure to the vaccination campaign affects educational outcomes. Panel B of \autoref{table_vaccine_intensity_education_height} presents the estimates of \autoref{eq:vaccine2}, additionally distinguishing between whether the exposure to the vaccination campaign occurred during ages 1--2, 3--4 or 5--6. Columns (1)-(4) present the estimates for years of schooling, showing no significant effects of exposure to vaccination for any of the age groups once we account for district and school-year of birth fixed effects. \autoref{table_vaccine_intensity_quals}, Panels A.2, B.2 and C.2, also finds no significant effects of vaccination exposure during these age bands on the probability of having any qualification, an upper secondary or a degree qualification.

Columns (5)-(8) of \autoref{table_vaccine_intensity_education_height} examine the impact of the vaccine introduction on adult height. Column (5) of Panel A shows that full exposure to the post-vaccine period at ages 1--6 is associated with a height increase of 2.0 cm for each additional annual measles case per 100 population at the district-level prior to the introduction of the vaccination.\footnote{If fully exposed to the post-vaccine period, this is equivalent to a one standard deviation increase in the districts pre-vaccination measles rate being associated with a 0.38 cm increase in individuals' height.}  While controlling for district and school-year of birth fixed effects in column (7) attenuates this estimated impact on height to 0.45 cm, accounting for differential county-specific trends in column (8) renders the estimate close to zero and insignificant. Estimates in Panel B suggest any effects on height are driven by exposure to the vaccination programme at ages 1--2 and 5--6. However, these age-specific effects are not statistically significant when school-year of birth fixed effects are included in the estimations and are further attenuated when county-level trends are accounted for.

\subsection{Vaccine trial in 1966}\label{sec:vacc_trial}
We next show the difference-in-difference estimates that exploit the 1966/67 blanketing trial. This quasi-experiment caused a similar drop in measles cases, but in contrast to the nationwide introduction, the vaccine was only trialled in a selection of districts, leaving other districts unaffected. \autoref{table_trial_education_height} presents the estimates of \autoref{eq:trial} for years of education (columns 1-2) and height (columns 3-4). In Panel A, the interaction term of the trial dummy and the years of exposure captures the impact of the 1966/67 vaccine trial on individuals born in trial districts, compared to individuals exposed for the same period but born in control districts.

%==================BEGIN TABLE=================%
\begin{table}[tb!]\centering\scriptsize
  \begin{threeparttable}
    \caption{\label{table_trial_education_height}Long-term effects of the 1966 measles vaccine trial} %%TABLE TITLE
    
                       \begin{tabular*}{\linewidth}{@{\hskip\tabcolsep\extracolsep\fill}l*{5}{S[table-format=1.3  ] S[table-format=1.3  ] S[table-format=1.3  ] S[table-format=1.3  ] S[table-format=1.3  ] }}
                       \toprule
                       \input{"tables/height_education_largetrial_tablefragment.tex"}
                       \bottomrule
                       \end{tabular*}
                     
\begin{tablenotes}[para,flushleft]
{\footnotesize  Note: The explanatory variable of interest is an indicator for districts participating in the trial, interacted with the period of exposure (in years) to the vaccine trial period during the given age periods. We focus here on the trial  targeting susceptible children up to age 10 or 12. Individuals born in trial districts targeting children up to age 2 are excluded from the sample. The sample was furthermore restricted to individuals born in a district with a population density within one standard deviation of the mean among the trial districts. Standard errors clustered at the district of birth level are shown in parentheses. Significance levels are indicated as follows: * p$<$0.1, ** p$<$0.05, *** p$<$0.01}
\end{tablenotes}

  \end{threeparttable}
\end{table}
%==================END TABLE=================

Results in columns (1) and (2) show that exposure to the trial at age 1-6 had no differential impact on the educational attainment of individuals born in the trial districts compared to those born elsewhere. Indeed, the point estimates for overall exposure at ages 1--6 (Panel A) as well as separately for the age bands 1--2, 3--4 and 5--6 (Panel B) are close to zero. \autoref{table_trial_quals} furthermore finds no impact of the trial on the probabilities of completing any qualification, upper secondary or degree qualifications. This further supports our finding that measles vaccinations had little long-term impact on individuals' years of schooling.

In columns (3) and (4) of \autoref{table_trial_education_height} we examine the impact of the blanketing trial on adult height. In Panel A, exposure to a year of the trial period at ages 1--6 is estimated to increase the average height of those born in a trial district by 0.116 cm compared to those born in a control district. However, this effect is not found to be statistically different from zero. Results in Panel B of \autoref{table_trial_education_height} suggest that exposure to the trial at ages 1--2 positively impacts adult height: one year of trial exposure at ages 1--2 is estimated to increase height by 0.38 cm, with no evidence of any effects of exposures at other ages. 

\subsection{Complementarity between endowments and investments}
\label{sec:heterogeneity}
We next explore the extent to which public health investments and individuals' endowments are complements in the production of human capital and health, focusing on the 1968 national roll-out. \autoref{table_vaccine_intensity_1to6_education_height_GxE_EA3_GIANT} presents estimation results of \autoref{eq:vaccine2}, including an additional interaction between our explanatory variable of interest $Post_{age_i=a} \times PreRate_{d}$ and the PGI for the relevant outcome as well as the PGI main effect.\footnote{Our results are robust to including interactions between the covariates and the genetic and environmental measures, as suggested by \citet{Keller2014}.} Finding a positive interaction term would indicate that public health investments are more effective for those with higher endowments, consistent with the theoretical prediction of such complementarities. Columns (1)-(4) show little evidence of such complementarities for educational attainment.

%==================BEGIN TABLE=================%
\begin{table}[tb!]\centering\tiny
  \begin{threeparttable}
    \caption{\label{table_vaccine_intensity_1to6_education_height_GxE_EA3_GIANT}Gene-environment interplay: Long-term effects of the measles vaccine introduction} %%TABLE TITLE
    \setlength\tabcolsep{5.5pt}
                       \begin{tabular*}{\linewidth}{@{\hskip\tabcolsep\extracolsep\fill}l*{8}{S[table-format=1.3]}}
                       \toprule
     					\input{"tables/height_education_vacc1to6_GxE_GIANT_EA3_tablefragment.tex"}
                       \bottomrule
                       \end{tabular*}

\begin{tablenotes}[para,flushleft]
{\tiny  Note: The explanatory variable of interest is the share of the age period from 1 to 6 years during which the individual was exposed to the vaccination program, interacted with the measles cases per 100 people prior to the vaccination program. This measure of treatment intensity for the vaccine introduction is furthermore interacted with the polygenic index (PGI) for education (columns 1-4) / height (columns 5-8). The measure of genetic propensity for education is based on summary statistics from \citet{lee2018gene}, the measure for height is based on summary statistics from \citet{Wood2014GIANT}. Individuals born in districts that participated in the 1966 trial are excluded from the samples. Standard errors clustered at the district of birth level are shown in parentheses. Significance levels are indicated as follows: * p$<$0.1, ** p$<$0.05, *** p$<$0.01}
\end{tablenotes}
  \end{threeparttable}
\end{table}
%==================END TABLE=================

The results for adult height in columns (5)-(8) of  \autoref{table_vaccine_intensity_1to6_education_height_GxE_EA3_GIANT} suggest a stronger impact of the vaccine introduction among those with a higher genetic endowment. The estimate for the interaction effect is positive and of a similar magnitude across all specifications. An increase in the genetic endowment for height by one standard deviation is associated with an increase in the impact of full exposure to the post-vaccine period at age 1--6 by 0.08 cm for one additional annual case of measles per 100 population prior to the vaccine introduction.\footnote{Throughout the paper we do not focus on causal effects of actual measles infections but rather focus on the vaccination policy although the two are of course closely related. This also applies to the interaction effects of this subsection; see e.g. \cite{berg2022rationing} for a discussion on the relation between interaction effects of a policy and genetic endowments versus interaction effects of individual events (as triggered by the policy) and genetic endowments.}

To explore the robustness of these findings, we report separate estimates of \autoref{eq:vaccine2} for the sub-samples with an above- and below-median PGI for the respective outcome. The results, shown in columns (1) and (4) of \autoref{table_vaccine_intensity_1to6_education_height_pgisplit}, are in line with those shown in \autoref{table_vaccine_intensity_1to6_education_height_GxE_EA3_GIANT}. The point estimates for height suggest a positive impact of the vaccination programme on the height of those with a high PGI compared to a negative impact among those with a low PGI. We repeat these analyses using an alternative PGI obtained from the PGI repository \citep{Becker2021}\footnote{This provides separately constructed PGIs for three partitions of the UK Biobank. Since the GWAS discovery samples for each partition include the other two partitions, \citet{Becker2021} advise against jointly analysing data from multiple partitions as this will result in biased standard errors. We therefore only report the split-sample analysis of \autoref{eq:vaccine2} for those with a genetic endowment that is above and below the median.} in columns (2) and (5) of \autoref{table_vaccine_intensity_1to6_education_height_pgisplit}, and show similar results to those reported above. 
We also replicate the results using our own tailor-made PGI that is constructed using the summary statistics from a GWAS on individuals from the UK Biobank, excluding our analysis sample and their relatives. These are presented in columns (3) and (6) of \autoref{table_vaccine_intensity_1to6_education_height_pgisplit}, showing similar positive interaction effects for height. Finally, the estimates from the interacted model are shown in \autoref{table_vaccine_intensity_1to6_education_height_GxE_UKB}, again indicating that our results are robust to the use of differentially-constructed polygenic indices. 

\subsection{A sibling analysis to estimate complementarities}\label{sec:sibs}

The above results suggest complementarities between public health investments and genetic endowments in the production of health as proxied by adult height. While the estimation exploits exogenous variation in the exposure to and benefits from the measles vaccination campaign, the genetic endowments are not exogenous as they are dependent on parental genes. To address this, we use the sample of siblings in the UK Biobank to exploit random within-sibling variation in genetic endowments, estimating causal gene-environment interplay. Specifically, we use the PGI sibling mean deviation to capture the random genetic differences among siblings who share the same biological parents.

%==================BEGIN TABLE=================%
\begin{table}[tb!]\centering\tiny
  \begin{threeparttable}
    \caption{\label{table_vaccine_intensity_1to6_education_height_withinsibGxE_EA3_GIANT}Causal gene-environment interplay in the sibling sample: Long-term effects of the measles vaccine introduction} %%TABLE TITLE
    \setlength\tabcolsep{5.5pt}
                       \begin{tabular*}{\linewidth}{@{\hskip\tabcolsep\extracolsep\fill}l*{8}{S[table-format=1.3]}}
                       \toprule
     					\input{"tables/height_education_vacc1to6_withinsibGxE_GIANT_EA3_tablefragment.tex"}
                       \bottomrule
                       \end{tabular*}

\begin{tablenotes}[para,flushleft]
{\tiny  Note: The sample was restricted to full siblings - identified based on genetic relatedness. The explanatory variable of interest is the share of the age period from 1 to 6 years during which the individual was exposed to the vaccination program, interacted with the measles cases per 100 people prior to the vaccination program. This measure of treatment intensity for the vaccine introduction is furthermore interacted with the deviation of the polygenic index (PGI) for education (columns 1-4) / height (columns 5-8) from the sibling mean. The measure of genetic propensity for education is based on summary statistics from \citet{lee2018gene}, the measure for height is based on summary statistics from \citet{Wood2014GIANT}. Individuals born in districts that participated in the 1966 trial are excluded from the samples. Standard errors clustered at the district of birth level are shown in parentheses. Significance levels are indicated as follows: * p$<$0.1, ** p$<$0.05, *** p$<$0.01}
\end{tablenotes}
  \end{threeparttable}
\end{table}
%==================END TABLE=================

\autoref{table_vaccine_intensity_1to6_education_height_withinsibGxE_EA3_GIANT} shows estimation results of \autoref{eq:vaccine2} for the sibling sample ($n \sim 10,000$), including an interaction between our explanatory variable of interest $Post_{age_i=a} \times PreRate_{d}$ and the PGI sibling mean deviation (PGI-SMD) for the respective outcome of interest. The main effects of the vaccination campaign and the PGI are positive. Since PGI-SMD accounts for parental genetic effects, its coefficient is smaller than that in \autoref{table_vaccine_intensity_1to6_education_height_GxE_EA3_GIANT}, as expected \cite[see e.g.,][]{lee2018gene, kong2018nature}. The effect of the introduction of the vaccination campaign on years of education is of similar magnitude, but it is larger for height.\footnote{This is driven by the sample. Repeating the estimations reported in \autoref{table_vaccine_intensity_1to6_education_height_GxE_EA3_GIANT} for the sibling sample (but using PGI rather than PGI-SMD) also shows a larger main effect for height, see \autoref{table_vaccine_intensity_1to6_education_height_sibsample_GxE_EA3_GIANT}. We find no strong differences in the estimated complementarities.} As in the results reported above, there is no evidence of any complementarities for educational attainment (columns 1-4). However, the results for height (columns 5-8) provide additional support for the presence of causal complementarities between the vaccination campaign and genetic endowments. While the interaction term is not statistically significant (the sample size has reduced substantially), the point estimate is in the same direction and indeed of a larger magnitude than that reported in \autoref{table_vaccine_intensity_1to6_education_height_GxE_EA3_GIANT}.

\section{Robustness analysis}\label{sec:robustness} 
Our estimates show that exposure to the measles vaccine had no systematic long-lasting effect on individuals' educational attainment, but led to a small increase in adult height among the affected cohorts who are genetically predisposed to be taller. We next explore the robustness of these findings to a range of sensitivity checks, where we again focus on the 1968 national vaccination campaign. We start by exploring potential heterogeneous effects by gender. Second, we explore whether our analyses are robust to the inclusion of other childhood disease rates as control variables, measured prior to vaccination. Third, we control for regional socio-economic composition measured prior to the vaccination, and fourth, we examine the potential role of mean reversion. Fifth, we explore the sensitivity to alternative specifications of differential trends, and sixth, we investigate the robustness to alternative definitions of the pre-vaccination time period. Finally, we explore the use of measles rates at the \textit{county} instead of \textit{district} level, we run our analysis using a binary indicator for \textit{any} exposure to the post-vaccination period (rather than the \textit{share} of exposure), as well as a dummy for experiencing a \textit{high} pre-vaccination rate of measles infections (instead of the continuous pre-vaccination rate) and investigate the importance of clustering the standard errors. Our findings are robust to these different specifications. 

\subsection{Heterogeneity by gender}
In \autoref{table_vaccine_intensity_1to6_education_height_gendersplit} we explore whether the long-term effects of the 1968 vaccine introduction differed between men and women. Results in columns (1) and (3) show that we cannot reject a null effect for either sub-sample for both education and height. We also find no evidence of gender-specific effects of the vaccination on the completion of different qualification levels (results not shown here but available from the authors upon request). 

In columns (2) and (4) of \autoref{table_vaccine_intensity_1to6_education_height_gendersplit}, we explore whether the complementarities between endowments and investments differ between men and women. We find no evidence of complementarities for educational attainment among men or women (column 2). However, the findings suggest that the complementarity between the measles vaccination and the genetic endowment for height is mainly driven by the female sub-sample. The point estimate for the gene-environment interaction in women is more than double of the estimate in men, and only in women can we reject a null effect (at the 10\% significance level).

\subsection{Measles and other diseases}
One issue that may affect our identification is unobservable characteristics correlating with measles infection rates as well as our outcomes of interest. For example, rates of measles infections are likely to be correlated with other childhood diseases that may also have longer-term effects on individuals' educational attainment or physical development. Indeed, there is evidence that measles infections increase susceptibility to other diseases, so vaccinating children against measles is likely to also have reduced the prevalence of other illness. However, this is a \textit{direct} consequence of measles vaccination, and any changes in educational attainment or height that are driven by improvements in the disease environment are part of the causal effect we aim to capture. We therefore do not control for changes in these other diseases in our specification. Instead, in our sensitivity analysis, we control for \textit{pre-vaccination} district-level rates of scarlet fever, whooping cough, diphtheria, pneumonia, respiratory tuberculosis, and polio, interacted with $Post_{age_i=a}$: the share of the relevant age period during which the individual was exposed to the measles vaccination programme. This allows us to account for trends in the local disease environment which may coincide with the introduction of the measles vaccine. 

Our results are presented in \autoref{figure_robustness_addit_controls}, with Panel (a) showing the findings for years of education, and Panel (b) for height. The vertical axis shows the coefficient estimate, with 90\% as well as 95\% confidence intervals. The legend below the horizontal axis identifies the empirical specification used. The left-most specification in each graph (the red diamond) is the main specification, corresponding to columns (4) and (8) in \autoref{table_vaccine_intensity_education_height}. The next specification accounts for the interaction between pre-vaccination rates of scarlet fever and $Post_{age_i=a}$, showing slightly larger effects for education, but smaller for height. In the following five specifications, we control for interactions between $Post_{age_i=a}$ and pre-vaccination rates of pertussis (whooping cough), diphtheria, pneumonia, respiratory TB, and polio, showing little difference in the effect estimates. Including all diseases simultaneously does also not substantially alter the estimates of interest. \autoref{figure_robustness_GxE_addit_controls} examines the robustness of the $G \times E$ results to the same specifications, also showing very similar estimates across the board.

\subsection{Controlling for regional socio-economic composition}
\label{ss:6.3}
Measles infection rates are correlated with the socio-economic composition of a district, with lower socio-economic areas on average having higher measles infection rates (see \autoref{table_descriptives_lowhighdistricts}). This socio-economic composition may itself affect individuals' educational attainment or height. Although we control for district fixed effects, accounting for any time-invariant differences in the socio-economic environment between districts, this may not capture changes in district-level socio-economic environments over time. In our sensitivity analysis, we therefore additionally control for proxies of this socio-economic composition measured prior to vaccination, including district-level infant mortality rates (IMR), illegitimacy rates, birth, death, and stillbirth rates, each interacted with $Post_{age_i=a}$. 

In addition to showing the alternative specifications accounting for the prevalence of other diseases, \autoref{figure_robustness_addit_controls} and \autoref{figure_robustness_GxE_addit_controls} present the estimates which control for proxies of socio-economic composition for the main analysis and the $G \times E$ interplay. The estimates for both years of education and height change little when including these additional controls.

\subsection{Mean reversion}
Our identification may be affected by mean reversion. For example, a pre-vaccination shock may have temporarily increased or decreased measles infections whilst simultaneously affecting educational attainment in some districts, with both reverting back to normal levels after (but not due to) the campaign. This would suggest a change in educational attainment post-vaccination, even if the vaccination programme had no effect. In our robustness analysis, we follow \citet{Butikofer2020} and additionally control for the district-level housing density, the proportion of individuals leaving education by age 14 and the proportion of low social class individuals, again interacted with $Post_{age_i=a}$. We obtain these data from the 1951 UK Census  \citep{Census1951,VisionOfBritain} and merge them with the UK Biobank. Results in \autoref{figure_robustness_addit_controls} and \autoref{figure_robustness_GxE_addit_controls} for the main and $G \times E$ analyses respectively show that the estimates for both years of education and height are unaffected.

\subsection{Differential trends}
Our analysis highlights the importance of accounting for differential trends in two ways. First, we show that accounting for the differential effect of the increase in the minimum school leaving age affects our estimates (see \autoref{appendix_no_rosla}). Second, we show that the positive estimates for height disappear when we additionally account for county-specific trends (see \autoref{table_vaccine_intensity_education_height}). We therefore next explore the robustness of our results to different specifications of these differential trends. 

In columns (1) and (5) of \autoref{table_vaccine_intensity_education_height_difftrends} we replicate our preferred specification from  \autoref{table_vaccine_intensity_education_height} which accounts for county-specific trends ($n=62$). Columns (2) and (6) instead account for a trend in the year of birth, interacted with the pre-vaccination measles rate ($PreRate_d$), columns (3) and (7) control for administrative-county-specific trends ($n=240$), and columns (4) and (8) for district-specific trends ($n=1472$). The results are robust across the different trend specifications, with estimates for both years of education and height that are not significantly different from zero. 

\autoref{table_vaccine_intensity_education_height_GxE_difftrends} investigates the robustness of the $G \times E$ results to the use of differential trends. These confirm our findings above: we find evidence of complementarity between endowments and investments in individuals' height, but not for years of education. Although the effect is not significant in all specifications, the point estimates remain in a similar ballpark throughout.

\subsection{Choice of pre-vaccination time window}
In \autoref{figure_robustness_prevacc_periods} and \autoref{figure_robustness_GxE_prevacc_periods} we explore alternative definitions of the time window used to construct the pre-vaccination measles rate ($PreRate_d$) as a measure of treatment intensity. In our main estimations, we use the 10-year period from September 1950 to August 1960. In this robustness check we use alternative 4-year periods, each spanning from September to August. While the period choice does affect the coefficient estimate to an extent, the null result on years of education and height in the main analysis are similar across alternative empirical specifications. Furthermore, the positive GxE coefficient for height is not substantially affected by changes in the time window used.

\subsection{Further sensitivity checks}
We conduct three final sensitivity checks. First, we re-run our analysis using geographic variation in the treatment intensity at the county level ($n=62$) instead of the district level ($n=1472$). Specifically, we construct our variable of interest $Post_{age_i=a} \times PreRate_{county}$ at the \textit{county} instead of \textit{district} level, and control for \textit{county} instead of \textit{district} fixed effects. This specification is more similar to that used in the existing literature on the effects of measles on later life outcomes in the US and Mexico. The estimates for both education and height, presented in \autoref{table_vaccine_intensity_education_height_county_cid_yob_cluster}, are substantially larger and significantly different from zero until we account for differential county-specific trends in columns (4) and (8), suggesting that the level of aggregation matters, especially when not accounting for differential trends.\footnote{In these estimations we cluster standard errors at the county and schoolyear of birth level to match the approach used by \citet{Atwood2022} and \citet{AtwoodPearlman2022}. When alternatively clustering standard errors at the county level, the standard errors are larger, rendering the estimates in columns (3) and (7) statistically insignificant.} However, our conclusions are the same, in that we find no significant impact of measles on either education or height once we account for area-specific trends.

Second, we use a binary indicator for \textit{any} exposure to the post-vaccination period instead of the continuous share of exposure $Post_{age_i=a}$. Similarly, we replace the continuous pre-vaccination rate $PreRate_d$ with a dummy variable indicating a high rate of measles infections, defined as higher than the median. The results using these binary measures of vaccination exposure and intensity (separately and combined) are presented in \autoref{table_vaccine_intensity_binary_education_height} and \autoref{table_vaccine_intensity_binary_education_height_GxE} for the main and $G \times E$ results respectively, showing similar estimates to those above.\footnote{Note that the coefficient estimates in \autoref{table_vaccine_intensity_binary_education_height} and \autoref{table_vaccine_intensity_binary_education_height_GxE} are not directly comparable to those above, since they use a binary specification.}

Finally, we explore the sensitivity of our results to different levels of standard error clustering. \autoref{figure_robustness_standard_errors} and \autoref{figure_robustness_GxE_standard_errors} present the findings for the main and $G \times E$ results respectively, where we show the importance of controlling for schoolyear of birth effects; not only for the magnitude of the coefficient, but also for its standard error. Indeed, not controlling for schoolyear of birth fixed effects leads to quantitatively much larger estimates of interest and smaller standard errors; similar to what we show in \autoref{table_vaccine_intensity_education_height}. For specifications controlling for district and schoolyear of birth fixed effects as well as county-level trends (the final five estimates on the right of each figure), the standard errors remain largely unchanged when clustering by year-month of birth or schoolyear of birth (instead of the district level), or when using two-way clustering by district and year-month / schoolyear of birth.

\section{Conclusion}
\label{sec:conclusion}
We find a positive impact of the measles vaccination programme on adult height among individuals with a higher genetic endowment, confirming the existence of complementarities between public health investments and individual's health endowments. However, we find no such evidence for education, neither of an average effect of public health investments, nor of any complentarities with educational genetic endowments. Also, exposure to the nationwide vaccination programme does not affect mean adult height. 

Our null findings for the average effects on human capital and health differ from the concurrent literature on the long-term effects of measles vaccinations in the United States \citep{Atwood2022, barteska2022mass, chuard2022economic} and Mexico \citep{AtwoodPearlman2022}, reporting positive long-term effects on education, employment, earnings and health. 
One reason for this discrepancy is that the health, economic and welfare policies in the UK differ substantially from those in the US and Mexico \citep[which has been shown to matter for the long-term effect of measles;][]{AtwoodPearlman2022}. An additional issue is that the take-up of the measles vaccine was near universal in the United States, driven by a combination of a mass media campaign, federal funding, as well as school immunization laws, with measles rates dropping to almost zero in the years after the vaccine introduction. In contrast, while measles rates reduced dramatically in England and Wales, they did not reach levels close to zero until many years later. As herd immunity only occurs when 90--95\% of the population is protected, this suggests a substantially different post-vaccine disease environment in England and Wales compared to the United States. As such, the existing studies on the US and Mexico estimate the effect of \textit{no} versus \textit{full} vaccine compliance, whereas our results reflect the effect of \textit{no} versus \textit{partial} compliance. We argue that the latter may have more external validity. First, the introduction of measles vaccines, especially in developing countries, may take time to reach the full eligible population, as it requires a well-functioning infrastructure. Second, the Covid vaccination campaign has shown that there is substantial vaccine hesitancy in the population, suggesting that a setting with partial compliance may be more policy relevant. 

The finding that the vaccination campaign increased individuals' height among those who are genetically predisposed to be taller has implications for the effect of vaccinations on societal inequality. The improvement in the childhood disease environment and avoidance of the inflammatory response associated with infections has helped individuals reach their genetic height potential. Hence, the vaccination campaign may have had the consequence of increasing inequalities in childhood health based on genetic endowments.

Finally, it should be emphasized that the absence of long-run average gains driven by vaccination does not entail the absence of short-run gains at younger ages. The substantial drop in measles cases post-vaccination implies that it did avoid ill health among young children and the substantial stress associated with it. These may directly concern measles but also later childhood health impacts due to the immunosuppression and increased vulnerability to other childhood diseases.

\vspace{1cm}

{\onehalfspacing {\small

\bibliographystyle{tandfx_max6names}
\bibliography{measles_references}

}}

\clearpage

%--------------------------------------------------------------------------------------------%
%--------------------------------------------------------------------------------------------%

\begin{appendices}

\counterwithin{figure}{section}
\counterwithin{table}{section}

\begin{center}
{\LARGE\textbf{Appendix For Online Publication}}
\end{center}

\section{Supplementary tables and figures} \label{appendix_tables_figures}

\begin{figure}[h!]
   \caption{Measles vaccination rates at two years of age, 1971-1986}
   \label{figure_vaccination_rate_timeseries} 
     {\centering \includegraphics[scale=0.9, trim=15 15 0 10, clip]{"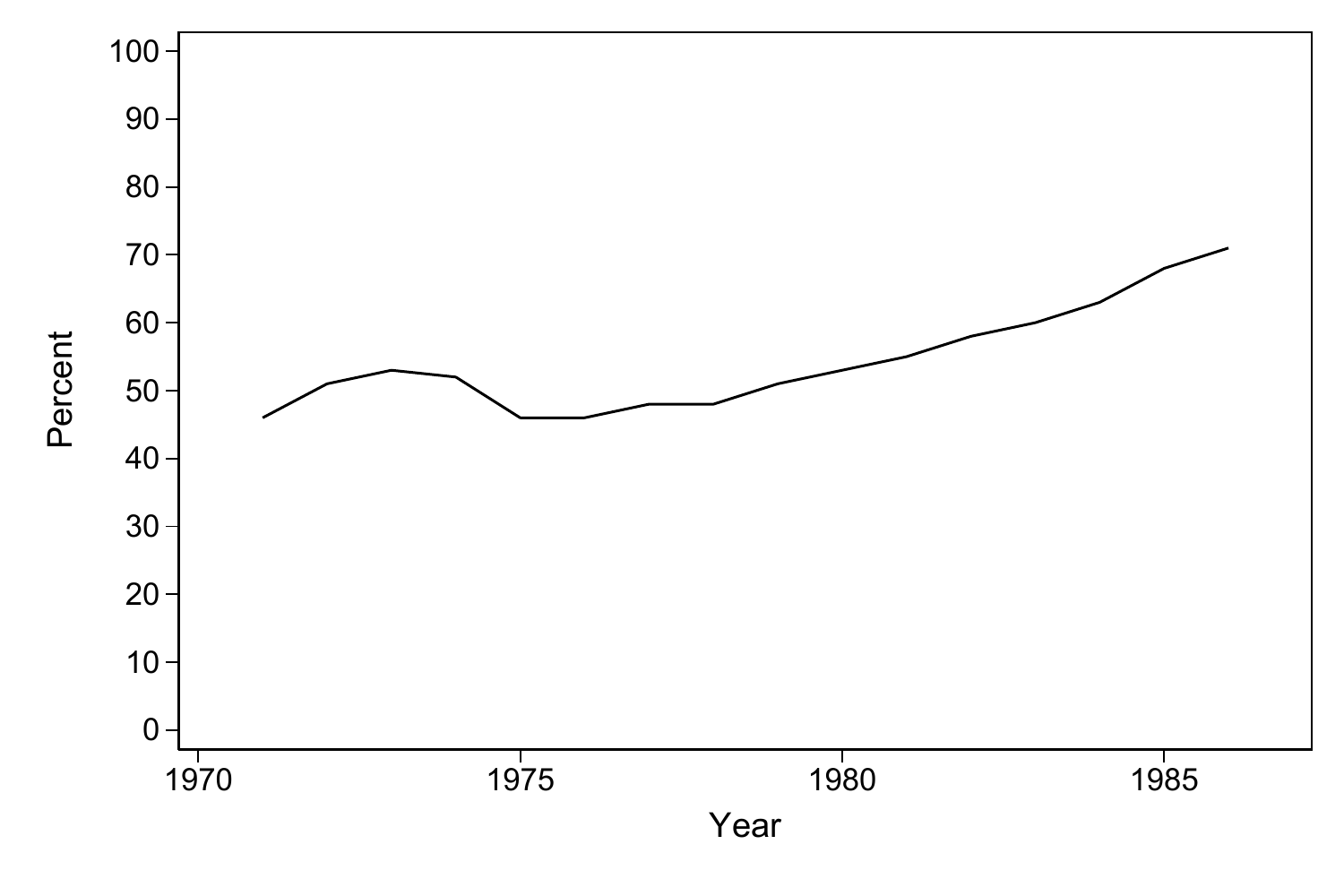"} \par}
     {\scriptsize \singlespacing Note: The figure shows the rate of children  aged two years who had completed the primary course of the measles vaccination by the end of the given year. Data source: \citet{PHE2014}. 
\par}
\end{figure}
\FloatBarrier
%

%==================BEGIN TABLE=================%
\begin{table}[h!]\centering\normalsize
    \caption{\label{table_education_mapping}Mapping between highest reported qualification and derived years of education} %%TABLE TITLE
\begin{tabular}{|l|c|}
\hline 
\textbf{Highest qualification(s)} & \textbf{Derived years of education} \\ 
\hline 
College or university degree & 16 \\ 
\hline 
Combination of: & \\
Other professional qualifications eg: nursing, teaching & 15 \\
\textit{and} A levels/AS levels or equivalent &  \\ 
\hline 
Combination of: & \\
NVQ or HND or HNC or equivalent & 14 \\
\textit{and} A levels/AS levels or equivalent &  \\ 
\hline 
Other professional qualifications eg: nursing, teaching & 13 \\ 
\hline 
A levels/AS levels or equivalent & 13 \\ 
\hline 
NVQ or HND or HNC or equivalent & 12 \\ 
\hline 
CSEs or equivalent & 11 \\ 
\hline 
O levels/GCSEs or equivalent & 11 \\ 
\hline 
None of the above & 10 \\ 
\hline 
Prefer not to answer & -- \\ 
\hline 
\end{tabular} 
\end{table}
\FloatBarrier
%==================END TABLE=================

%
%
\begin{figure}[h!]
	\captionsetup[subfigure]{aboveskip=0pt,belowskip=-1.5pt}
   \caption{Post-vaccination share ($Post_{age_i=a}$)}
   \label{figure_postvacc_share} 
   {\centering 
		\begin{subfigure}{0.475\linewidth}
		\caption{for ages 1 to 6}
     	\includegraphics[width=\linewidth, trim=15 15 10 10, clip]{"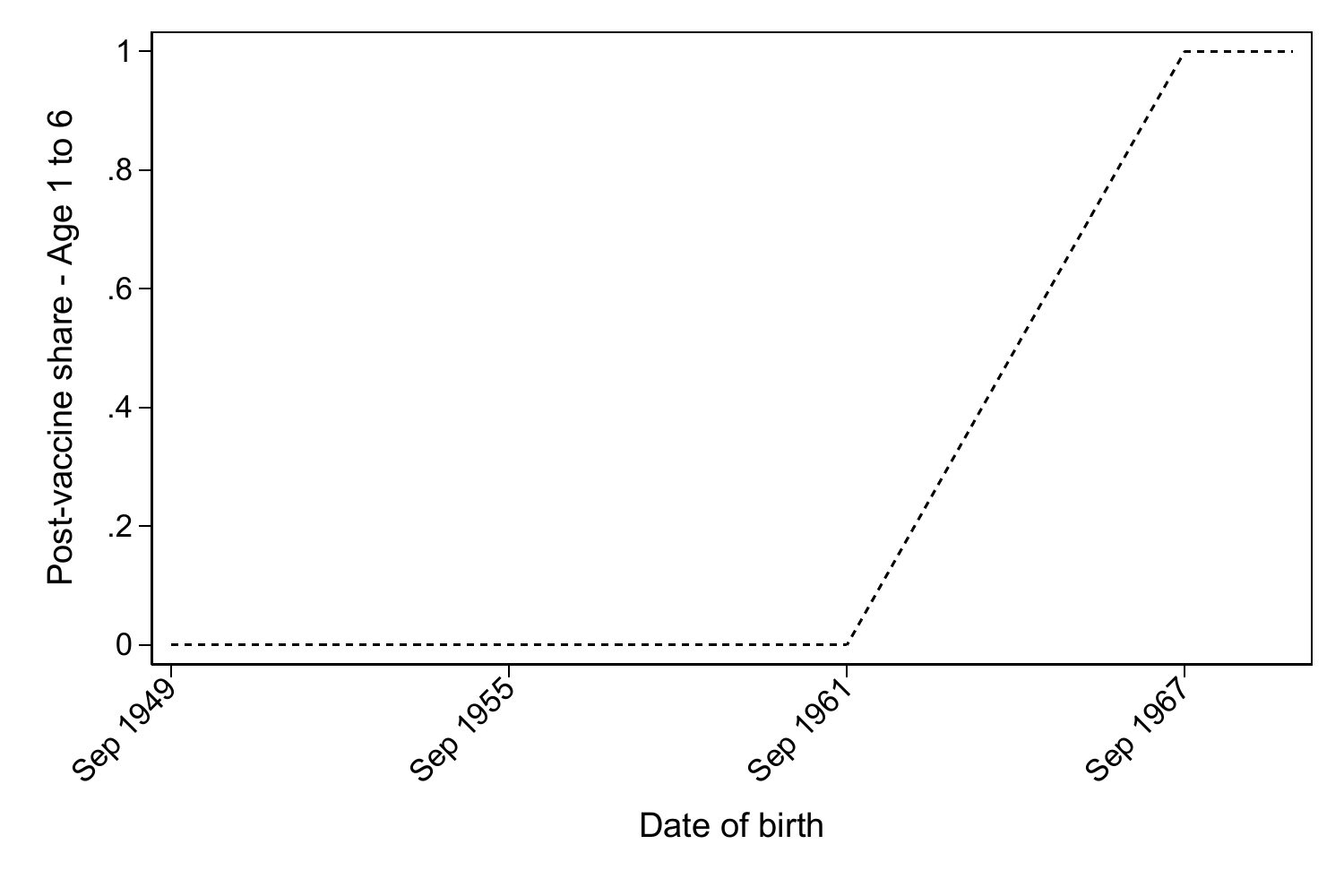"}
		\end{subfigure}   
		\hfill
		\begin{subfigure}{0.475\linewidth}
		\caption{for ages 1--2, 3--4, 5--6}  
		\includegraphics[width=\linewidth, trim=15 15 10 10, clip]{"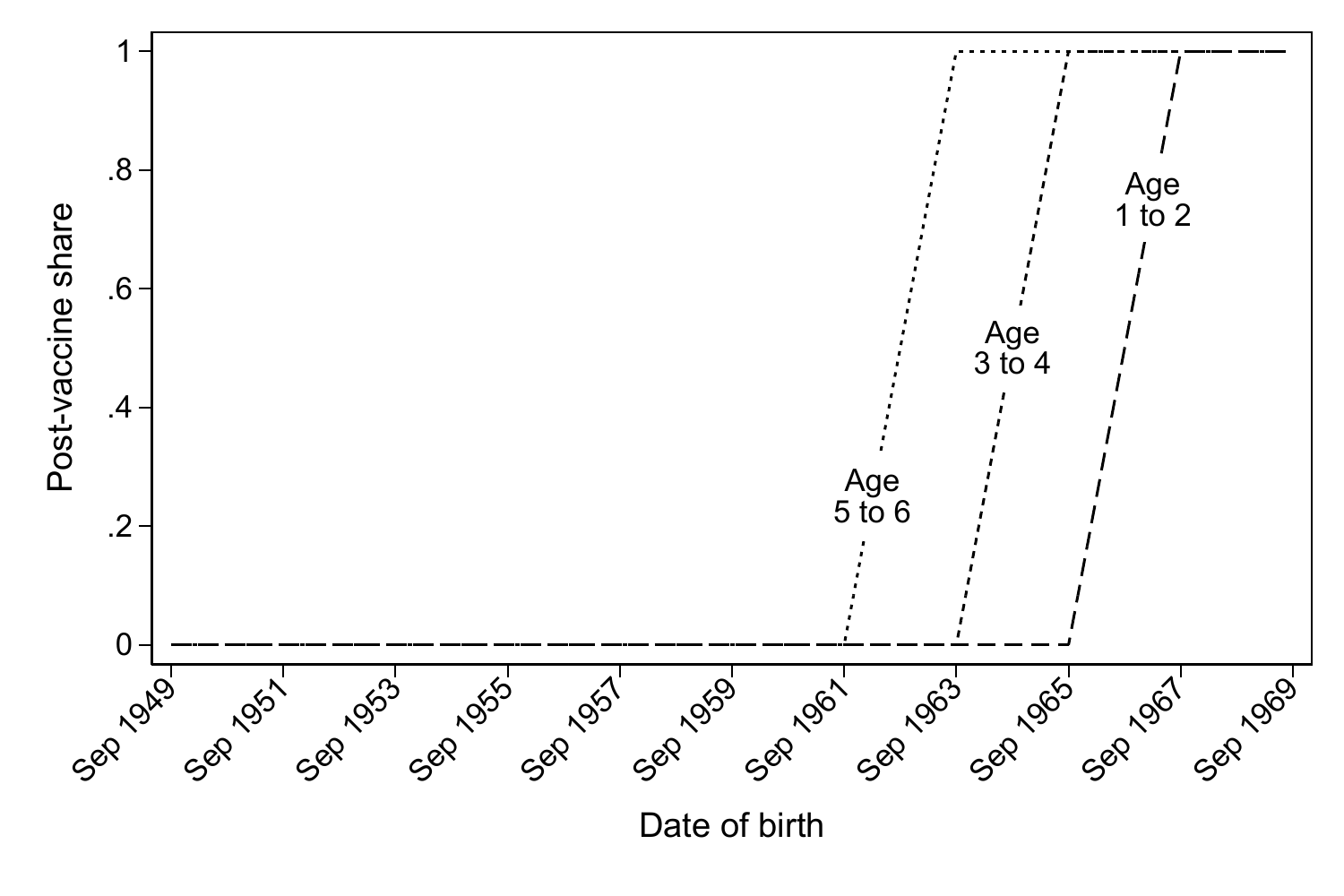"}
		\end{subfigure} \par}
\end{figure}
\FloatBarrier
\begin{figure}[h!]
        \captionsetup[subfigure]{aboveskip=0pt,belowskip=-1.5pt}
   \caption{Measles rates across England and Wales}
   \label{figure_measles_maps} 
   {\centering 
  \begin{minipage}{0.75\linewidth}
  	\begin{subfigure}{0.49\linewidth}
	  \centering
	  \caption{Sep 1950 -- Aug 1951}
	  \includegraphics[width=\linewidth, trim=75 0 25 0, clip]{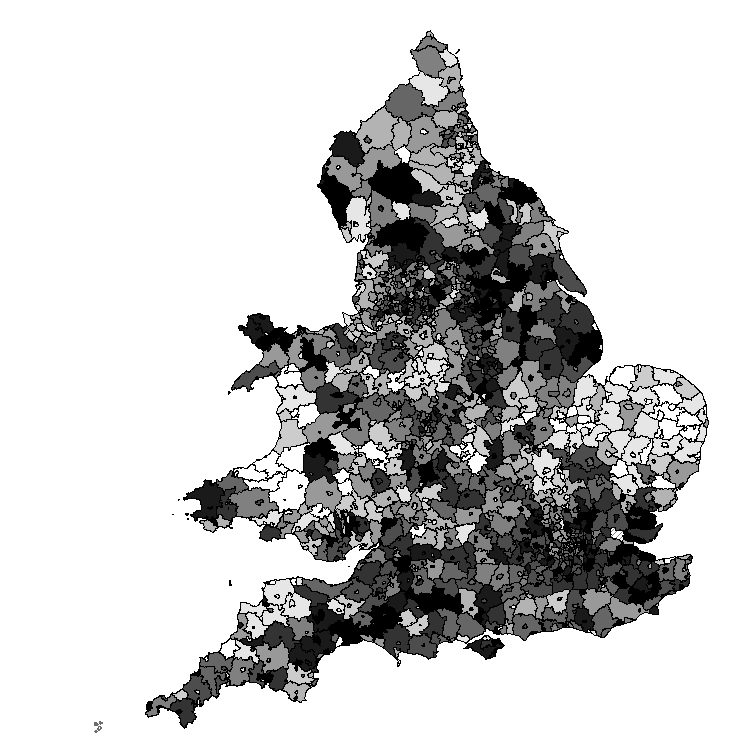}
	\end{subfigure}
	\begin{subfigure}{0.49\linewidth}
	  \centering
	  \caption{Sep 1951 -- Aug 1952}
	  \includegraphics[width=\linewidth, trim=75 0 25 0, clip]{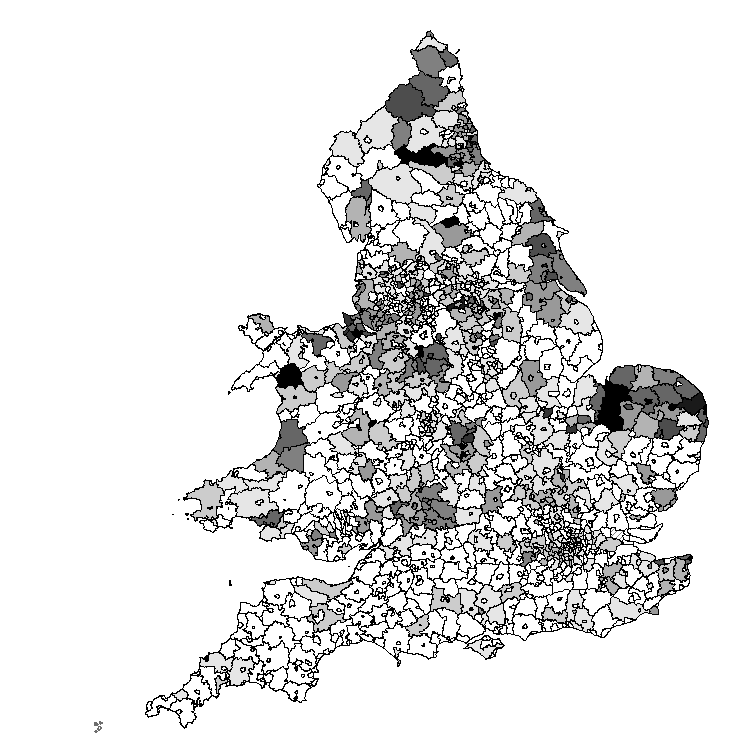}
	\end{subfigure}
	\par\bigskip
	\begin{subfigure}{0.49\linewidth}
	  \centering
	  \caption{Sep 1960 -- Aug 1961}
	  \includegraphics[width=\linewidth, trim=75 0 25 0, clip]{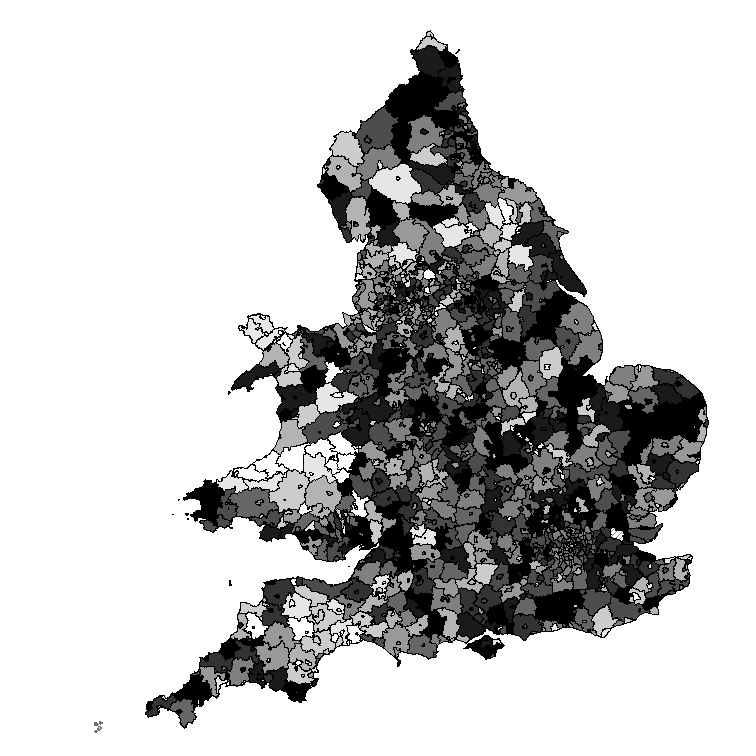}
	\end{subfigure}
	\begin{subfigure}{0.49\linewidth}
	  \centering
	  \caption{Sep 1961 -- Aug 1962}
	  \includegraphics[width=\linewidth, trim=75 0 25 0, clip]{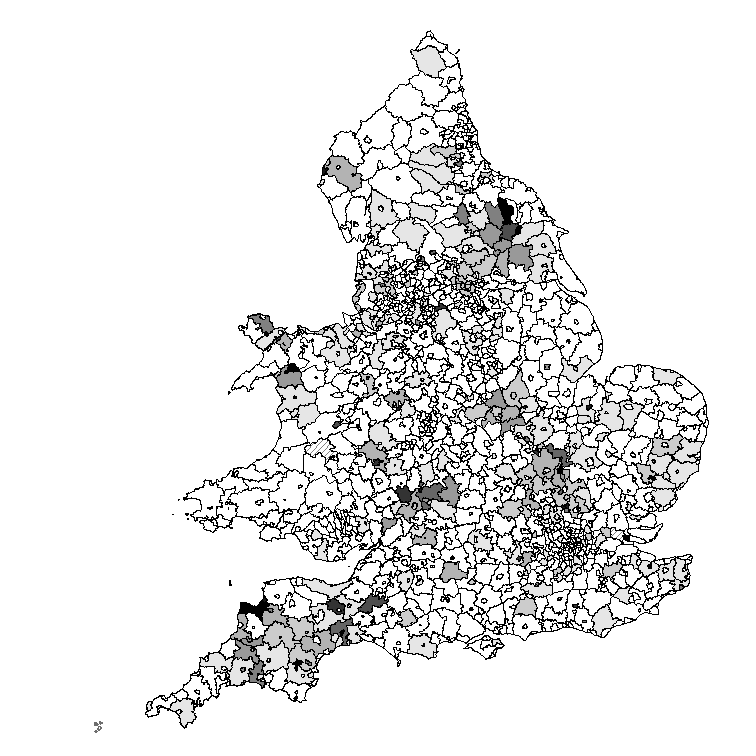}
	\end{subfigure}
  \end{minipage}%
  \hspace{0.05\linewidth}%
  \begin{minipage}{0.18\linewidth}
	  \centering
	  \includegraphics[width=\linewidth, trim=10 290 435 65, clip]{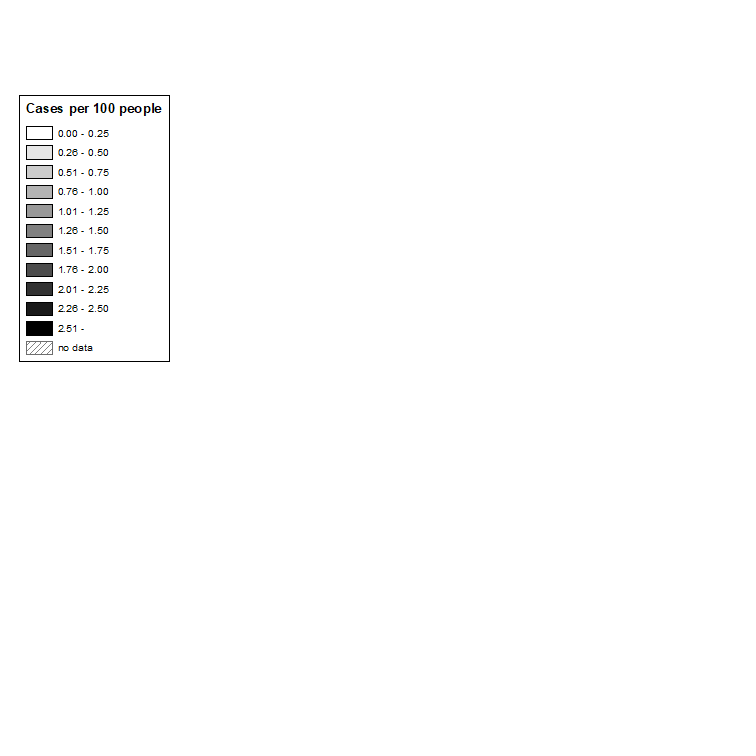}
  \end{minipage} \par}
     {\scriptsize \singlespacing Note: The maps shows the district-level annual measles rates in cases per 100 population. \par}
\end{figure}
\FloatBarrier
\begin{figure}[h!]
   \caption{Average annual measles rates for deciles of the pre-vaccination rate distribution}
   \label{figure_measles_tseries_by_prevacc_decile} 
     {\centering \includegraphics[scale=0.9, trim=15 15 0 10, clip]{"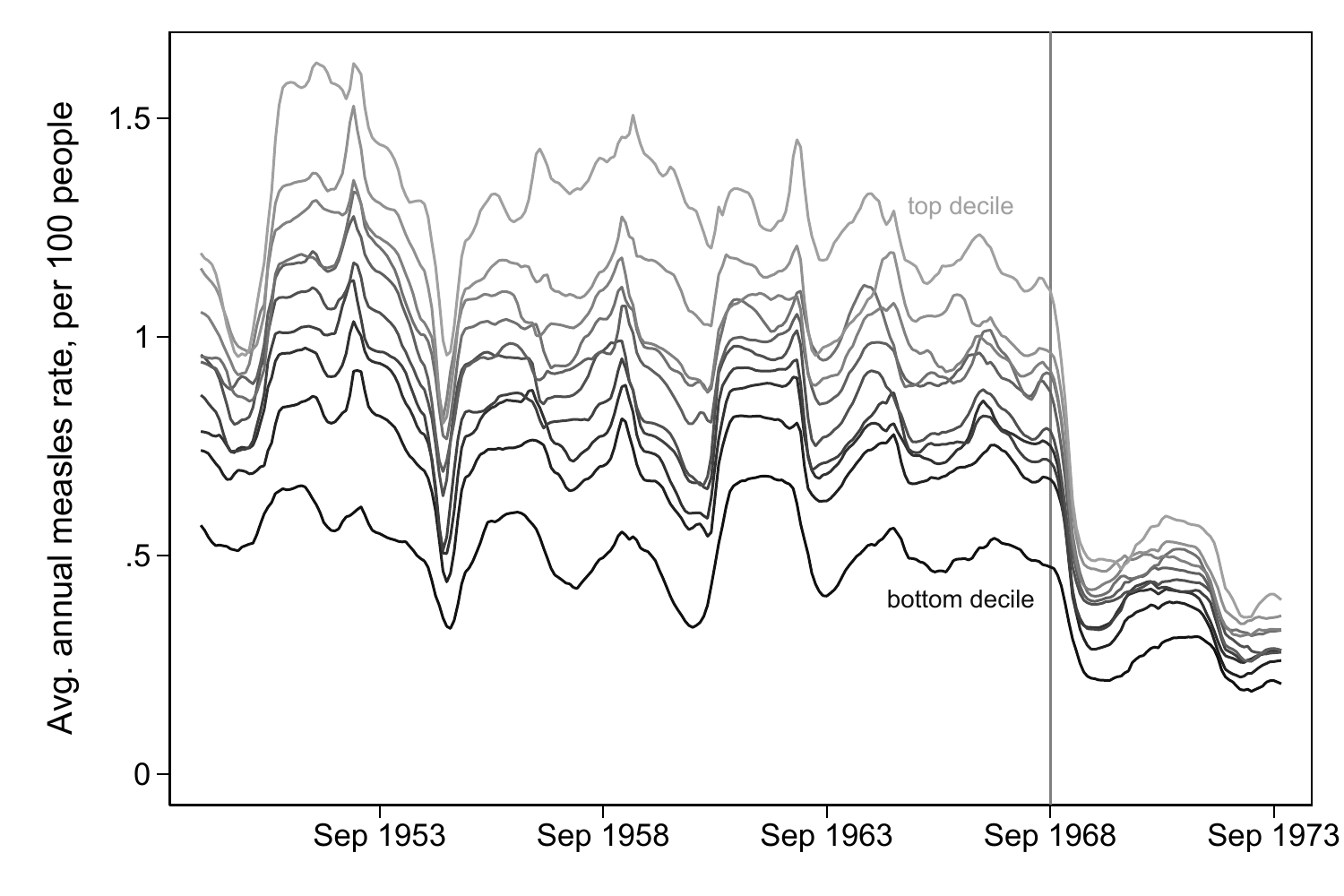"} \par}
     {\scriptsize \singlespacing Note: The grey vertical line represents the beginning of the vaccine roll-out in September 1968. Each monthly observation corresponds to the average annual measles rate (per 100 people) over the preceding 24 months. 11 out of 1472 districts were excluded due to (partially) missing data on measles cases or population size. Districts participating in the 1966 trial are excluded from the figure.
\par}
\end{figure}
\FloatBarrier
\begin{figure}[h!]
   \caption{Pre-vaccination measles rates and the post-vaccination change in measles rates}
   \label{figure_measles_prevacc_vs_change} 
     {\centering \includegraphics[scale=0.55, trim=5 15 5 5, clip]{"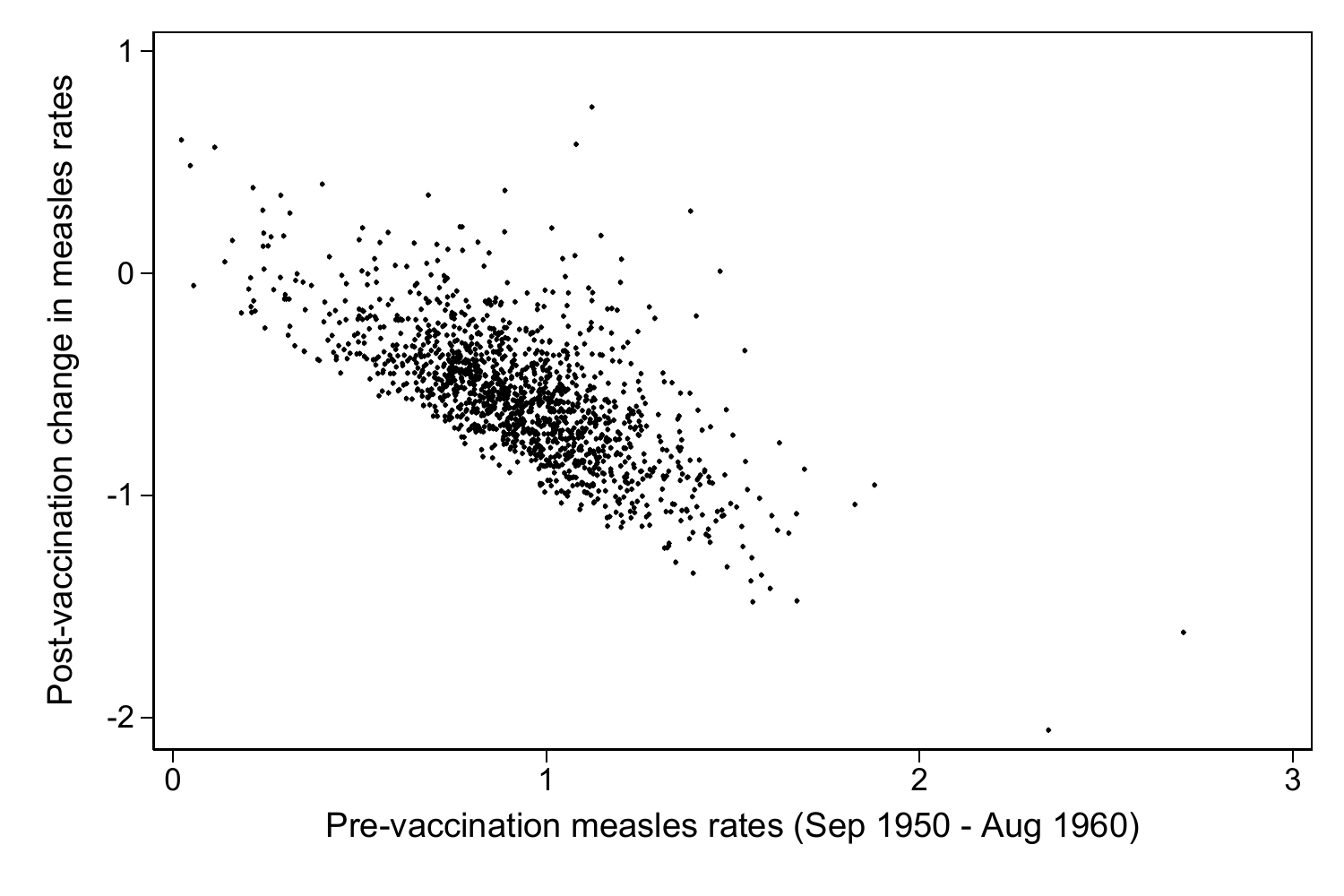"}\hfill
     \includegraphics[scale=0.55, trim=5 15 5 5, clip]{"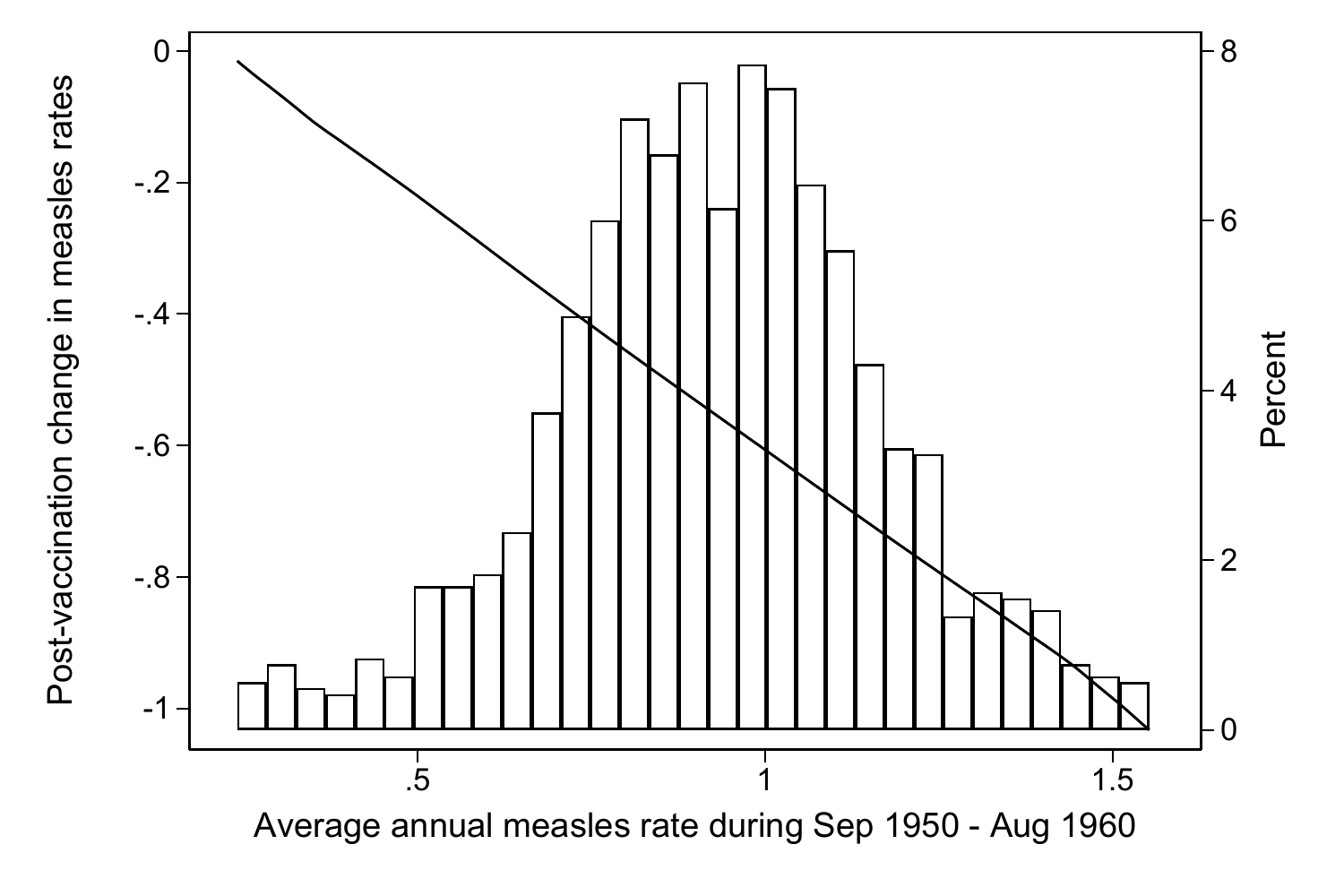"} \par}
     {\scriptsize \singlespacing Note: The horizontal axis measures the average annual measles rate between September 1950 and August 1960, in cases per 100 people. The vertical axis measures the change in the average annual measles rate following the introduction of the measles vaccine, comparing the post-vaccination period between September 1968 and August 1972 with the pre-vaccination period between September 1950 and August 1960. In the left panel, each data point corresponds to a district, excluding any districts that participated in the 1966 trial. In the panel on the right, the histogram shows the distribution of district-level measles rates during the pre-vaccination period. The solid line represents the conditional mean change in district measles rates between 1950-60 and 1968-72 from a local linear regression on the districts' pre-vaccination rates (bandwidth: 0.25 cases per 100 people). Districts in the top and bottom 1\% of the pre-vaccination measles rate distribution and districts participating in the 1966 trial are excluded from this graph. \par}
\end{figure}
\FloatBarrier
%

%==================BEGIN TABLE=================%
\begin{table}[h!]\centering\small
  \begin{threeparttable}
    \caption{\label{table_descriptives_lowhighdistricts}Descriptive statistics for districts with above- and below-median measles infection rates prior to vaccination} %%TABLE TITLE
    
                       \begin{tabular*}{\linewidth}{@{\hskip\tabcolsep\extracolsep\fill}l*{2}{S[table-format=1.3  ] S[table-format=1.3  ] }}
                       \toprule
                       \input{"tables/1951census_district_descriptives_tablefragment.tex"}
                       \bottomrule
                       \end{tabular*}
                    
\begin{tablenotes}[para,flushleft]
{\footnotesize  Note: Data on social class, housing density and age left education is obtained from the 1951 UK Census \citep{Census1951,VisionOfBritain}.}
\end{tablenotes}
   
  \end{threeparttable}
\end{table}
\FloatBarrier
%==================END TABLE=================

%
%
\begin{figure}[h!]
   \caption{Measles rates, 1948-1973}
   \label{figure_monthly_measles} 
     {\centering \includegraphics[scale=0.9, trim=15 15 0 10, clip]{"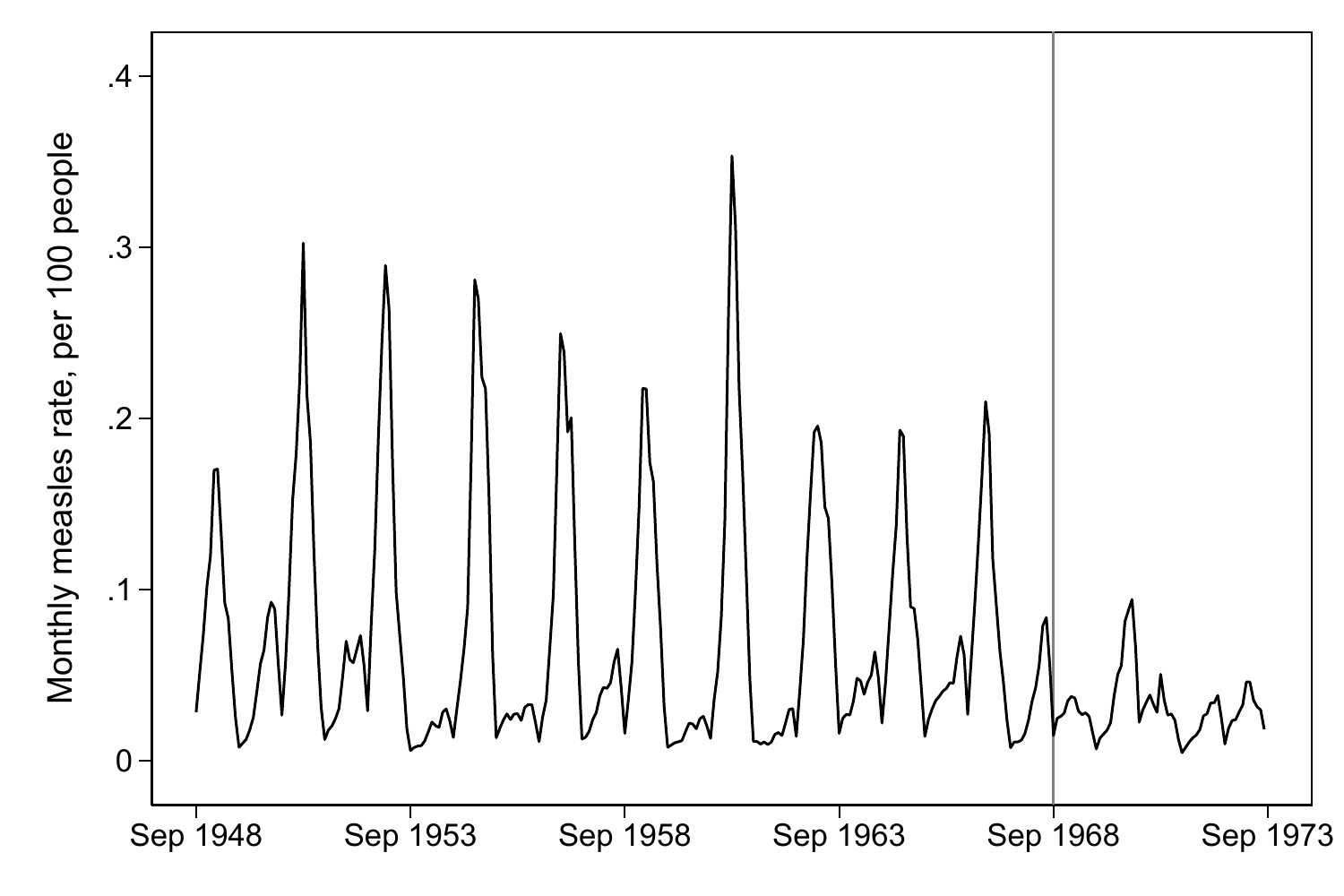"} \par}
     {\scriptsize \singlespacing Note: The figure shows monthly measles infection rates between 1948 and 1973. 11 out of 1472 districts were excluded due to (partially) missing data on measles cases or population size. The grey vertical line represents the last observation prior to the vaccine roll-out. Districts participating in the 1966 trial are excluded from the figure. 
\par}
\end{figure}
\FloatBarrier
%

%==================BEGIN TABLE=================%
\begin{table}[h!]\centering\small
  \begin{threeparttable}
    \caption{\label{table_vaccine_intensity_quals}Long-term effects of the measles vaccine introduction -- Other educational outcomes} %%TABLE TITLE
    
                       \begin{tabular*}{\linewidth}{@{\hskip\tabcolsep\extracolsep\fill}l*{4}{S[table-format=1.3  ] S[table-format=1.3  ] S[table-format=1.3  ] S[table-format=1.3  ]}}
                       \toprule
                       \input{"tables/quals_combined_measles_tablefragment.tex"}
                       \bottomrule
                       \end{tabular*}
                     
\begin{tablenotes}[para,flushleft]
{\footnotesize  Note: The dependent variables are indicators for completion of any qualification, an upper secondary qualification and a degree qualification respectively.  The explanatory variables of interest are the share of the given age periods during which the individual was exposed to the vaccination program, interacted with the measles cases per 100 people prior to the vaccination program. Individuals born in districts that participated in the 1966 trial are excluded from the sample. Standard errors clustered at the district of birth level are shown in parentheses. Significance levels are indicated as follows: * p$<$0.1, ** p$<$0.05, *** p$<$0.01}
\end{tablenotes}

  \end{threeparttable}
\end{table}
\FloatBarrier
%==================END TABLE=================

%==================BEGIN TABLE=================%
\begin{table}[h!]\centering\scriptsize
  \begin{threeparttable}
    \caption{\label{table_trial_quals}Long-term effects of the 1966 measles vaccine trial on the completion of qualifications} %%TABLE TITLE
    
                       \begin{tabular*}{\linewidth}{@{\hskip\tabcolsep\extracolsep\fill}l*{5}{S[table-format=1.3  ] S[table-format=1.3  ] S[table-format=1.3  ] S[table-format=1.3  ] S[table-format=1.3  ] }}
                       \toprule
                       \input{"tables/quals_largetrial_tablefragment.tex"}
                       \bottomrule
                       \end{tabular*}
                     
\begin{tablenotes}[para,flushleft]
{\scriptsize  Note: The dependent variables are indicators for completion of any qualification, an upper secondary qualification and a degree qualification respectively. The explanatory variable of interest is an indicator for districts participating in the trial, interacted with the period of exposure (in years) to the vaccine trial period during the given age periods. We focus here on the trial  targeting susceptible children up to age 10 or 12. Individuals born in trial districts targeting children up to age 2 are excluded from the sample. The sample was furthermore restricted to individuals born in a district with a population density within one standard deviation of the mean among the trial districts. Standard errors clustered at the district of birth level are shown in parentheses. Significance levels are indicated as follows: * p$<$0.1, ** p$<$0.05, *** p$<$0.01}
\end{tablenotes}

  \end{threeparttable}
\end{table}
\FloatBarrier
%==================END TABLE=================

%==================BEGIN TABLE=================%
\begin{table}[h!]\centering\tiny
  \begin{threeparttable}
    \caption{\label{table_vaccine_intensity_1to6_education_height_pgisplit}Gene-environment interplay: Long-term effects of the measles vaccine introduction -- Sample split by PGI} %%TABLE TITLE
    \setlength\tabcolsep{5.5pt}
                       \begin{tabular*}{\linewidth}{@{\hskip\tabcolsep\extracolsep\fill}l*{6}{S[table-format=1.3]}}
                       \toprule
     					\input{"tables/height_education_vacc1to6_pgisplit_tablefragment.tex"}
                       \bottomrule
                       \end{tabular*}

\begin{tablenotes}[para,flushleft]
{\tiny  Note: The explanatory variable of interest is the share of the age period from 1 to 6 years during which the individual was exposed to the vaccination program, interacted with the measles cases per 100 people prior to the vaccination program. Panel A is for the sub-sample with an above-median genetic propensity for education (columns 1-3) / height (columns 4-6), Panel B for the sub-sample with a below-median propensity. The measures of genetic propensity are based on: EA3 = \citet{lee2018gene}, Repository = \citet{Becker2021}, UKB = GWAS using the UK Biobank (excluding the sample used for our main analysis), GIANT = \citet{Wood2014GIANT}. Standard errors clustered at the district of birth level are shown in parentheses. Significance levels are indicated as follows: * p$<$0.1, ** p$<$0.05, *** p$<$0.01}
\end{tablenotes}
  \end{threeparttable}
\end{table}
\FloatBarrier
%==================END TABLE=================

%==================BEGIN TABLE=================%
\begin{table}[h!]\centering\tiny
  \begin{threeparttable}
    \caption{\label{table_vaccine_intensity_1to6_education_height_GxE_UKB}Gene-environment interplay: Long-term effects of the measles vaccine introduction -- PGIs based on a UK Biobank GWAS} %%TABLE TITLE
    \setlength\tabcolsep{5.5pt}
                       \begin{tabular*}{\linewidth}{@{\hskip\tabcolsep\extracolsep\fill}l*{8}{S[table-format=1.3]}}
                       \toprule
     					\input{"tables/height_education_vacc1to6_GxE_UKB_tablefragment.tex"}
                       \bottomrule
                       \end{tabular*}

\begin{tablenotes}[para,flushleft]
{\tiny  Note: The explanatory variable of interest is the share of the age period from 1 to 6 years during which the individual was exposed to the vaccination program, interacted with the measles cases per 100 people prior to the vaccination program. This measure of treatment intensity for the vaccine introduction is furthermore interacted with the polygenic index (PGI) for education (columns 1-4) / height (columns 5-8). The PGIs were constructed
using the summary statistics from a GWAS on individuals from the UK Biobank (excluding the sample used for our main analysis). Individuals born in districts that participated in the 1966 trial are excluded from the samples. Standard errors clustered at the district of birth level are shown in parentheses. Significance levels are indicated as follows: * p$<$0.1, ** p$<$0.05, *** p$<$0.01}
\end{tablenotes}
  \end{threeparttable}
\end{table}
\FloatBarrier
%==================END TABLE=================

%==================BEGIN TABLE=================%
\begin{table}[h!]\centering\tiny
  \begin{threeparttable}
    \caption{\label{table_vaccine_intensity_1to6_education_height_sibsample_GxE_EA3_GIANT}Gene-environment interplay in the sibling sample: Long-term effects of the measles vaccine introduction} %%TABLE TITLE
    \setlength\tabcolsep{5.5pt}
                       \begin{tabular*}{\linewidth}{@{\hskip\tabcolsep\extracolsep\fill}l*{8}{S[table-format=1.3]}}
                       \toprule
     					\input{"tables/height_education_vacc1to6_sibsample_GxE_GIANT_EA3_tablefragment.tex"}
                       \bottomrule
                       \end{tabular*}

\begin{tablenotes}[para,flushleft]
{\tiny  Note: The sample was restricted to full siblings - identified based on genetic relatedness. The explanatory variable of interest is the share of the age period from 1 to 6 years during which the individual was exposed to the vaccination program, interacted with the measles cases per 100 people prior to the vaccination program. This measure of treatment intensity for the vaccine introduction is furthermore interacted with the polygenic index (PGI) for education (columns 1-4) / height (columns 5-8). The measure of genetic propensity for education is based on summary statistics from \citet{lee2018gene}, the measure for height is based on summary statistics from \citet{Wood2014GIANT}. Individuals born in districts that participated in the 1966 trial are excluded from the samples. Standard errors clustered at the district of birth level are shown in parentheses. Significance levels are indicated as follows: * p$<$0.1, ** p$<$0.05, *** p$<$0.01}
\end{tablenotes}
  \end{threeparttable}
\end{table}
\FloatBarrier
%==================END TABLE=================

%==================BEGIN TABLE=================%
\begin{table}[h!]\centering\scriptsize
  \begin{threeparttable}
    \caption{\label{table_vaccine_intensity_1to6_education_height_gendersplit}Heterogeneity by gender: Long-term effects of the measles vaccine introduction and gene-environment interplay} %%TABLE TITLE
    \setlength\tabcolsep{5.5pt}    
                       \begin{tabular*}{\linewidth}{@{\hskip\tabcolsep\extracolsep\fill}l*{4}{S[table-format=1.3  ] }}
                       \toprule
     					\input{"tables/height_education_vacc1to6_gendersplit_tablefragment.tex"}
                       \bottomrule
                       \end{tabular*}

\begin{tablenotes}[para,flushleft]
{\tiny  Note: The explanatory variable of interest is the share of the age period from 1 to 6 years during which the individual was exposed to the vaccination program, interacted with the measles cases per 100 people prior to the vaccination program. This measure of treatment intensity for the vaccine introduction is furthermore interacted with the polygenic index (PGI) for education (column 2) / height (column 4). The measure of genetic propensity for education is based on summary statistics from \citet{lee2018gene}, the measure for height is based on summary statistics from \citet{Wood2014GIANT}. Individuals born in districts that participated in the 1966 trial are excluded from the samples. Standard errors clustered at the district of birth level are shown in parentheses. Significance levels are indicated as follows: * p$<$0.1, ** p$<$0.05, *** p$<$0.01}
\end{tablenotes}
  \end{threeparttable}
\end{table}
\FloatBarrier
%==================END TABLE=================

%
%
\begin{figure}[h!]
	\captionsetup[subfigure]{aboveskip=0pt,belowskip=-1.5pt}
   	\caption{Robustness of results to inclusion of other pre-vaccination disease rates and socio-economic measures interacted with the post-vaccine share}
  	\label{figure_robustness_addit_controls} 
     {\centering 
		\begin{subfigure}{0.475\linewidth}
		\caption{Years of education}
		\includegraphics[width=\linewidth, trim=5 5 5 5, clip]{"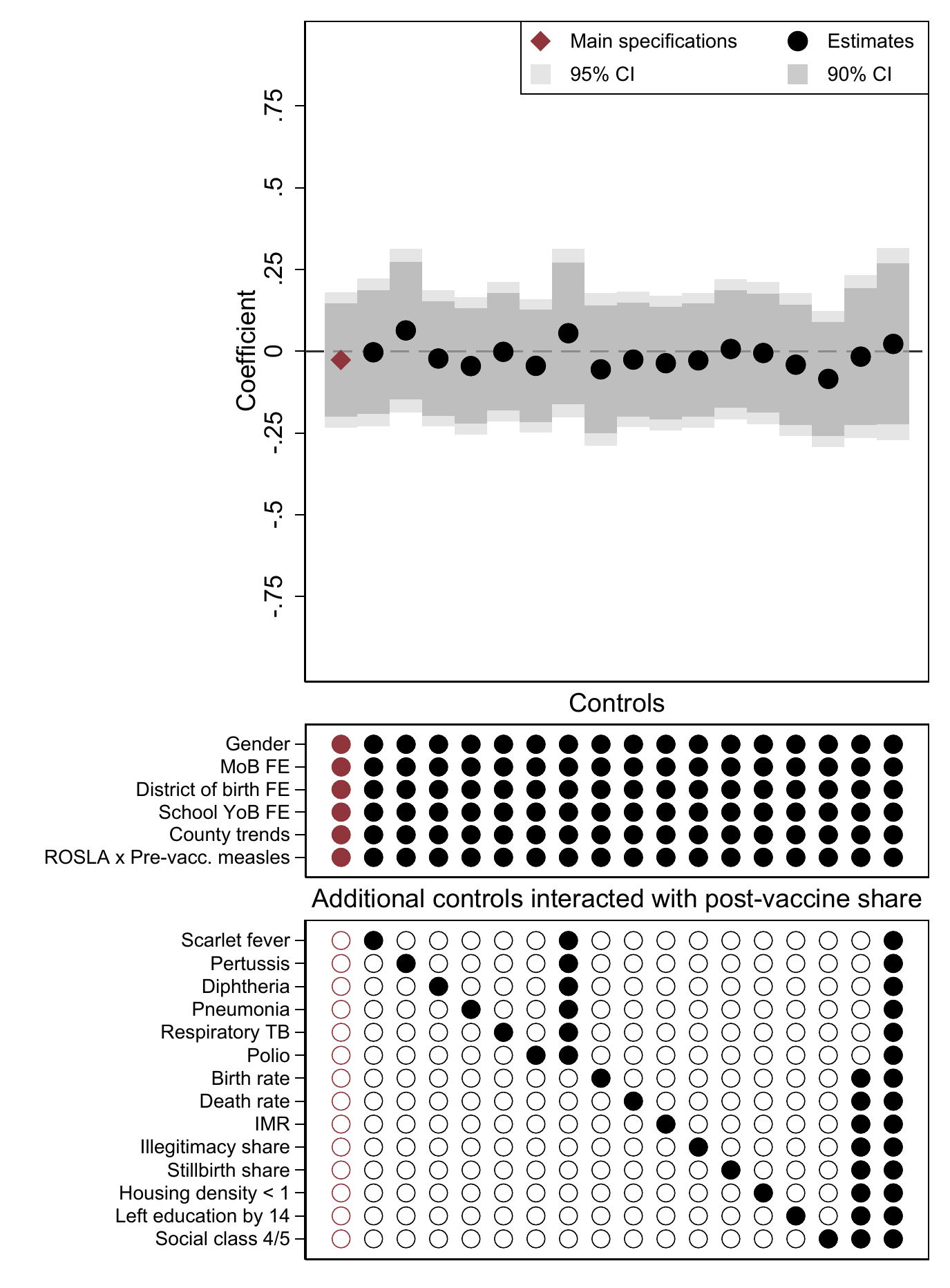"}
		\end{subfigure}   
		\hfill
		\begin{subfigure}{0.475\linewidth}
		\caption{Height}  
      	\includegraphics[width=\linewidth, trim=5 5 5 5, clip]{"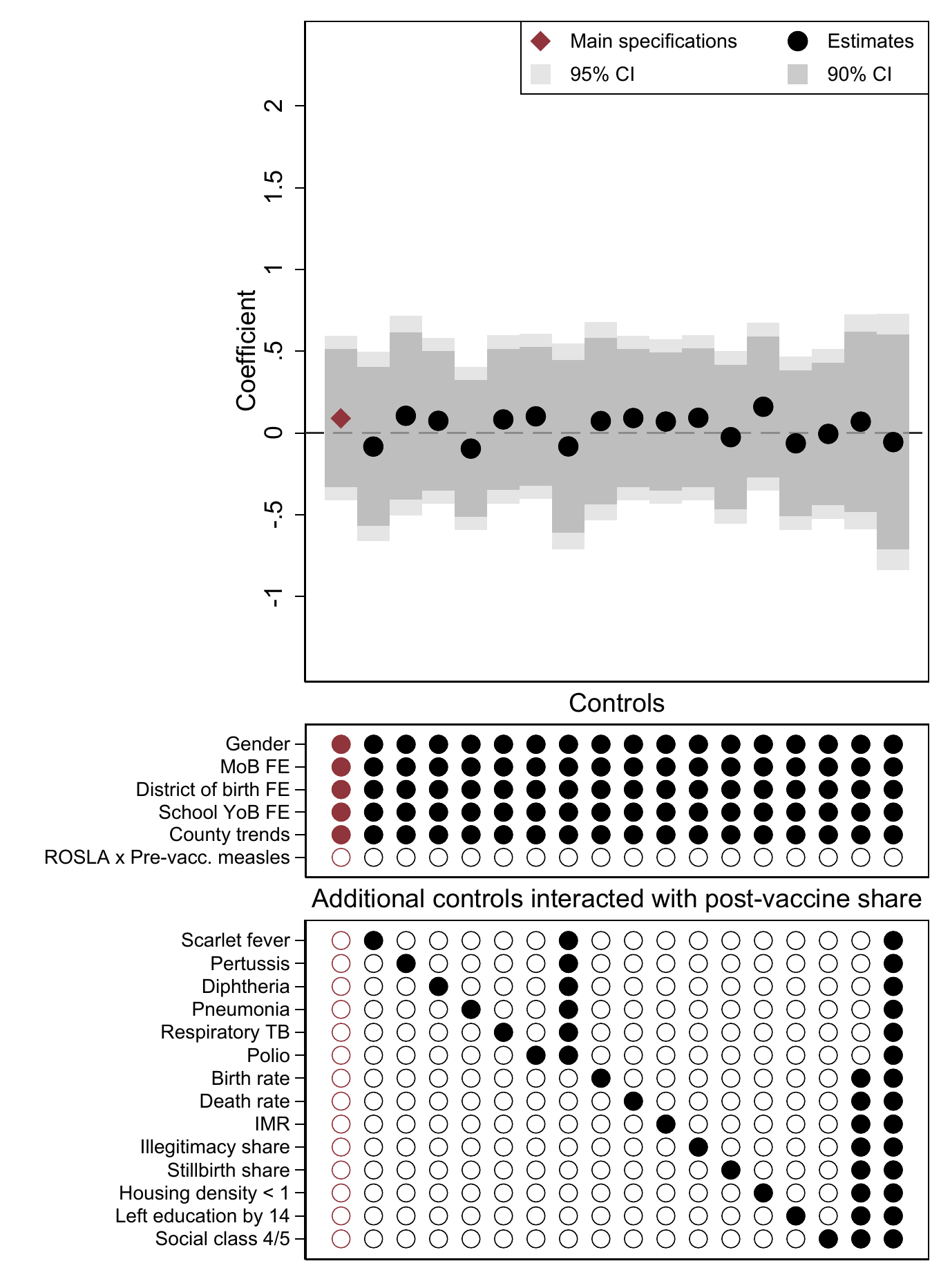"}
		\end{subfigure} \par}
     {\scriptsize \singlespacing Note: The figure plots the estimated coefficient for the explanatory variable of interest in Panel A of \autoref{table_vaccine_intensity_education_height}, i.e. for the share of the age period from 1 to 6 years during which the individual was exposed to the vaccination program interacted with the measles cases per 100 people prior to the vaccination program. We focus on specifications (4) and (8) in \autoref{table_vaccine_intensity_education_height} and include several pre-vaccination disease rates and socio-economic measures, all interacted with the share of the age period from 1 to 6 years during which the individual was exposed to the vaccination program. \par}
\end{figure}
\FloatBarrier
\begin{figure}[h!]
	\captionsetup[subfigure]{aboveskip=0pt,belowskip=-1.5pt}
   	\caption{Robustness of gene-environment interaction results to inclusion of other pre-vaccination disease rates and socio-economic measures interacted with the post-vaccine share}
  	\label{figure_robustness_GxE_addit_controls} 
     {\centering 
		\begin{subfigure}{0.475\linewidth}
		\caption{Years of education}
		\includegraphics[width=\linewidth, trim=5 5 5 5, clip]{"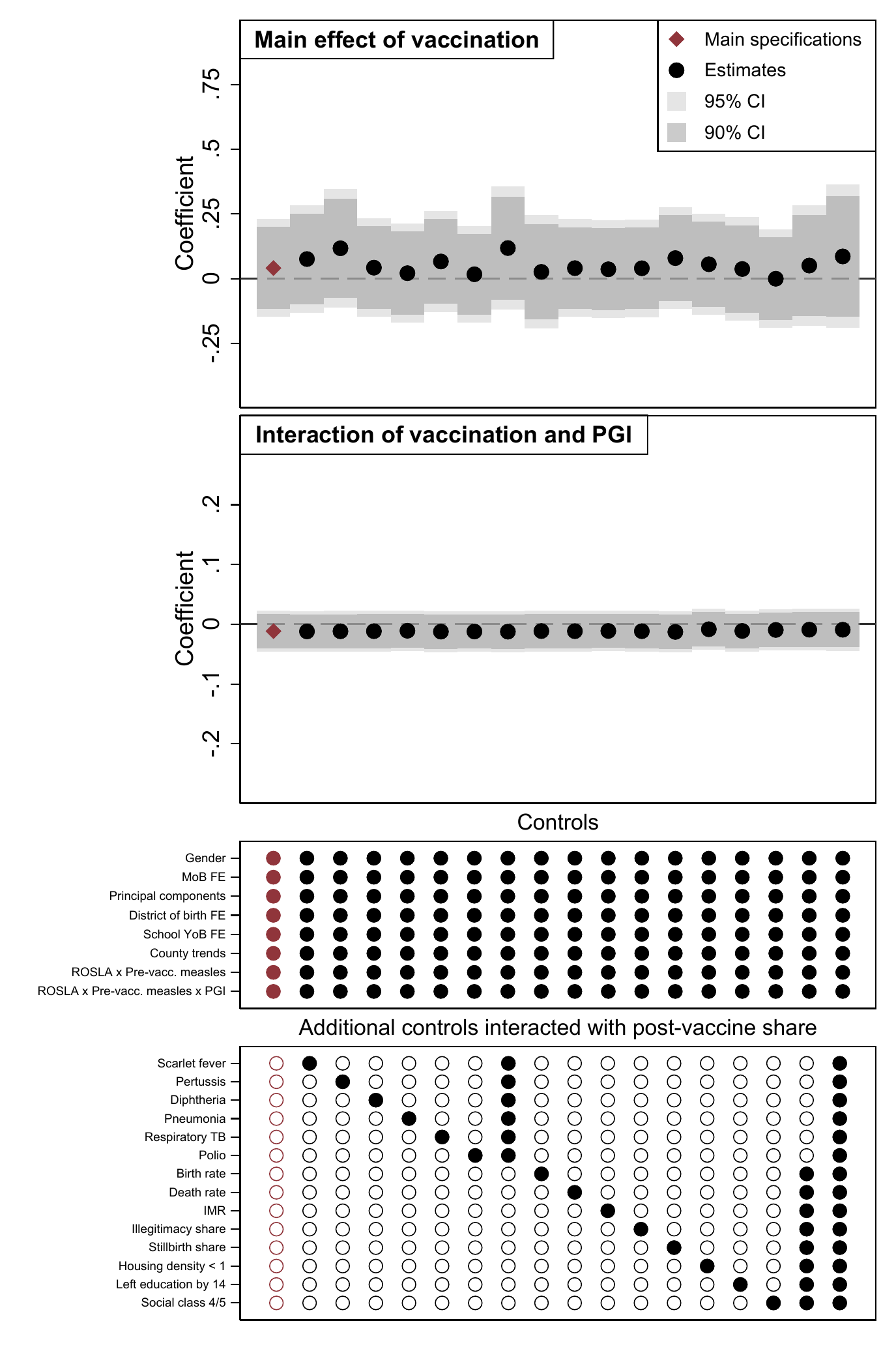"}
		\end{subfigure}   
		\hfill
		\begin{subfigure}{0.475\linewidth}
		\caption{Height}  
      	\includegraphics[width=\linewidth, trim=5 5 5 5, clip]{"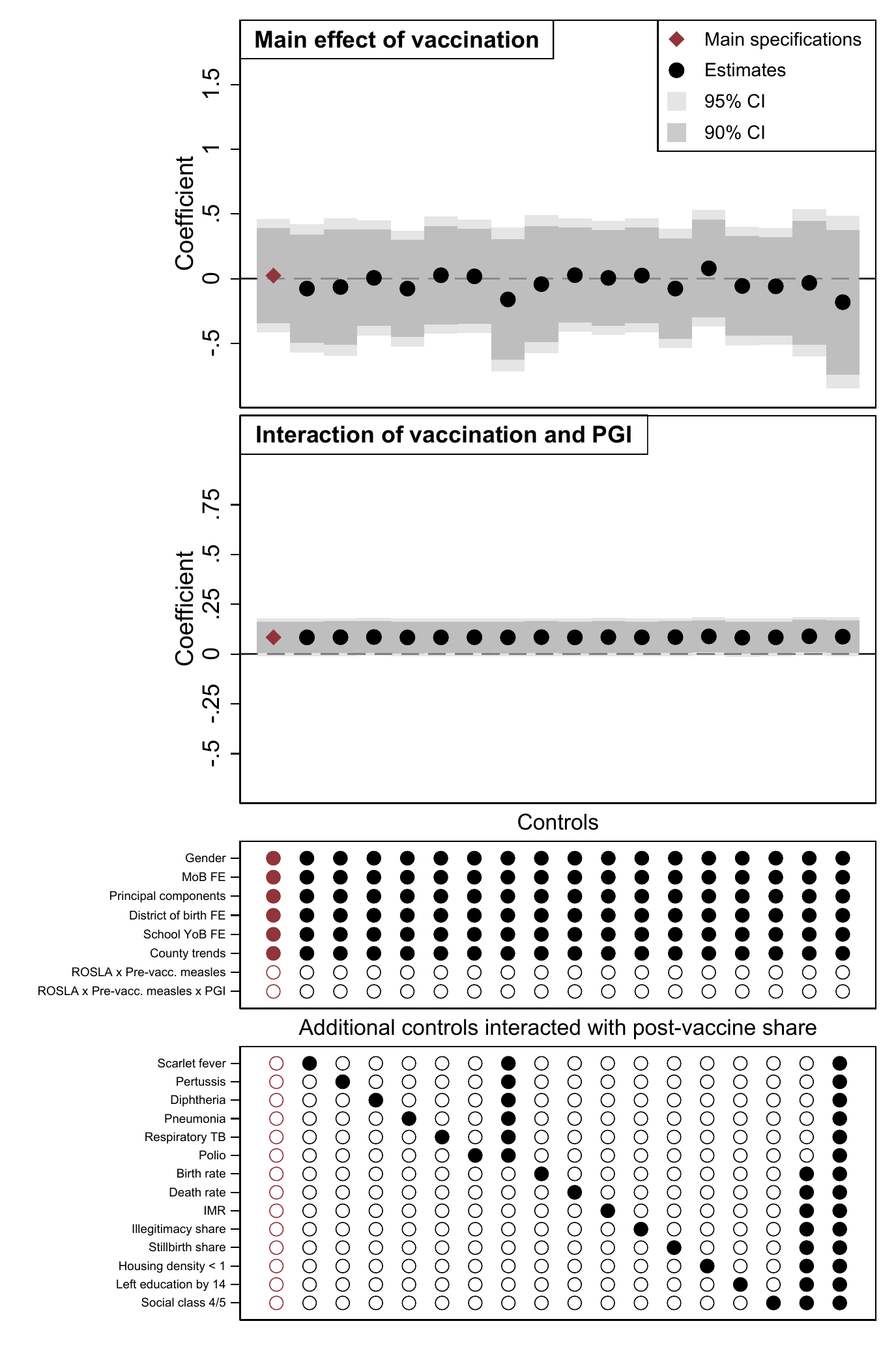"}
		\end{subfigure} \par}
     {\scriptsize \singlespacing Note: The figure plots the estimated coefficients for the explanatory variables of interest in  \autoref{table_vaccine_intensity_1to6_education_height_GxE_EA3_GIANT}, i.e. for the share of the age period from 1 to 6 years during which the individual was exposed to the vaccination program interacted with the measles cases per 100 people prior to the vaccination program and for its interaction with the PGI. We focus on specifications (4) and (8) in \autoref{table_vaccine_intensity_1to6_education_height_GxE_EA3_GIANT} and include several pre-vaccination disease rates and socio-economic measures, all interacted with the share of the age period from 1 to 6 years during which the individual was exposed to the vaccination program. \par}
\end{figure}
\FloatBarrier
%
%

%==================BEGIN TABLE=================%
\begin{table}[h!]\centering\scriptsize
  \begin{threeparttable}
    \caption{\label{table_vaccine_intensity_education_height_difftrends}Long-term effects of the measles vaccine introduction -- Differential trends} %%TABLE TITLE
    
                       \begin{tabular*}{\linewidth}{@{\hskip\tabcolsep\extracolsep\fill}l*{8}{S[table-format=1.3  ] S[table-format=1.3  ] S[table-format=1.3  ] S[table-format=1.3  ] S[table-format=1.3  ] S[table-format=1.3  ] S[table-format=1.3  ]}}
                       \toprule
                       \input{"tables/height_education_measlesvacc_difftrends_tablefragment.tex"}
                       \bottomrule
                       \end{tabular*}
                     
\begin{tablenotes}[para,flushleft]
{\tiny  Note: The explanatory variable of interest is the share of the age period from 1 to 6 years during which the individual was exposed to the vaccination program, interacted with the measles cases per 100 people prior to the vaccination program. Individuals born in districts that participated in the 1966 trial are excluded from the sample. Standard errors clustered at the district of birth level are shown in parentheses. Significance levels are indicated as follows: * p$<$0.1, ** p$<$0.05, *** p$<$0.01}
\end{tablenotes}

  \end{threeparttable}
\end{table}
\FloatBarrier
%==================END TABLE=================

%==================BEGIN TABLE=================%
\begin{table}[h!]\centering\tiny
  \begin{threeparttable}
    \caption{\label{table_vaccine_intensity_education_height_GxE_difftrends}Gene-environment interplay: Long-term effects of the measles vaccine introduction -- Differential trends} %%TABLE TITLE
    
                       \begin{tabular*}{\linewidth}{@{\hskip\tabcolsep\extracolsep\fill}l*{8}{S[table-format=1.3  ] S[table-format=1.3  ] S[table-format=1.3  ] S[table-format=1.3  ] S[table-format=1.3  ] S[table-format=1.3  ] S[table-format=1.3  ]}}
                       \toprule
                       \input{"tables/height_education_vacc1to6_GxE_difftrends_tablefragment.tex"}
                       \bottomrule
                       \end{tabular*}
                     
\begin{tablenotes}[para,flushleft]
{\tiny  Note: The explanatory variable of interest is the share of the age period from 1 to 6 years during which the individual was exposed to the vaccination program, interacted with the measles cases per 100 people prior to the vaccination program. This measure of treatment intensity for the vaccine introduction is furthermore interacted with the polygenic index (PGI) for education (columns 1-4) / height (columns 5-8). The measure of genetic propensity for education is based on summary statistics from \citet{lee2018gene}, the measure for height is based on summary statistics from \citet{Wood2014GIANT}. Individuals born in districts that participated in the 1966 trial are excluded from the samples. Standard errors clustered at the district of birth level are shown in parentheses. Significance levels are indicated as follows: * p$<$0.1, ** p$<$0.05, *** p$<$0.01}
\end{tablenotes}

  \end{threeparttable}
\end{table}
\FloatBarrier
%==================END TABLE=================

%
%
\begin{figure}[h!]
	\captionsetup[subfigure]{aboveskip=0pt,belowskip=-1.5pt}
   	\caption{Robustness of results to the use of alternative pre-vaccination time windows}
  	\label{figure_robustness_prevacc_periods} 
     {\centering 
		\begin{subfigure}{0.475\linewidth}
		\caption{Years of education}
		\includegraphics[width=\linewidth, trim=5 5 5 5, clip]{"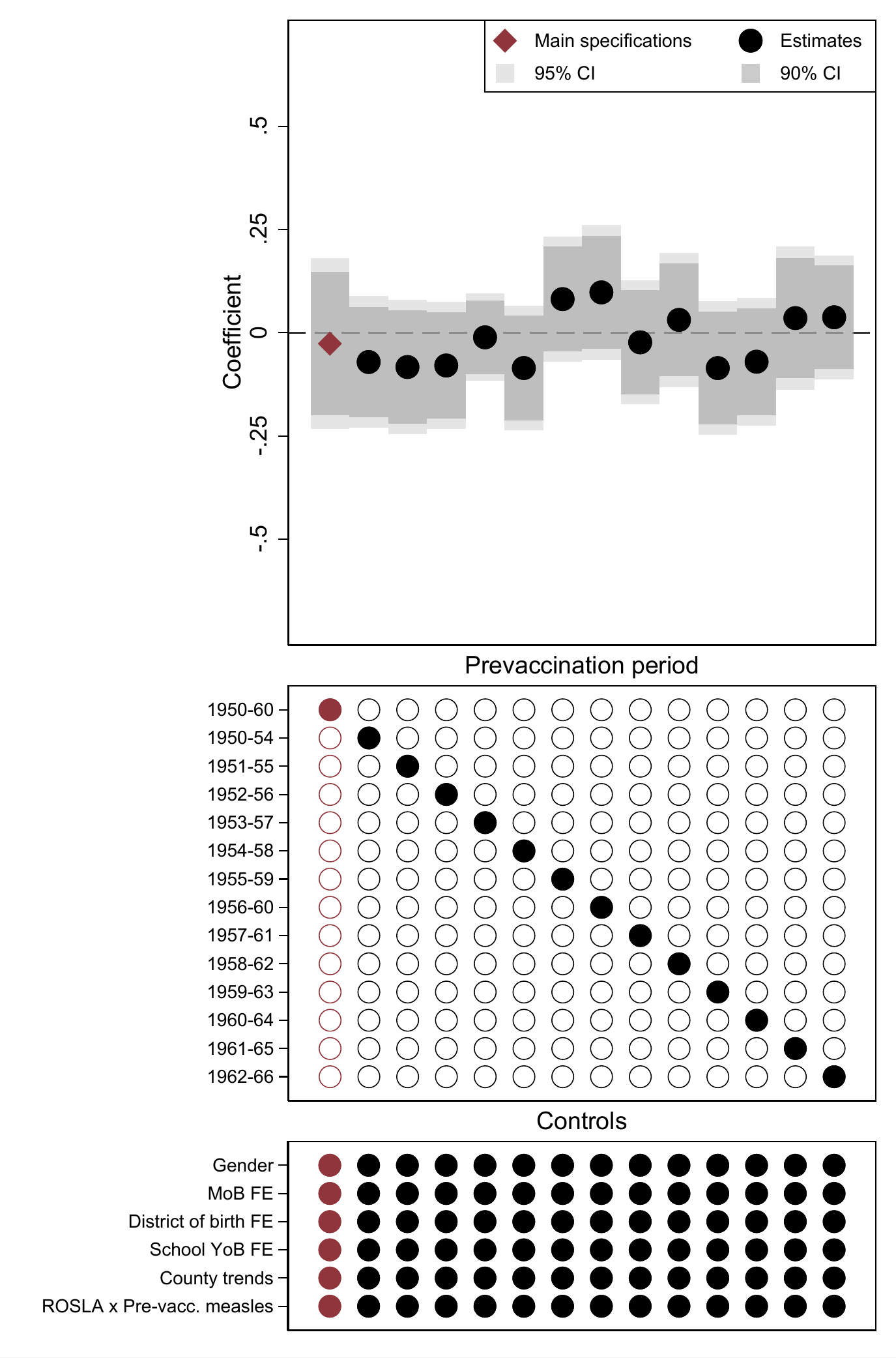"}
		\end{subfigure}   
		\hfill
		\begin{subfigure}{0.475\linewidth}
		\caption{Height}  
      	\includegraphics[width=\linewidth, trim=5 5 5 5, clip]{"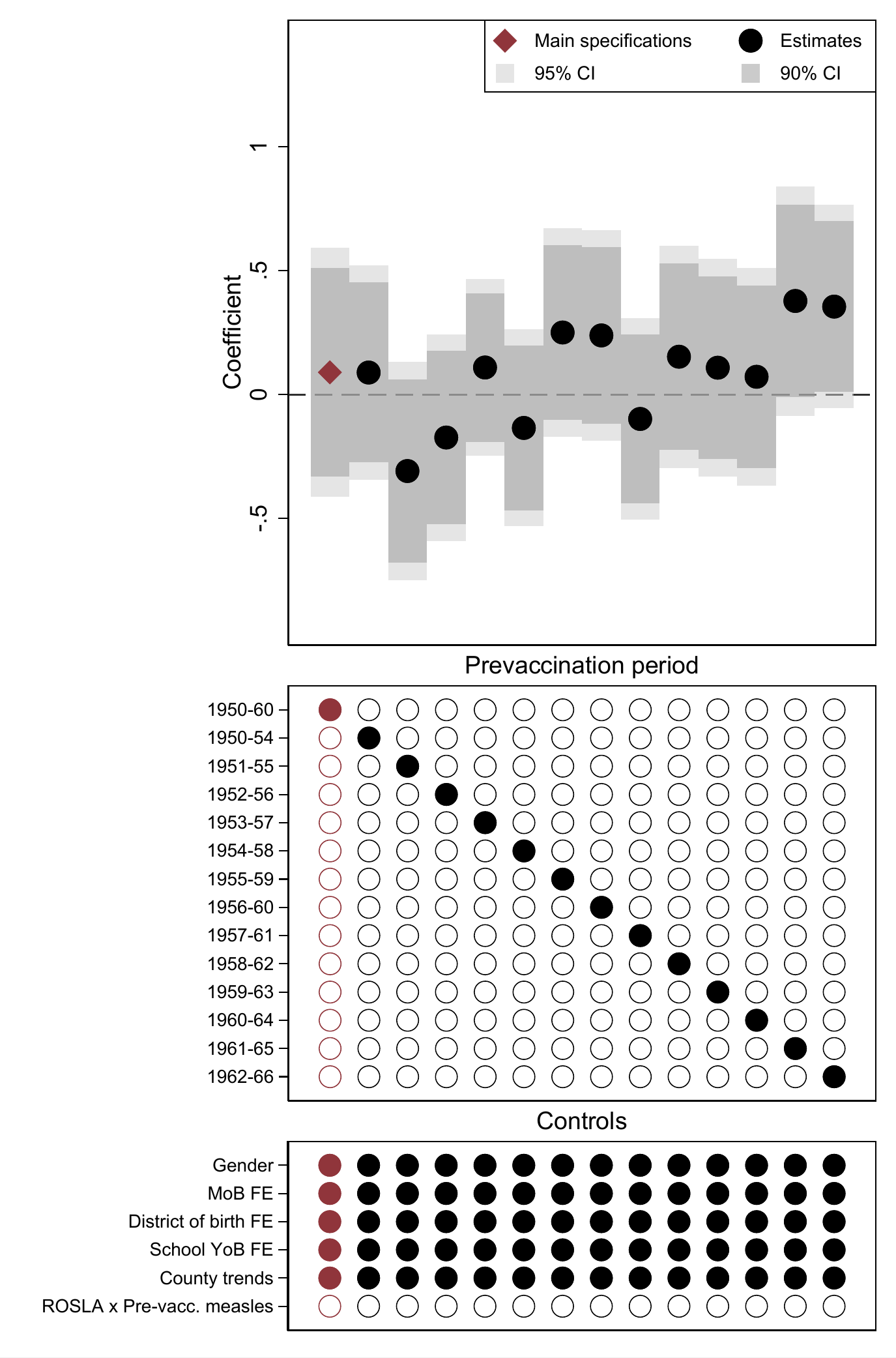"}
		\end{subfigure} \par}
     {\scriptsize \singlespacing Note: The figure plots the estimated coefficient for the explanatory variable of interest in Panel A of \autoref{table_vaccine_intensity_education_height}, i.e. for the share of the age period from 1 to 6 years during which the individual was exposed to the vaccination program interacted with the measles cases per 100 people prior to the vaccination program. We focus on specifications (4) and (8) in \autoref{table_vaccine_intensity_education_height} and use different time windows for the pre-vaccination measles rates. \par}
\end{figure}
\FloatBarrier
\begin{figure}[h!]
	\captionsetup[subfigure]{aboveskip=0pt,belowskip=-1.5pt}
   	\caption{Robustness of gene-environment interaction results to the use of alternative pre-vaccination time windows}
  	\label{figure_robustness_GxE_prevacc_periods} 
     {\centering 
		\begin{subfigure}{0.475\linewidth}
		\caption{Years of education}
		\includegraphics[width=\linewidth, trim=5 5 5 5, clip]{"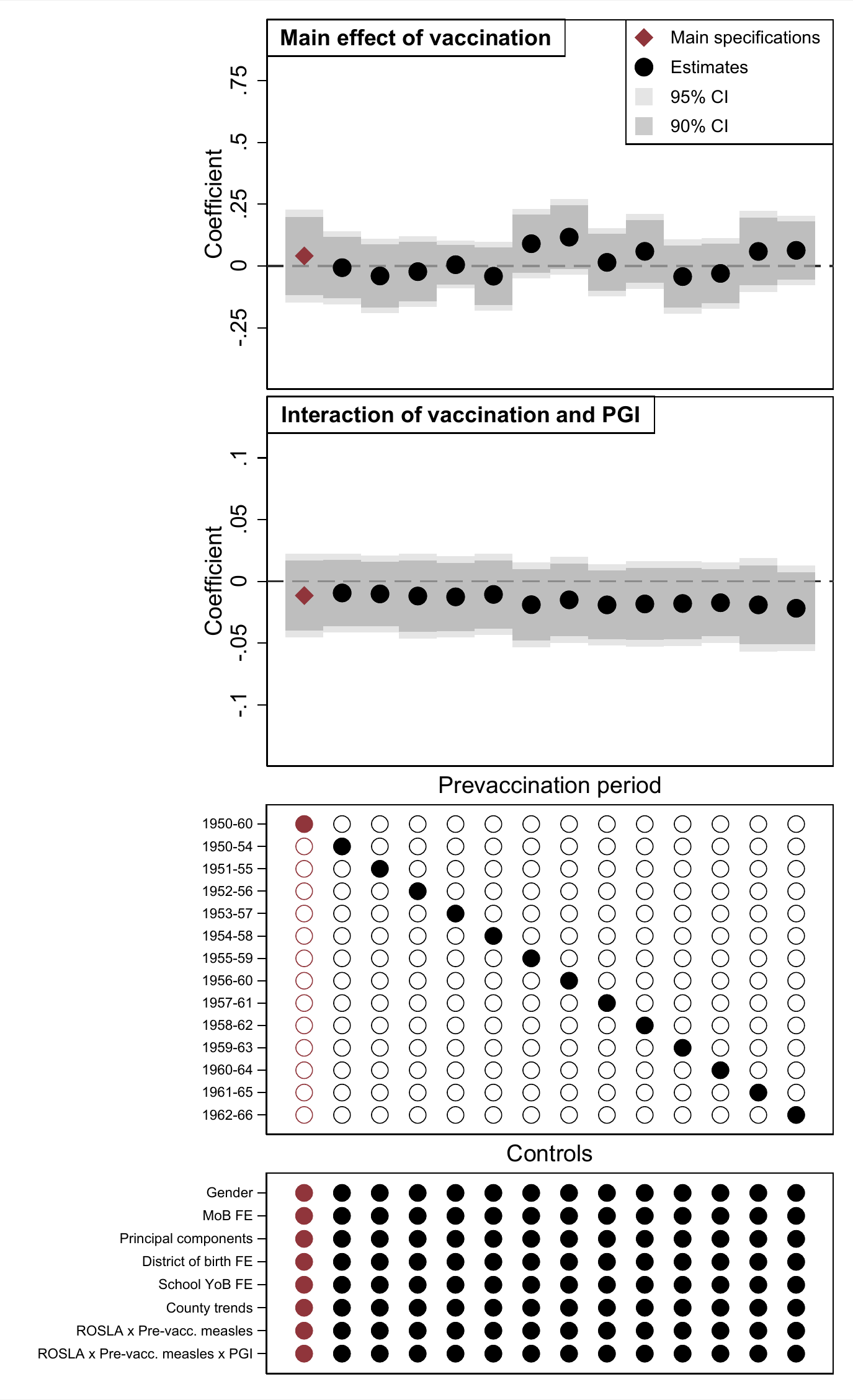"}
		\end{subfigure}   
		\hfill
		\begin{subfigure}{0.475\linewidth}
		\caption{Height}  
      	\includegraphics[width=\linewidth, trim=5 5 5 5, clip]{"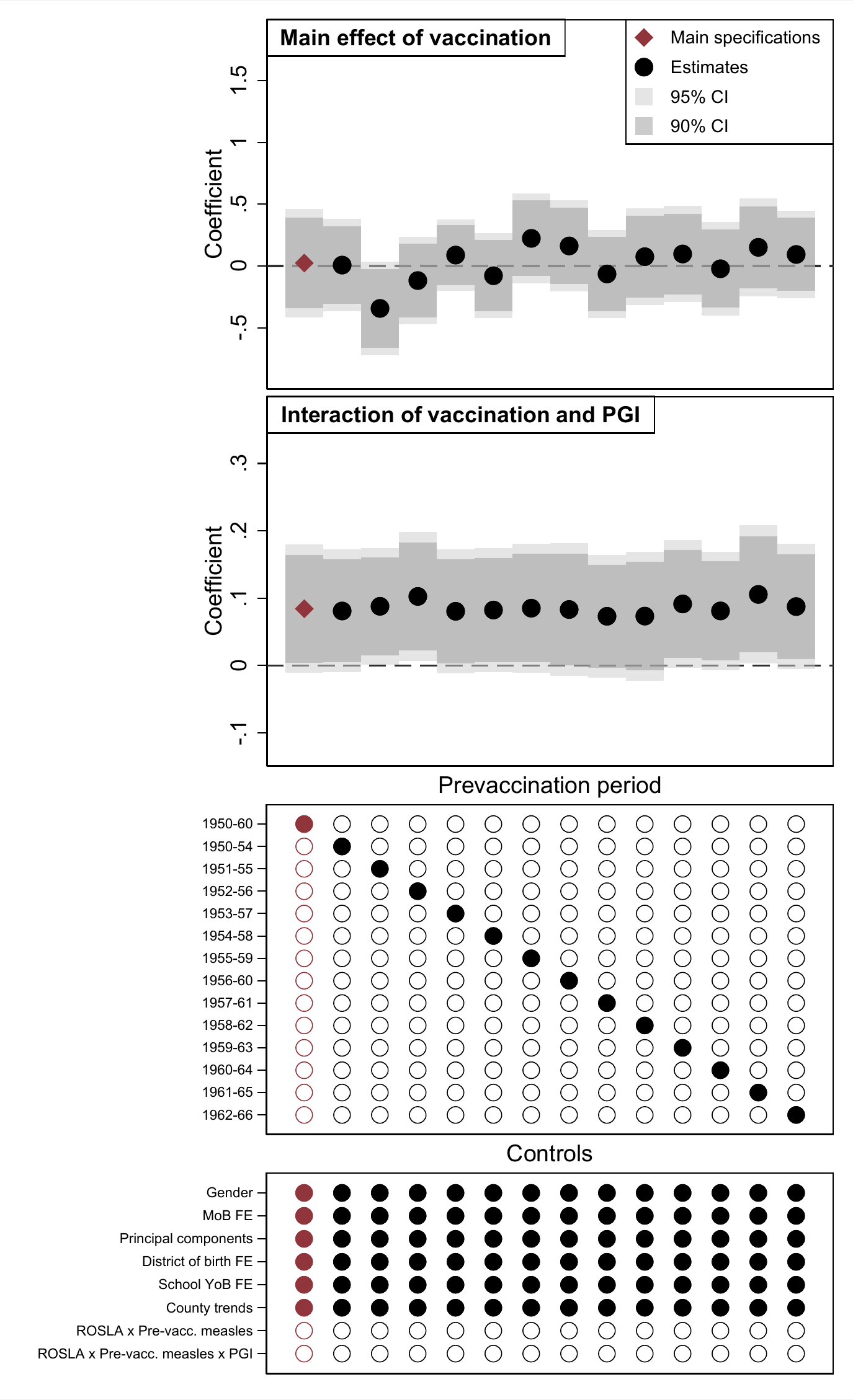"}
		\end{subfigure} \par}
     {\scriptsize \singlespacing Note: The figure plots the estimated coefficients for the explanatory variables of interest in  \autoref{table_vaccine_intensity_1to6_education_height_GxE_EA3_GIANT}, i.e. for the share of the age period from 1 to 6 years during which the individual was exposed to the vaccination program interacted with the measles cases per 100 people prior to the vaccination program and for its interaction with the PGI. We focus on specifications (4) and (8) in \autoref{table_vaccine_intensity_1to6_education_height_GxE_EA3_GIANT} and use different time windows for the pre-vaccination measles rates. \par}
\end{figure}
\FloatBarrier
%
%

%==================BEGIN TABLE=================%
\begin{table}[h!]\centering\scriptsize
  \begin{threeparttable}
    \caption{\label{table_vaccine_intensity_education_height_county_cid_yob_cluster}Long-term effects of the measles vaccine introduction - Analysis at the county level} %%TABLE TITLE
    \setlength\tabcolsep{5.5pt}
                       \begin{tabular*}{\linewidth}{@{\hskip\tabcolsep\extracolsep\fill}l*{8}{S[table-format=1.3]}}
                       \toprule
                       \input{"tables/height_education_measlesvacc_county_cid_yob_cluster_tablefragment.tex"}
                       \bottomrule
                       \end{tabular*}
                     
\begin{tablenotes}[para,flushleft]
{\tiny  Note: The explanatory variables of interest are the share of the given age periods during which the individual was exposed to the vaccination program, interacted with the measles cases per 100 people prior to the vaccination program. Individuals born in counties that participated in the 1966 trial are excluded from the sample. Standard errors clustered at the county and school year of birth level are shown in parentheses. Significance levels are indicated as follows: * p$<$0.1, ** p$<$0.05, *** p$<$0.01}
\end{tablenotes}

  \end{threeparttable}
\end{table}
\FloatBarrier
%==================END TABLE=================

%==================BEGIN TABLE=================%
\begin{table}[h!]\centering\scriptsize
  \begin{threeparttable}
    \caption{\label{table_vaccine_intensity_binary_education_height}Long-term effects of the measles vaccine introduction -- Binary measures of vaccine exposure and intensity} %%TABLE TITLE
    \setlength\tabcolsep{5.5pt}
                       \begin{tabular*}{\linewidth}{@{\hskip\tabcolsep\extracolsep\fill}l*{6}{S[table-format=1.3]}}
                       \toprule
                       \input{"tables/height_education_measlesvacc_binary_tablefragment.tex"}
                       \bottomrule
                       \end{tabular*}
                     
\begin{tablenotes}[para,flushleft]
{\tiny  Note: The explanatory variable of interest is the share of the age period from 1 to 6 years during which the individual was exposed to the vaccination program interacted with an indicator for above-median measles cases (per 100 people) prior to the vaccination program (columns 1\&4), indicator for any exposure to the vaccination program during ages 1 to 6 interacted with the measles cases per 100 people prior to the vaccination program (columns 2\&5), indicator for any exposure to the vaccination program during ages 1 to 6 interacted with an indicator for above-median measles cases (per 100 people) prior to the vaccination program (columns 3\&6). Individuals born in districts that participated in the 1966 trial are excluded from the sample. Standard errors clustered at the district of birth level are shown in parentheses. Significance levels are indicated as follows: * p$<$0.1, ** p$<$0.05, *** p$<$0.01}
\end{tablenotes}

  \end{threeparttable}
\end{table}
%==================END TABLE=================

%==================BEGIN TABLE=================%
\begin{table}[h!]\centering\scriptsize
  \begin{threeparttable}
    \caption{\label{table_vaccine_intensity_binary_education_height_GxE}Gene-environment interplay: Long-term effects of the measles vaccine introduction - Binary measures of vaccine exposure and intensity} %%TABLE TITLE
    \setlength\tabcolsep{5.5pt}
                       \begin{tabular*}{\linewidth}{@{\hskip\tabcolsep\extracolsep\fill}l*{6}{S[table-format=1.3]}}
                       \toprule
                       \input{"tables/height_education_measlesvacc_GxE_binary_tablefragment.tex"}
                       \bottomrule
                       \end{tabular*}
                     
\begin{tablenotes}[para,flushleft]
{\tiny  Note: The explanatory variable of interest is the share of the age period from 1 to 6 years during which the individual was exposed to the vaccination program interacted with an indicator for above-median measles cases (per 100 people) prior to the vaccination program (columns 1\&4), indicator for any exposure to the vaccination program during ages 1 to 6 interacted with the measles cases per 100 people prior to the vaccination program (columns 2\&5), indicator for any exposure to the vaccination program during ages 1 to 6 interacted with an indicator for above-median measles cases (per 100 people) prior to the vaccination program (columns 3\&6). This measure of treatment intensity for the vaccine introduction is furthermore interacted with the polygenic index (PGI) for education (columns 1-3) / height (columns 5-6). The measure of genetic propensity for education is based on summary statistics from \citet{lee2018gene}, the measure for height is based on summary statistics from \citet{Wood2014GIANT}. Individuals born in districts that participated in the 1966 trial are excluded from the samples. Standard errors clustered at the district of birth level are shown in parentheses. Significance levels are indicated as follows: * p$<$0.1, ** p$<$0.05, *** p$<$0.01}
\end{tablenotes}

  \end{threeparttable}
\end{table}
\FloatBarrier
%==================END TABLE=================

%
%
\begin{figure}[h!]
	\captionsetup[subfigure]{aboveskip=0pt,belowskip=-1.5pt}
   	\caption{Robustness of results to different levels of standard error clustering}
  	\label{figure_robustness_standard_errors} 
     {\centering 
		\begin{subfigure}{0.475\linewidth}
		\caption{Years of education}
		\includegraphics[width=\linewidth, trim=5 5 5 5, clip]{"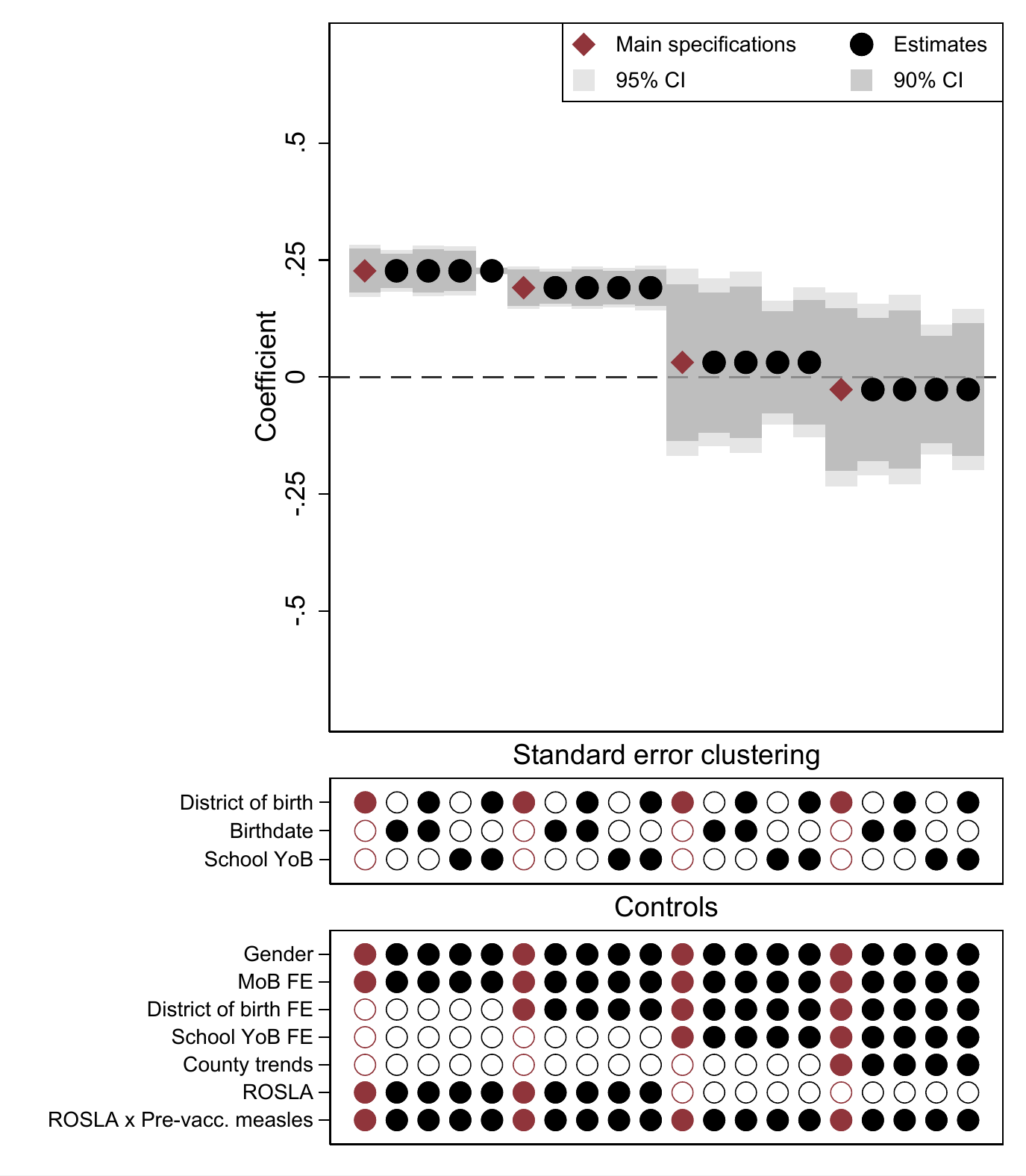"}
		\end{subfigure}   
		\hfill
		\begin{subfigure}{0.475\linewidth}
		\caption{Height}  
      	\includegraphics[width=\linewidth, trim=5 5 5 5, clip]{"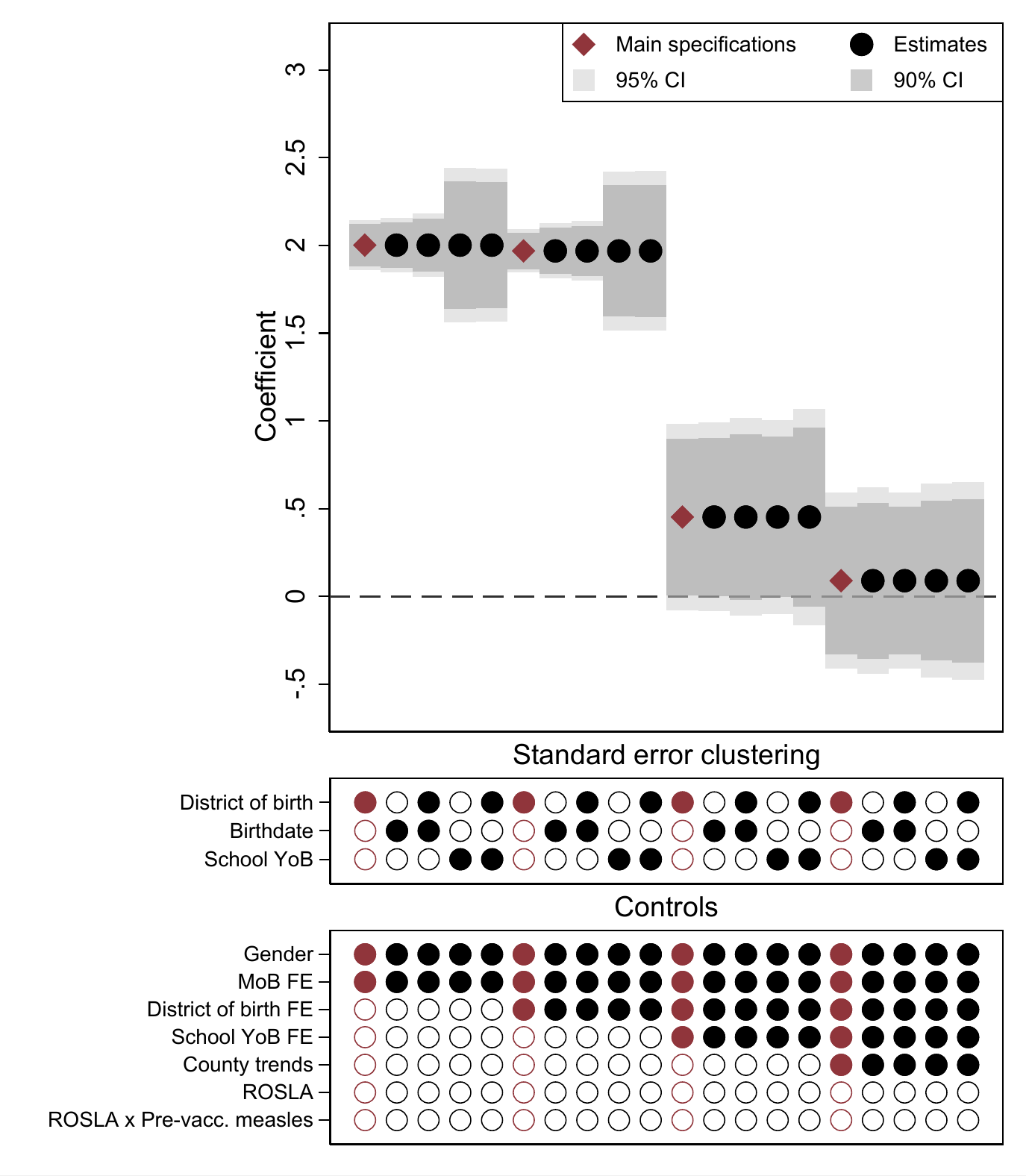"}
		\end{subfigure} \par}
     {\scriptsize \singlespacing Note: The figure plots the estimated coefficient for the explanatory variable of interest in Panel A of \autoref{table_vaccine_intensity_education_height}, i.e. for the share of the age period from 1 to 6 years during which the individual was exposed to the vaccination program interacted with the measles cases per 100 people prior to the vaccination program. \par}
\end{figure}
\FloatBarrier
\begin{figure}[h!]
	\captionsetup[subfigure]{aboveskip=0pt,belowskip=-1.5pt}
   	\caption{Robustness of gene-environment interaction results to different levels of standard error clustering}
  	\label{figure_robustness_GxE_standard_errors} 
     {\centering 
		\begin{subfigure}{0.475\linewidth}
		\caption{Years of education}
		\includegraphics[width=\linewidth, trim=5 5 5 5, clip]{"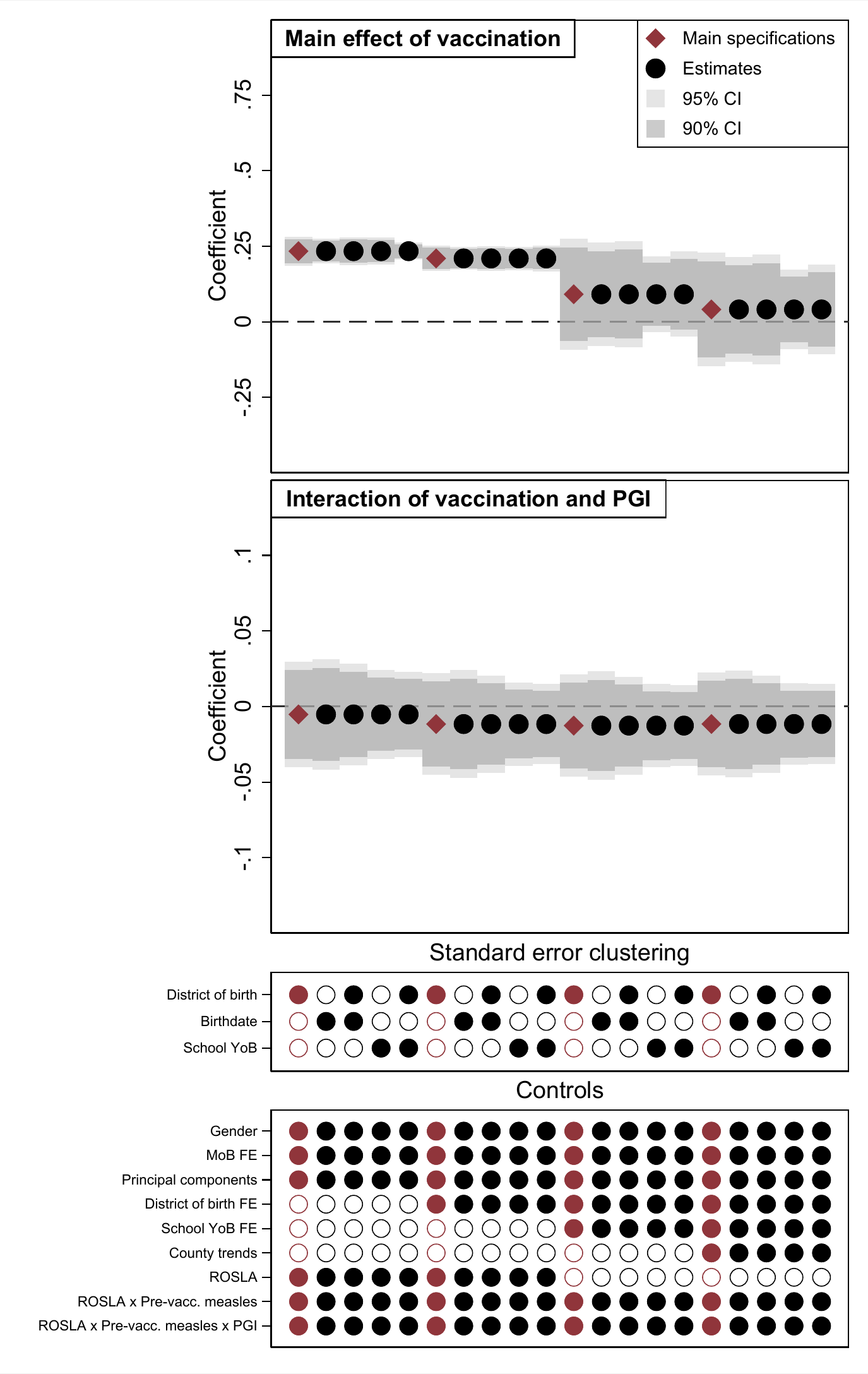"}
		\end{subfigure}   
		\hfill
		\begin{subfigure}{0.475\linewidth}
		\caption{Height}  
      	\includegraphics[width=\linewidth, trim=5 5 5 5, clip]{"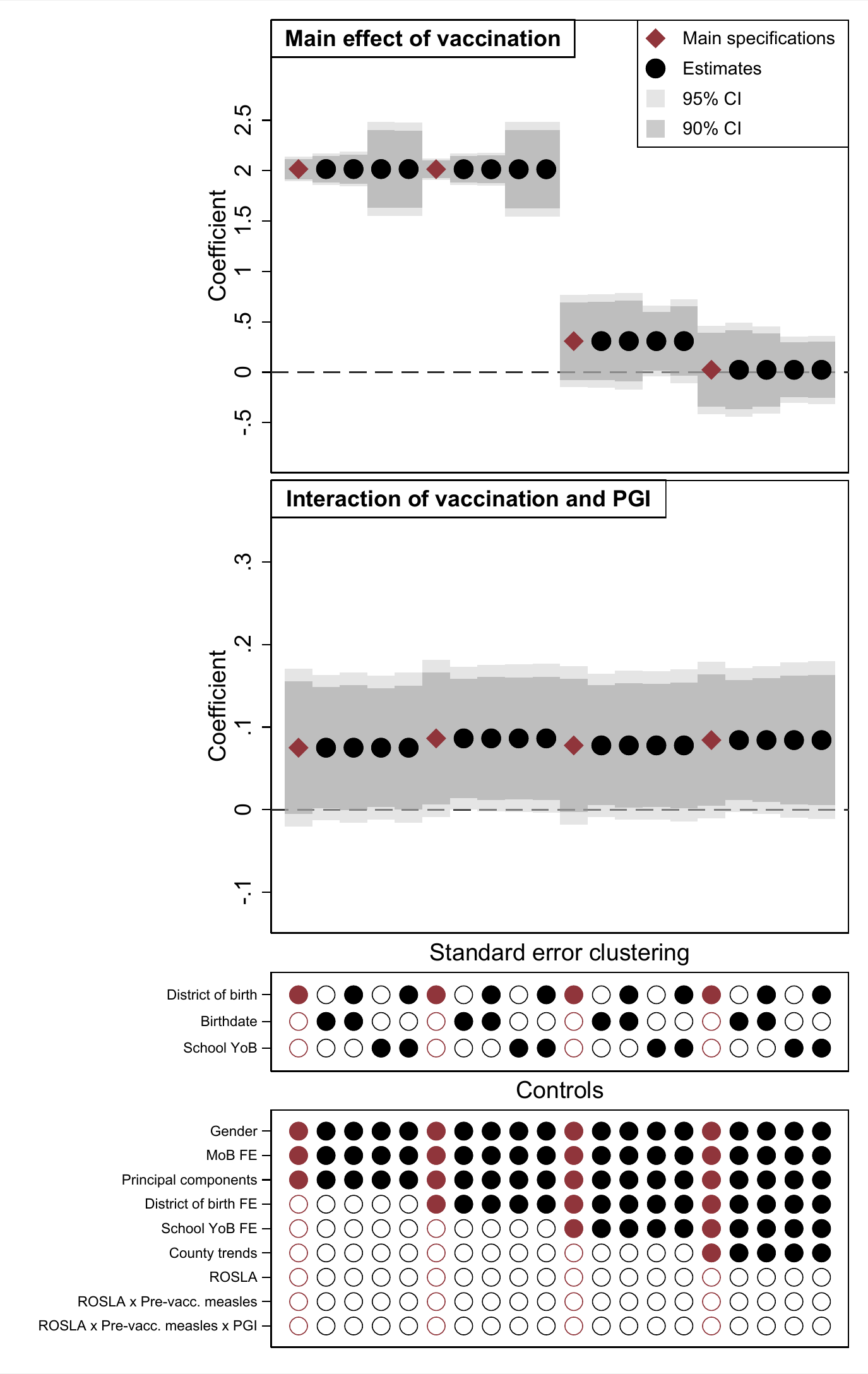"}
		\end{subfigure} \par}
     {\scriptsize \singlespacing Note: The figure plots the estimated coefficients for the explanatory variables of interest in  \autoref{table_vaccine_intensity_1to6_education_height_GxE_EA3_GIANT}, i.e. for the share of the age period from 1 to 6 years during which the individual was exposed to the vaccination program interacted with the measles cases per 100 people prior to the vaccination program and for its interaction with the PGI. \par}
\end{figure}
\FloatBarrier

\clearpage

%--------------------------------------------------------------------------------------------%
%--------------------------------------------------------------------------------------------%

\section{An introduction to genetics} \label{appendix_genetics}
Humans have 46 chromosomes stored in every cell of the human body other than gametes (sex-cells). Chromosomes come in pairs; one from one's biological mother and one from one's biological father. A single chromosome consists of a double-strand of deoxyribonucleic acid (DNA) containing a large number of `base pairs': pairs of nucleotide molecules (referred to as the `letters' A (adenine) that binds with T (thymine), and G (guanine) that binds with C (cytosine)) that together make up the human genome. In a population there will be variation in the base pairs at some locations. Such variation is known as a single nucleotide polymorphism (SNP, pronounced `snip') -- a change in the base pair at one particular locus (location) -- and is the most commonly studied genetic variation. When there are two possible base pairs at a given location (i.e., two alleles), the most frequent base pair is called the major allele, while the less frequent is called the minor allele. As humans have two copies of each chromosome, an individual can have either zero, one, or two copies of the minor allele. 

To identify specific SNPs that are robustly associated with a particular outcome of interest, so-called Genome-Wide Association Studies (GWAS) relate each SNP to the outcome in a hypothesis-free approach. As genetic datasets contain more SNPs than individuals, the SNP effects cannot be identified in a multivariate regression model. Instead, a GWAS runs a large number of univariate regressions of the outcome on each SNP. These analyses have shown that most outcomes of interest in the social sciences are `polygenic': they are affected by a large number of SNPs, each with a very small effect. To increase the predictive power of the SNPs, it is therefore customary to aggregate the individual SNPs into so-called polygenic indices (also known as polygenic scores), as:
$$
G_i = \sum_{j=1}^J \beta_j X_{ij},
$$
where $X_{ij}$ is a count of the number of minor alleles (i.e., 0, 1 or 2) at SNP $j$ for  individual $i$, and $\beta_j$ is its effect size obtained from an independent GWAS. Hence, the polygenic indices are weighted linear combinations of SNPs, where the weights are estimated in an independent GWAS. This is motivated by an additive genetic model where all SNPs contribute additively to the overall genetic endowment of an individual  \citep[see e.g.,][]{purcell2009prs}. 

In our main specification, we use polygenic indices of educational attainment and height, constructed from the summary statistics (i.e., the $\beta_j$'s) derived from a GWAS of educational attainment \citep{lee2018gene} and height \citep{Wood2014GIANT}. Note that the GWAS discovery samples for both PGIs exclude the UK Biobank to avoid over-fitting. We also examine the robustness of our results to the use of differentially-constructed polygenic indices from the polygenic repository \citep{Becker2021}. We account for linkage disequilibrium between SNPs using LDpred \citep{vilhjalmsson2015modeling}, setting the fraction of causal SNPs to 1. All polygenic indices are standardised to have zero mean and unit standard deviation in the analysis sample. 

Our main specification uses all UK Biobank individuals born between September 1949 and August 1969 in England and Wales. In the additional analysis, we re-run our analysis on a sample of siblings. Although the UK Biobank is not a family study and does not have self-reported relatedness, we observe a large number of siblings, whom we identify using the (genetic) kinship matrix. 

\clearpage

%--------------------------------------------------------------------------------------------%
%--------------------------------------------------------------------------------------------%

\section{Not accounting for the raising of the school leaving age} \label{appendix_no_rosla}
\autoref{table_vaccine_intensity_education_norosla} shows the estimates from \autoref{eq:vaccine2} for the analysis of the long-term impact of the introduction of the measles vaccination in 1968 on educational outcomes, exploiting the fact that the benefits of the vaccination campaign were stronger for districts with previously high rates of measles infections. However, we here do \textit{not} control for the additional dummy variable indicating whether the cohort was affected by the educational reform that raised the school leaving age from 15 to 16 years (see \autoref{sec:method_vacc_introduction}) and its interaction with $PreRate_d$. Hence, this analysis does not control for any differential effects this reform may have had on the educational outcomes in districts with low and high measles rates.

%==================BEGIN TABLE=================%
\begin{table}[b!]\centering\scriptsize
  \begin{threeparttable}
    \caption{\label{table_vaccine_intensity_education_norosla}Education and average measles exposure during early childhood -- The introduction of the measles vaccine; not accounting for differential trends in education that are driven by the increase in minimum school leaving age} %%TABLE TITLE
    
                       \begin{tabular*}{\linewidth}{@{\hskip\tabcolsep\extracolsep\fill}l*{4}{S[table-format=1.3]}}
                       \toprule
                       \input{"tables/ed_combined_measles_1to6_norosla_tablefragment.tex"}
                       \bottomrule
                       \end{tabular*}
                     
\begin{tablenotes}[para,flushleft]
{\scriptsize  Note: The dependent variables are years of education, indicators for completion of any qualification, an upper secondary qualification and a degree qualification respectively.  The explanatory variables of interest are the share of the given age periods during which the individual was exposed to the vaccination program, interacted with the measles cases per 100 people prior to the vaccination program. Individuals born in districts that participated in the 1966 trial are excluded from the sample. The analyses do \textit{not} control for the differential trends in education driven by the raising of the school leaving age. Standard errors clustered at the district of birth level are shown in parentheses. Significance levels are indicated as follows: * p$<$0.1, ** p$<$0.05, *** p$<$0.01}
\end{tablenotes}

  \end{threeparttable}
\end{table}
%==================END TABLE=================

Column (1) of Panel A shows that full exposure to the post-vaccine period at age 1--6 is associated with 0.32 more years of education in districts with one additional annual measles case per 100 population prior to the introduction of the vaccination, which is robust to the inclusion of district fixed effects (column 2), but not to school year of birth fixed effects (columns 3). Indeed, the estimate reduces in magnitude and with the increase in standard errors, it is no longer statistically significantly different from zero. However, the estimate remains large: in districts with 1 case of measles per 100 population prior to the vaccination campaign, individuals have 0.1 additional years of education compared to districts with no reports of measles. Once county-specific trends are included in column (4), the estimated impact on years of education attenuates to zero. 

Panels B, C and D of \autoref{table_vaccine_intensity_education_norosla} illustrate how the educational impact of the measles vaccination varied throughout the educational distribution. Panel B shows strong and statistically significant effects of the vaccine introduction on the probability of completing \textit{any} educational qualification. Even when controlling for school year, month and district of birth fixed effects, full exposure to the post-vaccine period is found to increase the probability of any qualification by 3.1 percentage points in districts with one additional annual measles case per 100 people prior to the introduction of the vaccination. However, controlling for county-specific trends again attenuates the estimate to zero (column 4). For higher levels of educational qualifications, the effects are smaller once school year of birth fixed effects are included in column (3).  

As we show in \autoref{sec:method_vacc_introduction} and \autoref{sec:results}, the estimate in columns (1)-(3) of \autoref{table_vaccine_intensity_education_norosla} for the impact on the probability of obtaining any qualification is likely to be driven by the differential trends in education for those in districts with high and low measles rates prior to vaccination which is caused by the education reform that increased the minimum school leaving age. Once we control for differential trends at the county level (column 4) or explicitly control for the differential impact of the educational reform (\autoref{table_vaccine_intensity_education_height} and \autoref{table_vaccine_intensity_quals}), the estimates reduce and are statistically insignificantly different from zero. 

\clearpage

%--------------------------------------------------------------------------------------------%
%--------------------------------------------------------------------------------------------%

\section{Difference-in-difference design with continuous treatment} \label{appendix_did}
Our empirical strategy follows a difference-in-differences (DiD), or two-way fixed effects (TWFE) approach with a continuous treatment. While the recent literature on DiD estimation has mainly focused on binary treatments, there are some studies analysing DiD settings with a continuous treatment \citep{Callaway2021,Chaisemartin2022,Haultfoeuille2021}. \citet{Callaway2021} propose two parameters that may be of interest in a setting with continuous treatment: a \textit{level} effect (which compares treatment with dose $d$ to not receiving any treatment) and a \textit{slope} effect (causal response to a marginal change in dose $d$). Under the ``standard'' parallel trends assumption, a dose-specific average treatment effect on the treated $ATT(d|d)$ can be identified in the presence of a never-treated control group.\footnote{Following the notation in \citet{Callaway2021}, $ATT(d|d)$ represents the average effect of treatment with dose $d$ among the group actually treated with dose $d$. } When no such untreated group exists (as is the case in our application), level effects such as the $ATT(d|d)$ cannot be identified. While slope effects can be identified even in the absence of an untreated group, additional stronger assumptions are required to identify slope effects compared to the setting with binary treatment. The authors show that under the ``standard'' parallel trends assumption, heterogeneous treatment effects that differ across the distribution of dose $d$ can cause a ``selection bias'' when comparing treatment effects across different dosage levels to infer slope effects. For example, in a two dose scenario with one group receiving a low and the other a high dose, a ``selection bias'' would be present if the high and low dose groups differ from each other in their potential benefits (or harm) from receiving the low or high dose. The difference between $ATT(high|high)$ and $ATT(low|low)$ would then be a combination of the causal response of a change in dose from low to high among the high group, $ATT(high|high)-ATT(low|high)$, and the selection bias term $ATT(low|high)-ATT(low|low)$.

The authors show that valid comparisons of level effects across different doses require an additional assumption, which they call ``strong'' parallel trends. This assumption requires that the average expected change in outcomes across all units from being untreated in $t-1$ to treatment with dose $d$ in $t$ is the same as the average expected change for those who actually experience that dose. Hence, ``strong'' parallel trends requires that on average across all doses there is no selection; a weaker assumption than requiring homogeneous treatment. In comparison to the ``standard'' parallel trends assumptions, ``strong'' parallel trends involves treated potential outcomes and can therefore not be tested in most circumstances. Under ``strong'' parallel trends, an average treatment effect $ATE(d)$ can be identified in the presence of an untreated control group. Comparisons of $ATE(d)$ across different dosage levels are not subject to selection bias, and the average causal response $ACR(d)$ can be identified; our parameter of interest.

A second issue with the use of the TWFE estimator in a DiD setting with continuous treatment relates to the weights assigned to the different underlying comparisons. In a setting without staggered treatment, these weight are positive and add up to one. However, the TWFE weights may not correspond to the distribution of the dose and instead put higher weights on observations with rare dose levels. Specifically, the TWFE weights are highest at the mean dose rather than the most common dose level. If the dose distribution is substantially different from a normal distribution (e.g. in the case of bi-modal, strongly skewed or uniform distributions), TWFE weights will differ substantially from the dose distribution. The authors hence propose using different weights in these circumstances.

In this paper, we use the interaction of the temporal exposure to the post-vaccination period (at ages 1 to 6) with the district-level measles rates prior to vaccination to measure treatment intensity. This interaction therefore represents the treatment dose in our context, and in the following we explore (1) whether we are likely to have ``selection bias'', and (2) whether the TWFE weights differ substantially from the dose distribution. 

Since the continuous treatment intensity measure in our analysis consists of two components, we consider the ``selection bias'' issues separately for each of them. First, the use of a continuous measure of temporal exposure to the post-vaccination period may lead to a selection bias if there are differences in $ATT(d|d)$ between birth cohorts that are exposed to a different degree to the post-vaccination period. For example, if those born in September 1962 (and therefore exposed to the post-vaccination for one year) would benefit more (or less) from exposure to the post-vaccination period for five years than those born in September 1966 who are \textit{actually} exposed for five years, this could create a selection bias.\footnote{Note, that differences in the effects of exposure to measles at different ages does not imply a selection issue. The selection issue is caused by differences in potential treated outcomes and cannot be observed directly in the data.} We are not aware of any reason why this type of cohort-level selection is likely to be a major issue. Furthermore, our results are robust to the use of a binary indicator for exposure to the post-vaccination period (instead of the continuous measure used in our main specifications), for which this cohort-level ``selection bias'' does not play a role, suggesting cohort-level selection does not impact on our analysis.

\begin{figure}[p!]
   \caption{ATT curves under different scenarios for the relationship between measles and outcomes}
   \label{figure_did_scenarios} 
{\centering 
  	\begin{subfigure}{0.49\linewidth}
	  \centering
	  \caption{Increasing marginal impact}
	  \includegraphics[width=\linewidth, trim=5 15 5 5, clip]{"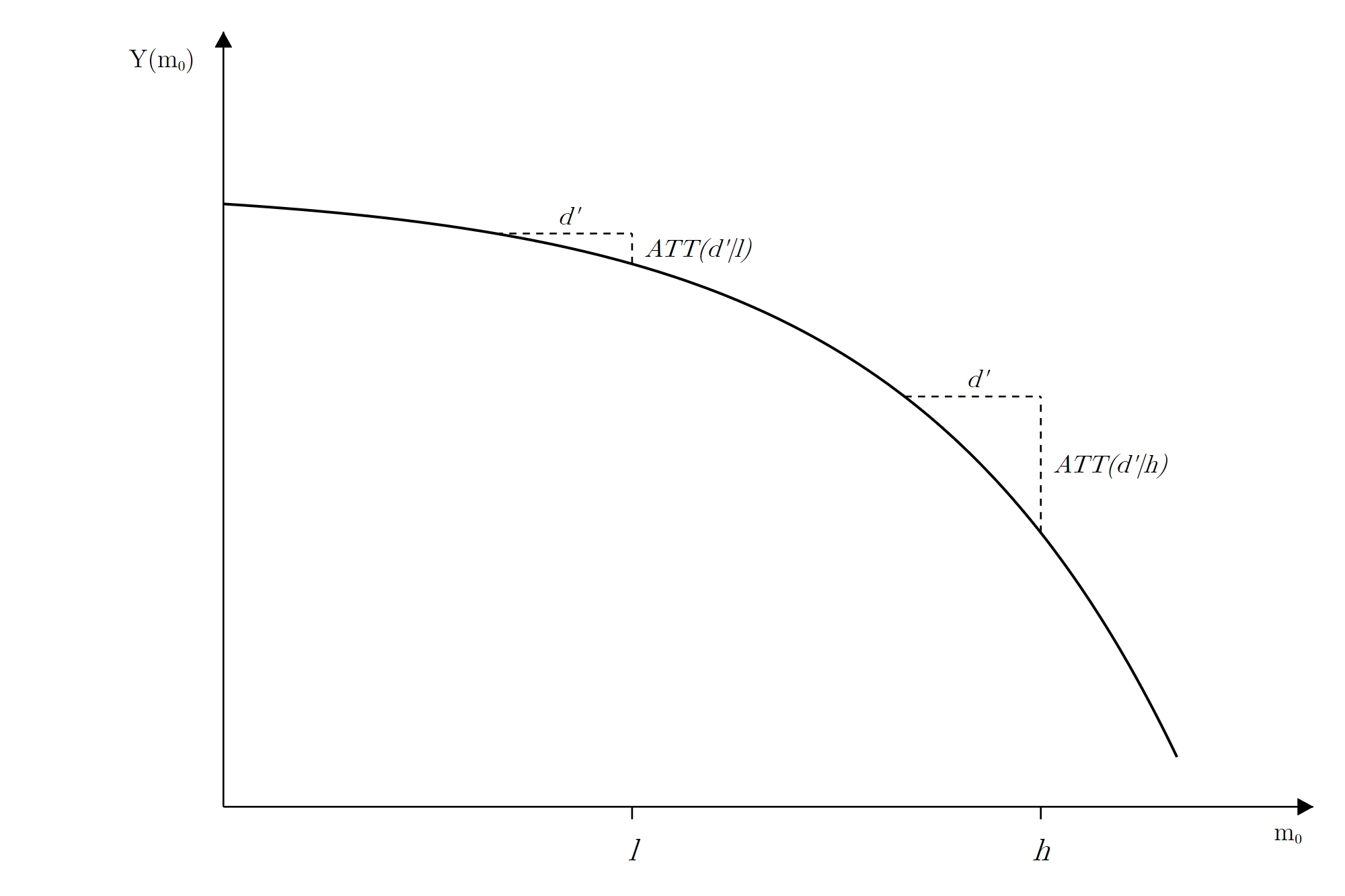"}
	\end{subfigure}
	\begin{subfigure}{0.49\linewidth}
	  \centering
	  \caption{ATT curves under increasing marginal impact}
	  \includegraphics[width=\linewidth, trim=5 15 5 5, clip]{"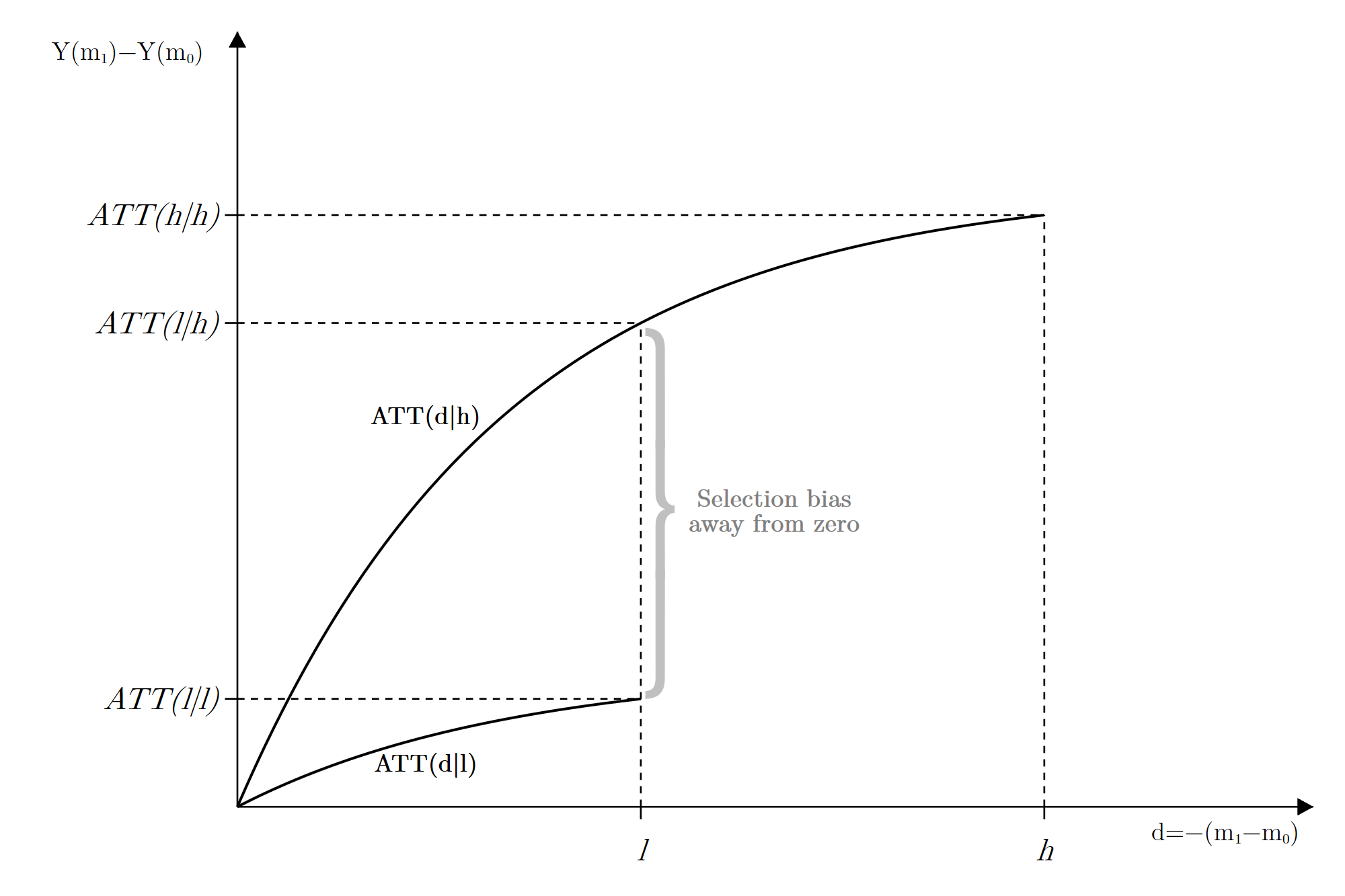"}
	\end{subfigure}
	\par\bigskip
  	\begin{subfigure}{0.49\linewidth}
	  \centering
	  \caption{Decreasing marginal impact}
	  \includegraphics[width=\linewidth, trim=5 15 5 5, clip]{"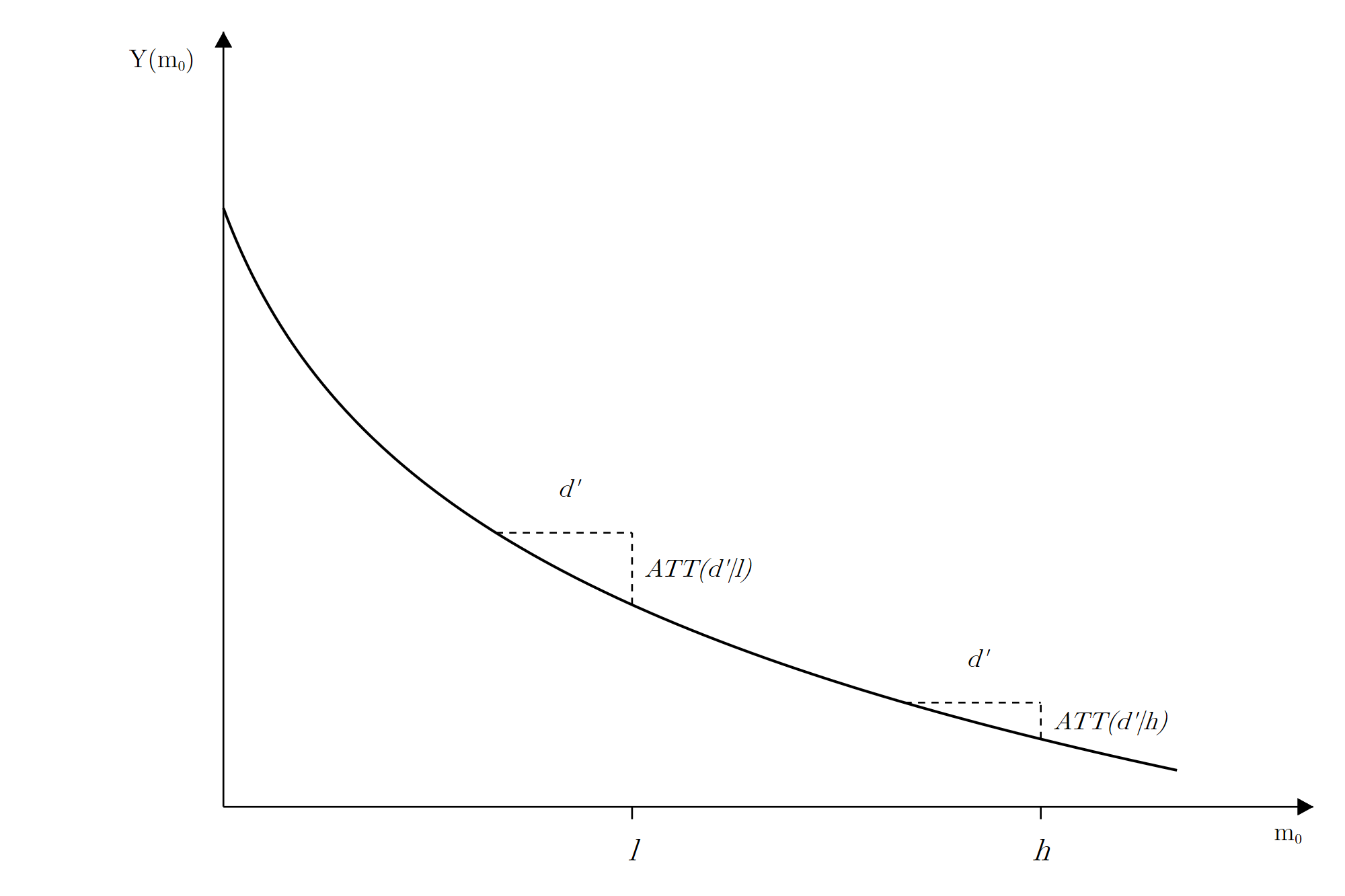"}
	\end{subfigure}
	\begin{subfigure}{0.49\linewidth}
	  \centering
	  \caption{ATT curves under decreasing marginal impact}
	  \includegraphics[width=\linewidth, trim=5 15 5 5, clip]{"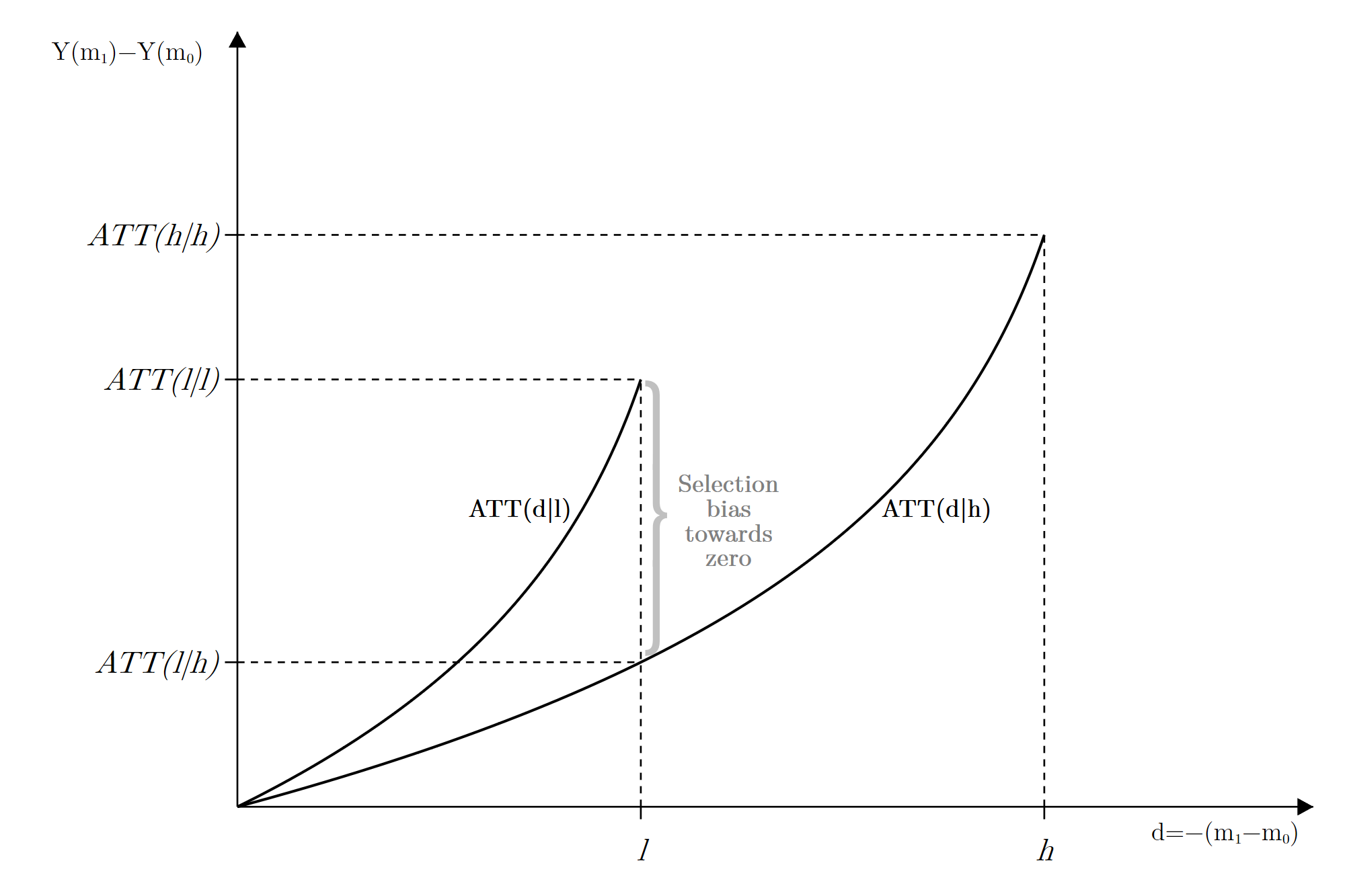"}
	\end{subfigure}
	\par\bigskip
  	\begin{subfigure}{0.49\linewidth}
	  \centering
	  \caption{Constant marginal impact}
	  \includegraphics[width=\linewidth, trim=5 15 5 5, clip]{"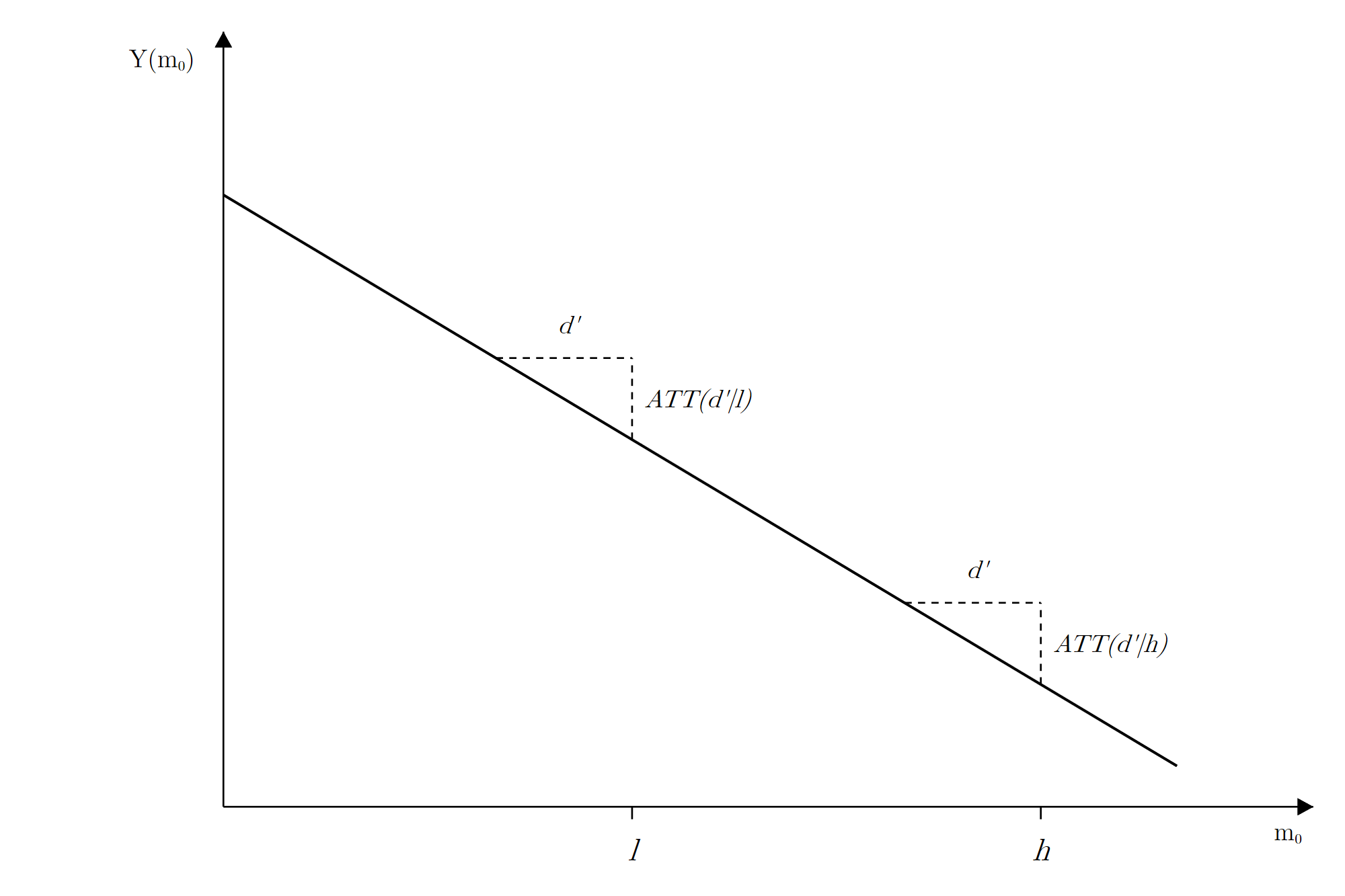"}
	\end{subfigure}
	\begin{subfigure}{0.49\linewidth}
	  \centering
	  \caption{ATT curves under constant marginal impact}
	  \includegraphics[width=\linewidth, trim=5 15 5 5, clip]{"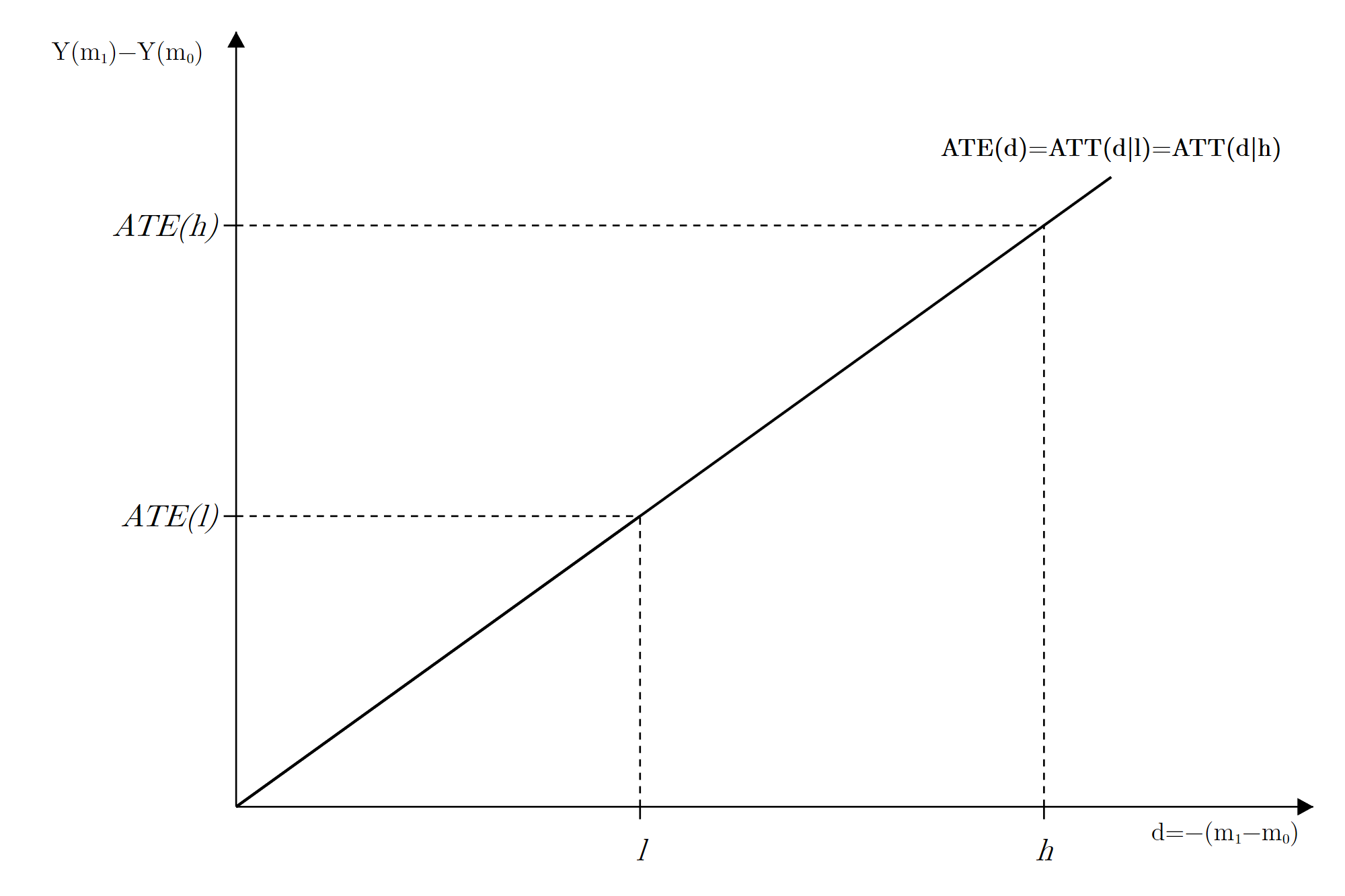"}
	\end{subfigure}
 \par}	
     {\scriptsize \singlespacing Note: The sub-figures on the left show examples of relationships between measles rates at baseline $m_0$ and an outcome $Y(m_0)$, with an increasing, decreasing and constant marginal impact of measles. The sub-figures on the right show the corresponding $ATT$ curves and illustrate the direction of any selection bias. The change in the outcome $Y(m_1) - Y(m_0)$ is plotted against the dose $d=-(m_1-m_0)$ which is the reduction in measles rates following the vaccine roll-out. For illustrative purposes, we assume that measles are fully eradicated so the dose (or measles reduction) corresponds to the prior measles rates.
\par}
\end{figure}

The second component of the continuous treatment intensity measure is the district-level measles rate prior to vaccination, capturing the potential benefit from the roll-out of the vaccination programme. A difference in the dose therefore corresponds to differences in prior measles exposure, and any selection would be driven by the shape of the relationship between measles exposure and the outcome. We illustrate this graphically in \autoref{figure_did_scenarios}. The sub-figures on the left (i.e., sub-figures a, c and e) show examples of relationships between measles rates at baseline $m_0$ (horizontal axis) and an outcome $Y(m_0)$ (vertical axis), with an increasing, decreasing and constant marginal impact.\footnote{For simplicity, in \autoref{figure_did_scenarios} we focus on cases with a negative impact of measles on the outcome. Our discussion, however, is in general terms and applies to both negative and positive relationships.} The sub-figures on the right (i.e., sub-figures b, d, and f) show the corresponding $ATT$ curves and illustrate the direction of any ``selection bias'', where the change in the outcome $Y(m_1) - Y(m_0)$ is plotted on the vertical axis against the dose $d=-(m_1-m_0)$ which is the reduction in measles rates following the vaccine roll-out. For illustrative purposes, we assume that measles are fully eradicated so the dose (or measles reduction) corresponds to the prior measles rates.

If the marginal impact of measles is increasing with the rate of measles (sub-figures a and b), then a reduction in measles rates by $d'$ from a high baseline level $h$ will have a larger impact than a reduction of the same size from a low baseline level $l$. This means $ATT(d|h)$ will be above $ATT(d|l)$ as illustrated in sub-figure b, and a comparison of these two $ATT$s will capture both the causal response to a change in dose from $l$ to $h$ and a selection term biasing estimates away from zero.

If the marginal impact of measles is decreasing (sub-figures c and d), then a reduction in measles rates by $d'$ from a high baseline level $h$ will have a smaller impact than a reduction of the same size from a low baseline level $l$. This means $ATT(d|h)$ will be below $ATT(d|l)$ as illustrated in sub-figure d, and a comparison of these two $ATT$s will capture both the causal response to a change in dose from $l$ to $h$ and a selection term biasing estimates towards zero.

If the relationship between measles and the outcome is linear and therefore the marginal impact constant (sub-figures e and f), then a reduction in measles rates by $d'$ from high and low baseline levels will have the same impact. This means average treatment effects are the same for any group, and hence $ATE(d)=ATT(d|l)=ATT(d|h)$. Comparisons of treatment effects across doses capture the causal response to a dose change, there is no selection bias.

Hence, the presence and direction of any district-level ``selection bias'' depends on whether the marginal impact of measles on the outcome differs between different measles rates. We next discuss the likely implications for our analysis. First, years of education: Measles rates may affect years of education in two ways: (1) contracting measles may disrupt the (biological) cognitive development of a child (i.e., a \textit{direct} impact), and (2) measles cases during the school year may cause disruption to the educational progress of schoolchildren due to complete suspension of lessons or due to slower teaching progress when many children are off sick (i.e., an \textit{indirect} impact). We expect the marginal \textit{direct} impact of measles on individuals' cognitive development to be independent of measles rates, since this is mainly driven by a biological mechanism which arguably is similar across children. One argument against this, however, is if severe measles cases (i.e., with larger direct impacts) are disproportionally (and non-linearly) concentrated in areas with high measles rates. If the risk of severe illness increases convexly form with the measles rate, the impact of an additional measles case would be larger in areas with higher measles rates compared to areas with lower rates, causing a ``selection bias'' in our estimates away from zero. Whether the marginal \textit{indirect} impact differs between different measles rates is unclear. A higher measles rate implies more children are affected, perhaps leading to school or class closures. However, its impact on children's educational outcomes depends on the `threshold' at which the disruption (e.g., the closure of schools/classes) occurs. Indeed, if schools/classes are suspended when, e.g., 5\% of pupils are infected, the disruption is likely to be similar in high and low measles areas, since even in low measles areas, this is an easy-to-reach threshold. In contrast, if schools/classes are only suspended when, e.g., 80\% of pupils are infected, this may be more likely to have higher impacts on areas with higher measles rates. Hence, \textit{a priori}, it is unclear to what extent (if at all) there would be any ``selection bias''. 

Next, height: The relationship between measles rates and health outcomes such as height can be expected to be largely due to the \textit{direct} effects of contracting measles, rather than any \textit{indirect} effects. Similar to the above, therefore, since this is mainly driven by a biological mechanism, we expect the marginal impact of measles on individuals' height to be independent of the measles rates. Hence, \textit{a priori}, it seems unlikely there would be any substantial selection bias in our estimates of the long-term impact on height.

The second potential issue of using TWFE is the weighting of the underlying comparisons. Specifically, the weights used by TWFE to aggregate the average causal responses on the treated $ACRT(d|d)$ across the dose distribution may not have any theoretical foundation and can differ substantially from the distribution of doses in the sample. If the dose distribution is symmetric and similar to the normal distribution, TWFE weights will be similar to the dose distribution \citep{Callaway2021}. However, if doses are not (approximately) normally distributed, the weights will differ substantially from the dose distribution which can result in rare doses receiving large weights and common doses receiving small weights.

\begin{figure}[tbp!]
   \caption{Distribution of pre-vaccination rates}
   \label{figure_pre_rates_distribution} 
{\centering 
	  \includegraphics[width=0.7\linewidth, trim=5 15 5 5, clip]{"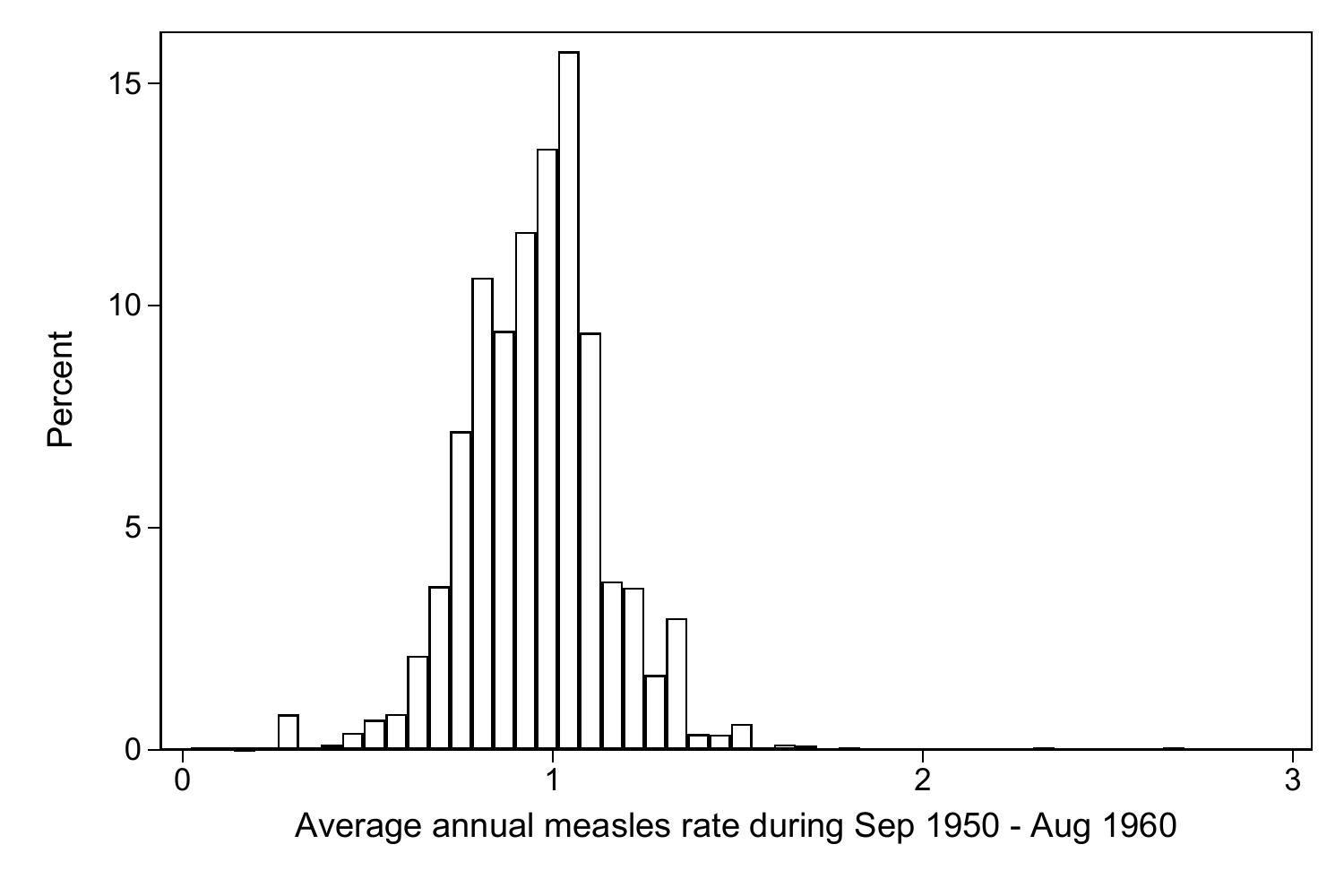"}
 \par}	
     {\scriptsize \singlespacing Note: Shown is the distribution of pre-vaccination measles rates among the sample used in our main analysis.
\par}
\end{figure}

In our main analysis, the dose distribution derives from the pre-vaccination rates of measles ($PreRate_d$), and the temporal exposure to the post-vaccination period ($Post_{age=a}$). In the robustness analysis, we also estimate the models using a binary indicator for exposure to the post-vaccination period (see \autoref{table_vaccine_intensity_binary_education_height}), in which case the dose only derives from the pre-vaccination measles rates.  \autoref{figure_pre_rates_distribution} presents the distribution of the rates. This shows that it does not differ substantially from a normal distribution. Hence, we can be confident that TWFE weights will be appropriate in these estimations.

\begin{figure}[tbp!]
   \caption{Distribution of doses and TWFE weights}
   \label{figure_doses_and_weights} 
{\centering 
	  \includegraphics[width=0.7\linewidth, trim=5 15 5 5, clip]{"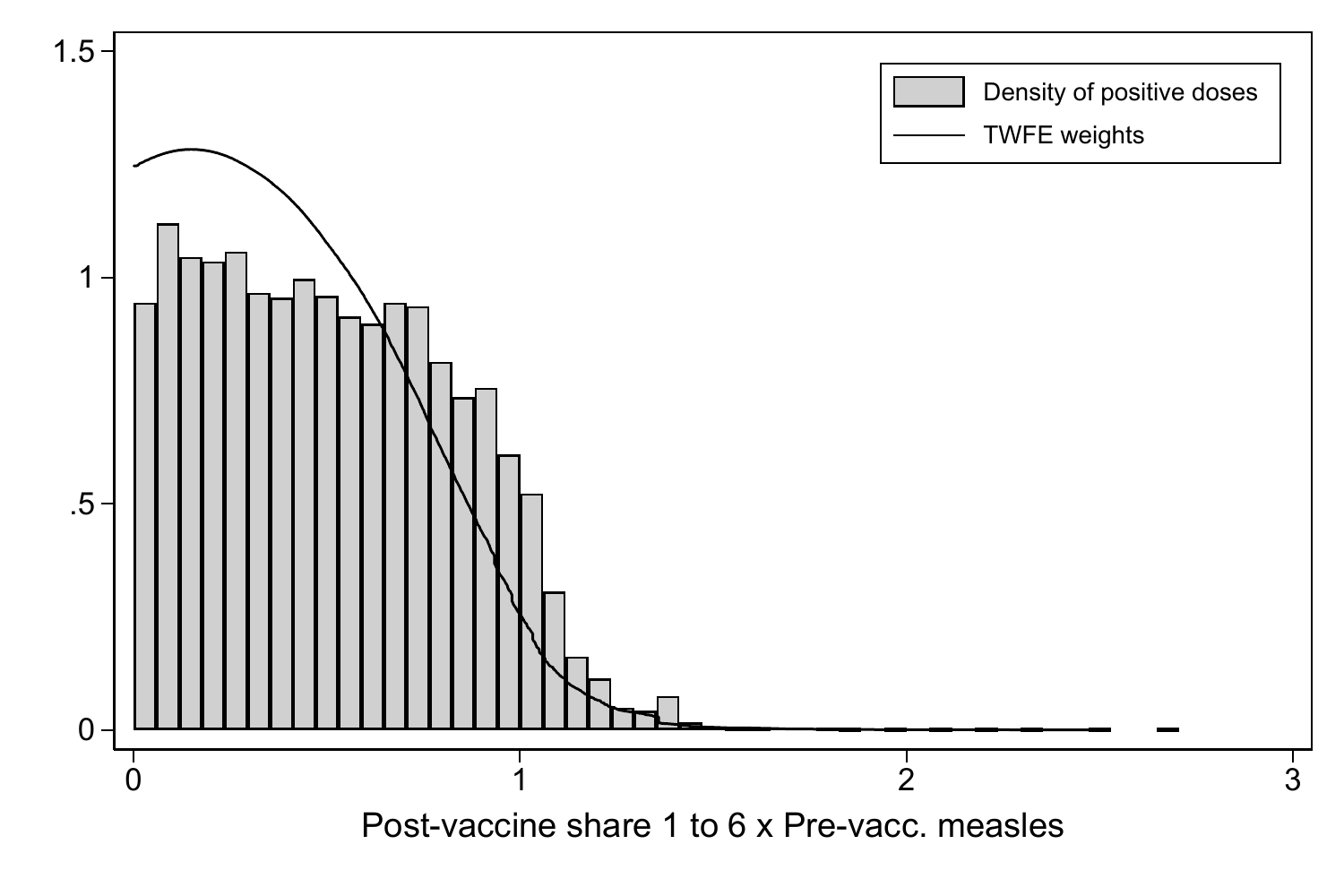"}
 \par}	
     {\scriptsize \singlespacing Note: The grey bars show the density distribution of positive doses, i.e. of the interaction term between the temporal exposure to the post-vaccination period at ages 1 to 6 and the the district-level measles rates prior to the vaccination programme. The solid line represents the weights that would be used to aggregate the average causal responses on the treated $ACRT(d|d)$ in a simple two-period TWFE estimation with this distribution of doses, as derived in Proposition 4 of \citet{Callaway2021}.
\par}
\end{figure}

\autoref{figure_doses_and_weights} shows the density distribution of the interaction term between the share of ages 1 to 6 during which an individual was exposed to the post-vaccination period and the district-level measles rates prior to the vaccination programme (i.e., $Post_{age=a} \times PreRate_d$). For comparison, \autoref{figure_doses_and_weights} also displays the weights that would be used to aggregate the average causal responses on the treated $ACRT(d|d)$ in a simple two-period TWFE estimation with this distribution of doses, based on Proposition 4 of \citet{Callaway2021}.\footnote{Weights cannot be easily derived for the more complex estimations used in our analysis, as they involve repeated cross-sections and control for district, school year of birth and month of birth fixed effects as well as other covariates.} In comparison to the dose distribution, TWFE puts slightly higher weights on observations with a lower dose, and slightly lower weights on observations with a higher dose. Overall, however, the differences between the dose distribution and the TWFE weights are not large and hence should not be a major concern in our analysis.

\end{appendices}

\end{document}

%% file: tables/descriptives_tablefragment.tex
          &   {Mean}&     {SD}&    {Min}&    {Max}&      {N}\\
\midrule
Measles rates (annual avg. per 100 people):&         &         &         &         &         \\
\quad - Age 1 to 6&     0.86&     0.25&     0.00&     5.48&  176,067\\
\quad - Age 1 to 2&     0.89&     0.36&     0.00&    13.56&  187,045\\
\quad - Age 3 to 4&     0.83&     0.37&     0.00&    13.82&  185,592\\
\quad - Age 5 to 6&     0.80&     0.37&     0.00&    13.63&  176,163\\
\quad - pre-vaccination (Sep 1950 - Aug 1960)&     0.96&     0.19&     0.02&     2.71&  187,078\\
\addlinespace Outcomes:&         &         &         &         &         \\
\quad - years of education&    13.38&     2.22&    10.00&    16.00&  186,104\\
\quad - any qualification&     0.91&     0.29&     0.00&     1.00&  186,104\\
\quad - upper secondary qualification&     0.72&     0.45&     0.00&     1.00&  186,104\\
\quad - degree qualification&     0.35&     0.48&     0.00&     1.00&  186,104\\
\quad - height in cm&   169.62&     9.24&   118.00&   209.00&  186,766\\
\addlinespace Demographic characteristics:&         &         &         &         &         \\
\quad - female&     0.55&     0.50&     0.00&     1.00&  187,078\\
\quad - year of birth&  1957.54&     5.44&  1949.00&  1969.00&  187,078\\

%% file: tables/height_education_measlesvacc_tablefragment.tex
          &\multicolumn{4}{c}{Years of education}             &\multicolumn{4}{c}{Height in cm}                   \\\cmidrule(lr){2-5}\cmidrule(lr){6-9}
          &\multicolumn{1}{c}{(1)}   &\multicolumn{1}{c}{(2)}   &\multicolumn{1}{c}{(3)}   &\multicolumn{1}{c}{(4)}   &\multicolumn{1}{c}{(5)}   &\multicolumn{1}{c}{(6)}   &\multicolumn{1}{c}{(7)}   &\multicolumn{1}{c}{(8)}   \\
\midrule
\textbf{Panel A:}&            &            &            &            &            &            &            &            \\
\midrule Post-vaccine share 1 to 6 $\times$ Pre-vacc. measles&    0.227***&    0.191***&    0.031   &   -0.027   &    2.002***&    1.969***&    0.453*  &    0.090   \\
          &  (0.029)   &  (0.023)   &  (0.102)   &  (0.105)   &  (0.073)   &  (0.063)   &  (0.271)   &  (0.256)   \\
\midrule
\textbf{Panel B:}&            &            &            &            &            &            &            &            \\
\midrule Post-vaccine share 1 to 2 $\times$ Pre-vacc. measles&    0.044   &    0.027   &   -0.039   &   -0.056   &    0.428***&    0.325***&    0.402   &    0.304   \\
          &  (0.041)   &  (0.034)   &  (0.092)   &  (0.093)   &  (0.115)   &  (0.111)   &  (0.325)   &  (0.320)   \\
Post-vaccine share 3 to 4 $\times$ Pre-vacc. measles&    0.082** &    0.062   &   -0.009   &   -0.027   &   -0.039   &   -0.052   &   -0.150   &   -0.215   \\
          &  (0.039)   &  (0.038)   &  (0.090)   &  (0.091)   &  (0.110)   &  (0.110)   &  (0.277)   &  (0.282)   \\
Post-vaccine share 5 to 6 $\times$ Pre-vacc. measles&    0.088***&    0.087***&    0.060   &    0.037   &    1.388***&    1.427***&    0.286   &    0.105   \\
          &  (0.029)   &  (0.027)   &  (0.083)   &  (0.082)   &  (0.076)   &  (0.075)   &  (0.245)   &  (0.237)   \\
\midrule
\textbf{Controls for:} \\ Gender&    {Yes}   &    {Yes}   &    {Yes}   &    {Yes}   &    {Yes}   &    {Yes}   &    {Yes}   &    {Yes}   \\
Month of birth FE&    {Yes}   &    {Yes}   &    {Yes}   &    {Yes}   &    {Yes}   &    {Yes}   &    {Yes}   &    {Yes}   \\
District of birth FE&     {No}   &    {Yes}   &    {Yes}   &    {Yes}   &     {No}   &    {Yes}   &    {Yes}   &    {Yes}   \\
School year of birth FE&     {No}   &     {No}   &    {Yes}   &    {Yes}   &     {No}   &     {No}   &    {Yes}   &    {Yes}   \\
County-specific birthdate trend &     {No}   &     {No}   &     {No}   &    {Yes}   &     {No}   &     {No}   &     {No}   &    {Yes}   \\
Compulsory schooling 16 &    {Yes}   &    {Yes}   &     {No}   &     {No}   &     {No}   &     {No}   &     {No}   &     {No}   \\
Compulsory schooling 16 $\times$ Pre-vacc. measles &    {Yes}   &    {Yes}   &    {Yes}   &    {Yes}   &     {No}   &     {No}   &     {No}   &     {No}   \\
\midrule
N         &\multicolumn{1}{c}{170,802}   &\multicolumn{1}{c}{170,778}   &\multicolumn{1}{c}{170,778}   &\multicolumn{1}{c}{170,778}   &\multicolumn{1}{c}{171,395}   &\multicolumn{1}{c}{171,370}   &\multicolumn{1}{c}{171,370}   &\multicolumn{1}{c}{171,370}   \\

%% file: tables/height_education_largetrial_tablefragment.tex
          &\multicolumn{2}{c}{Years of education}&\multicolumn{2}{c}{Height in cm}\\\cmidrule(lr){2-3}\cmidrule(lr){4-5}
          &\multicolumn{1}{c}{(1)}   &\multicolumn{1}{c}{(2)}   &\multicolumn{1}{c}{(3)}   &\multicolumn{1}{c}{(4)}   \\
\midrule
\textbf{Panel A:}&            &            &            &            \\
\midrule Trial district $\times$ Trial period exposure - Age 1 to 6&   -0.000   &   -0.003   &    0.134   &    0.116   \\
          &  (0.029)   &  (0.029)   &  (0.106)   &  (0.107)   \\
Trial period exposure - Age 1 to 6&    0.083***&    0.099** &    0.615***&    0.002   \\
          &  (0.011)   &  (0.049)   &  (0.032)   &  (0.201)   \\
\midrule
\textbf{Panel B:}&            &            &            &            \\
\midrule Trial district $\times$ Trial period exposure - Age 1 to 2&   -0.031   &   -0.033   &    0.383** &    0.374** \\
          &  (0.034)   &  (0.034)   &  (0.162)   &  (0.158)   \\
Trial district $\times$ Trial period exposure - Age 3 to 4&    0.024   &    0.023   &    0.058   &    0.037   \\
          &  (0.040)   &  (0.040)   &  (0.133)   &  (0.140)   \\
Trial district $\times$ Trial period exposure - Age 5 to 6&   -0.003   &   -0.007   &    0.025   &   -0.005   \\
          &  (0.066)   &  (0.066)   &  (0.108)   &  (0.106)   \\
Trial period exposure - Age 1 to 2&    0.125***&    0.104   &    0.898***&    0.027   \\
          &  (0.019)   &  (0.063)   &  (0.057)   &  (0.248)   \\
Trial period exposure - Age 3 to 4&    0.062***&    0.012   &    0.540***&   -0.121   \\
          &  (0.019)   &  (0.079)   &  (0.053)   &  (0.247)   \\
Trial period exposure - Age 5 to 6&    0.072***&    0.096   &    0.457***&   -0.023   \\
          &  (0.017)   &  (0.062)   &  (0.052)   &  (0.214)   \\
\midrule
\textbf{Controls for:} \\ Gender&    {Yes}   &    {Yes}   &    {Yes}   &    {Yes}   \\
Month of birth FE&    {Yes}   &    {Yes}   &    {Yes}   &    {Yes}   \\
District of birth FE&    {Yes}   &    {Yes}   &    {Yes}   &    {Yes}   \\
School year of birth FE&     {No}   &    {Yes}   &     {No}   &    {Yes}   \\
\midrule
N         &\multicolumn{1}{c}{95,774}   &\multicolumn{1}{c}{95,774}   &\multicolumn{1}{c}{96,148}   &\multicolumn{1}{c}{96,148}   \\

%% file: tables/height_education_vacc1to6_GxE_GIANT_EA3_tablefragment.tex
          &\multicolumn{4}{c}{Years of education}             &\multicolumn{4}{c}{Height in cm}                   \\\cmidrule(lr){2-5}\cmidrule(lr){6-9}
          &\multicolumn{1}{c}{(1)}   &\multicolumn{1}{c}{(2)}   &\multicolumn{1}{c}{(3)}   &\multicolumn{1}{c}{(4)}   &\multicolumn{1}{c}{(5)}   &\multicolumn{1}{c}{(6)}   &\multicolumn{1}{c}{(7)}   &\multicolumn{1}{c}{(8)}   \\
\midrule
Post-vaccine share 1 to 6 $\times$ Pre-vacc. measles&    0.234***&    0.210***&    0.091   &    0.041   &    2.016***&    2.014***&    0.309   &    0.024   \\
          &  (0.024)   &  (0.021)   &  (0.094)   &  (0.096)   &  (0.062)   &  (0.053)   &  (0.234)   &  (0.223)   \\
PGI       &    0.674***&    0.614***&    0.615***&    0.614***&    3.071***&    3.042***&    3.047***&    3.046***\\
          &  (0.009)   &  (0.009)   &  (0.009)   &  (0.009)   &  (0.015)   &  (0.015)   &  (0.015)   &  (0.015)   \\
Post-vacc. share 1 to 6 $\times$ Pre-vacc. measles $\times$ PGI&   -0.005   &   -0.012   &   -0.013   &   -0.012   &    0.075   &    0.086*  &    0.078   &    0.084*  \\
          &  (0.018)   &  (0.017)   &  (0.017)   &  (0.017)   &  (0.049)   &  (0.048)   &  (0.049)   &  (0.049)   \\
\midrule \textbf{Controls for:} \\ Gender&    {Yes}   &    {Yes}   &    {Yes}   &    {Yes}   &    {Yes}   &    {Yes}   &    {Yes}   &    {Yes}   \\
Month of birth FE&    {Yes}   &    {Yes}   &    {Yes}   &    {Yes}   &    {Yes}   &    {Yes}   &    {Yes}   &    {Yes}   \\
District of birth FE&     {No}   &    {Yes}   &    {Yes}   &    {Yes}   &     {No}   &    {Yes}   &    {Yes}   &    {Yes}   \\
School year of birth FE&     {No}   &     {No}   &    {Yes}   &    {Yes}   &     {No}   &     {No}   &    {Yes}   &    {Yes}   \\
County-specific birthdate trend &     {No}   &     {No}   &     {No}   &    {Yes}   &     {No}   &     {No}   &     {No}   &    {Yes}   \\
Compulsory schooling 16 &    {Yes}   &    {Yes}   &     {No}   &     {No}   &     {No}   &     {No}   &     {No}   &     {No}   \\
Compulsory schooling 16 $\times$ Pre-vacc. measles &    {Yes}   &    {Yes}   &    {Yes}   &    {Yes}   &     {No}   &     {No}   &     {No}   &     {No}   \\
Comp. schooling 16 $\times$ Pre-vacc. measles $\times$ PGI &    {Yes}   &    {Yes}   &    {Yes}   &    {Yes}   &     {No}   &     {No}   &     {No}   &     {No}   \\
Principal components 1-20 &    {Yes}   &    {Yes}   &    {Yes}   &    {Yes}   &    {Yes}   &    {Yes}   &    {Yes}   &    {Yes}   \\
\midrule
N         &\multicolumn{1}{c}{170,158}   &\multicolumn{1}{c}{170,134}   &\multicolumn{1}{c}{170,134}   &\multicolumn{1}{c}{170,134}   &\multicolumn{1}{c}{170,750}   &\multicolumn{1}{c}{170,725}   &\multicolumn{1}{c}{170,725}   &\multicolumn{1}{c}{170,725}   \\

%% file: tables/height_education_vacc1to6_withinsibGxE_GIANT_EA3_tablefragment.tex
          &\multicolumn{4}{c}{Years of education}             &\multicolumn{4}{c}{Height in cm}                   \\\cmidrule(lr){2-5}\cmidrule(lr){6-9}
          &\multicolumn{1}{c}{(1)}   &\multicolumn{1}{c}{(2)}   &\multicolumn{1}{c}{(3)}   &\multicolumn{1}{c}{(4)}   &\multicolumn{1}{c}{(5)}   &\multicolumn{1}{c}{(6)}   &\multicolumn{1}{c}{(7)}   &\multicolumn{1}{c}{(8)}   \\
\midrule
Post-vaccine share 1 to 6 $\times$ Pre-vacc. measles&    0.238***&    0.205** &    0.074   &    0.003   &    1.967***&    2.227***&    2.982** &    2.181   \\
          &  (0.091)   &  (0.091)   &  (0.425)   &  (0.443)   &  (0.267)   &  (0.268)   &  (1.250)   &  (1.433)   \\
PGI sibling mean deviation (PGI-SMD)&    0.378***&    0.387***&    0.385***&    0.387***&    2.575***&    2.587***&    2.598***&    2.578***\\
          &  (0.055)   &  (0.053)   &  (0.054)   &  (0.055)   &  (0.104)   &  (0.103)   &  (0.102)   &  (0.101)   \\
Post-vacc. share 1 to 6 $\times$ Pre-vacc. measles $\times$ PGI-SMD&   -0.085   &   -0.124   &   -0.123   &   -0.142   &    0.471   &    0.425   &    0.360   &    0.422   \\
          &  (0.179)   &  (0.178)   &  (0.179)   &  (0.179)   &  (0.449)   &  (0.451)   &  (0.438)   &  (0.438)   \\
\midrule \textbf{Controls for:} \\ Gender&    {Yes}   &    {Yes}   &    {Yes}   &    {Yes}   &    {Yes}   &    {Yes}   &    {Yes}   &    {Yes}   \\
Month of birth FE&    {Yes}   &    {Yes}   &    {Yes}   &    {Yes}   &    {Yes}   &    {Yes}   &    {Yes}   &    {Yes}   \\
District of birth FE&     {No}   &    {Yes}   &    {Yes}   &    {Yes}   &     {No}   &    {Yes}   &    {Yes}   &    {Yes}   \\
School year of birth FE&     {No}   &     {No}   &    {Yes}   &    {Yes}   &     {No}   &     {No}   &    {Yes}   &    {Yes}   \\
County-specific birthdate trend &     {No}   &     {No}   &     {No}   &    {Yes}   &     {No}   &     {No}   &     {No}   &    {Yes}   \\
Compulsory schooling 16 &    {Yes}   &    {Yes}   &     {No}   &     {No}   &     {No}   &     {No}   &     {No}   &     {No}   \\
Compulsory schooling 16 $\times$ Pre-vacc. measles &    {Yes}   &    {Yes}   &    {Yes}   &    {Yes}   &     {No}   &     {No}   &     {No}   &     {No}   \\
Comp. schooling 16 $\times$ Pre-vacc. measles $\times$ PGI-SMD &    {Yes}   &    {Yes}   &    {Yes}   &    {Yes}   &     {No}   &     {No}   &     {No}   &     {No}   \\
Principal components 1-20 &    {Yes}   &    {Yes}   &    {Yes}   &    {Yes}   &    {Yes}   &    {Yes}   &    {Yes}   &    {Yes}   \\
\midrule
N         &\multicolumn{1}{c}{10,748}   &\multicolumn{1}{c}{10,538}   &\multicolumn{1}{c}{10,538}   &\multicolumn{1}{c}{10,538}   &\multicolumn{1}{c}{10,788}   &\multicolumn{1}{c}{10,577}   &\multicolumn{1}{c}{10,577}   &\multicolumn{1}{c}{10,577}   \\

%% file: tables/1951census_district_descriptives_tablefragment.tex
          &\multicolumn{2}{c}{\shortstack{Pre-vaccination measles rate \\ (Sep 1950 - Aug 1960)}}\\\cmidrule(lr){2-3}
          &{below median}&{above median}\\
\midrule
Social class:&         &         \\
\quad - class 1 (professional)&    0.037&    0.030\\
\quad - class 2 (intermediate)&    0.156&    0.135\\
\quad - class 3 (skilled)&    0.526&    0.532\\
\quad - class 4 (partly skilled)&    0.160&    0.163\\
\quad - class 5 (unskilled)&    0.122&    0.140\\
Housing density:&         &         \\
\quad - 0 to 1 persons per room&    0.749&    0.731\\
\quad - 1 to 1.5 persons per room&    0.168&    0.177\\
\quad - 1.5 to 2 persons per room&    0.064&    0.068\\
\quad - 2 to 3 persons per room&    0.017&    0.020\\
\quad - 3+ persons per room&    0.003&    0.004\\
Age left education:&         &         \\
\quad - 0 to 14 years&    0.738&    0.770\\
\quad - 15 years&    0.104&    0.101\\
\quad - 16 years&    0.081&    0.072\\
\quad - 17 to 19 years&    0.051&    0.037\\
\quad - 20+ years&    0.026&    0.020\\

%% file: tables/quals_combined_measles_tablefragment.tex
          &\multicolumn{1}{c}{(1)}   &\multicolumn{1}{c}{(2)}   &\multicolumn{1}{c}{(3)}   &\multicolumn{1}{c}{(4)}   \\
\midrule
\textbf{Panel A.1 - Any qualification:}&            &            &            &            \\
\midrule Post-vaccine share 1 to 6 $\times$ Pre-vacc. measles&    0.029***&    0.026***&    0.004   &   -0.010   \\
          &  (0.002)   &  (0.002)   &  (0.008)   &  (0.010)   \\
\midrule
\textbf{Panel A.2 - Any qualification:}&            &            &            &            \\
\midrule Post-vaccine share 1 to 2 $\times$ Pre-vacc. measles&    0.003   &    0.001   &    0.001   &   -0.005   \\
          &  (0.003)   &  (0.003)   &  (0.008)   &  (0.009)   \\
Post-vaccine share 3 to 4 $\times$ Pre-vacc. measles&    0.011***&    0.009***&    0.007   &    0.004   \\
          &  (0.003)   &  (0.003)   &  (0.008)   &  (0.009)   \\
Post-vaccine share 5 to 6 $\times$ Pre-vacc. measles&    0.013***&    0.013***&   -0.004   &   -0.010   \\
          &  (0.003)   &  (0.003)   &  (0.007)   &  (0.007)   \\
\midrule
\textbf{Panel B.1 - Upper secondary qualification:}&            &            &            &            \\
\midrule Post-vaccine share 1 to 6 $\times$ Pre-vacc. measles&    0.034***&    0.029***&   -0.002   &   -0.014   \\
          &  (0.005)   &  (0.005)   &  (0.020)   &  (0.021)   \\
\midrule
\textbf{Panel B.2 - Upper secondary qualification:}&            &            &            &            \\
\midrule Post-vaccine share 1 to 2 $\times$ Pre-vacc. measles&   -0.004   &   -0.007   &   -0.024   &   -0.028   \\
          &  (0.008)   &  (0.008)   &  (0.020)   &  (0.021)   \\
Post-vaccine share 3 to 4 $\times$ Pre-vacc. measles&    0.012*  &    0.009   &    0.003   &   -0.000   \\
          &  (0.007)   &  (0.007)   &  (0.018)   &  (0.018)   \\
Post-vaccine share 5 to 6 $\times$ Pre-vacc. measles&    0.020***&    0.019***&    0.010   &    0.005   \\
          &  (0.006)   &  (0.005)   &  (0.017)   &  (0.017)   \\
\midrule
\textbf{Panel C.1 - Degree qualification:}&            &            &            &            \\
\midrule Post-vaccine share 1 to 6 $\times$ Pre-vacc. measles&    0.044***&    0.038***&    0.014   &    0.008   \\
          &  (0.006)   &  (0.005)   &  (0.024)   &  (0.024)   \\
\midrule
\textbf{Panel C.2 - Degree qualification:}&            &            &            &            \\
\midrule Post-vaccine share 1 to 2 $\times$ Pre-vacc. measles&    0.015   &    0.012   &    0.002   &    0.001   \\
          &  (0.010)   &  (0.008)   &  (0.022)   &  (0.022)   \\
Post-vaccine share 3 to 4 $\times$ Pre-vacc. measles&    0.014   &    0.010   &    0.006   &    0.003   \\
          &  (0.009)   &  (0.008)   &  (0.020)   &  (0.020)   \\
Post-vaccine share 5 to 6 $\times$ Pre-vacc. measles&    0.016** &    0.015** &    0.006   &    0.003   \\
          &  (0.006)   &  (0.006)   &  (0.020)   &  (0.019)   \\
\midrule
\textbf{Controls for:} \\ Gender&    {Yes}   &    {Yes}   &    {Yes}   &    {Yes}   \\
Month of birth FE&    {Yes}   &    {Yes}   &    {Yes}   &    {Yes}   \\
District of birth FE&     {No}   &    {Yes}   &    {Yes}   &    {Yes}   \\
School year of birth FE&     {No}   &     {No}   &    {Yes}   &    {Yes}   \\
County-specific birthdate trend &     {No}   &     {No}   &     {No}   &    {Yes}   \\
Compulsory schooling 16 &    {Yes}   &    {Yes}   &     {No}   &     {No}   \\
Compulsory schooling 16 $\times$ Pre-vacc. measles &    {Yes}   &    {Yes}   &    {Yes}   &    {Yes}   \\
\midrule
N         &\multicolumn{1}{c}{170,802}   &\multicolumn{1}{c}{170,778}   &\multicolumn{1}{c}{170,778}   &\multicolumn{1}{c}{170,778}   \\

%% file: tables/quals_largetrial_tablefragment.tex
          &\multicolumn{2}{c}{Any qualification}&\multicolumn{2}{c}{Upper secondary}&\multicolumn{2}{c}{Degree}\\\cmidrule(lr){2-3}\cmidrule(lr){4-5}\cmidrule(lr){6-7}
          &\multicolumn{1}{c}{(1)}   &\multicolumn{1}{c}{(2)}   &\multicolumn{1}{c}{(3)}   &\multicolumn{1}{c}{(4)}   &\multicolumn{1}{c}{(5)}   &\multicolumn{1}{c}{(6)}   \\
\midrule
\textbf{Panel A:}&            &            &            &            &            &            \\
\midrule Trial district $\times$ Trial period exposure - Age 1 to 6&    0.006   &    0.005   &    0.001   &    0.001   &   -0.005   &   -0.005   \\
          &  (0.007)   &  (0.007)   &  (0.006)   &  (0.006)   &  (0.004)   &  (0.004)   \\
Trial period exposure - Age 1 to 6&    0.040***&    0.005   &    0.017***&    0.027** &    0.007***&    0.008   \\
          &  (0.002)   &  (0.005)   &  (0.002)   &  (0.012)   &  (0.002)   &  (0.011)   \\
\midrule
\textbf{Panel B:}&            &            &            &            &            &            \\
\midrule Trial district $\times$ Trial period exposure - Age 1 to 2&   -0.004   &   -0.004   &    0.004   &    0.004   &   -0.010   &   -0.011   \\
          &  (0.011)   &  (0.011)   &  (0.005)   &  (0.005)   &  (0.008)   &  (0.008)   \\
Trial district $\times$ Trial period exposure - Age 3 to 4&    0.008   &    0.008   &   -0.008   &   -0.008   &    0.006   &    0.005   \\
          &  (0.006)   &  (0.006)   &  (0.005)   &  (0.005)   &  (0.009)   &  (0.010)   \\
Trial district $\times$ Trial period exposure - Age 5 to 6&    0.012   &    0.011   &    0.009   &    0.009   &   -0.012   &   -0.012   \\
          &  (0.007)   &  (0.008)   &  (0.016)   &  (0.016)   &  (0.010)   &  (0.010)   \\
Trial period exposure - Age 1 to 2&    0.052***&    0.001   &    0.024***&    0.029** &    0.012***&    0.011   \\
          &  (0.003)   &  (0.005)   &  (0.004)   &  (0.015)   &  (0.004)   &  (0.014)   \\
Trial period exposure - Age 3 to 4&    0.031***&   -0.002   &    0.015***&    0.005   &    0.005   &   -0.007   \\
          &  (0.002)   &  (0.007)   &  (0.003)   &  (0.016)   &  (0.004)   &  (0.018)   \\
Trial period exposure - Age 5 to 6&    0.042***&    0.008   &    0.015***&    0.024*  &    0.005   &    0.007   \\
          &  (0.003)   &  (0.006)   &  (0.003)   &  (0.015)   &  (0.004)   &  (0.014)   \\
\midrule
\textbf{Controls for:} \\ Gender&    {Yes}   &    {Yes}   &    {Yes}   &    {Yes}   &    {Yes}   &    {Yes}   \\
Month of birth FE&    {Yes}   &    {Yes}   &    {Yes}   &    {Yes}   &    {Yes}   &    {Yes}   \\
District of birth FE&    {Yes}   &    {Yes}   &    {Yes}   &    {Yes}   &    {Yes}   &    {Yes}   \\
School year of birth FE&     {No}   &    {Yes}   &     {No}   &    {Yes}   &     {No}   &    {Yes}   \\
\midrule
N         &\multicolumn{1}{c}{95,774}   &\multicolumn{1}{c}{95,774}   &\multicolumn{1}{c}{95,774}   &\multicolumn{1}{c}{95,774}   &\multicolumn{1}{c}{95,774}   &\multicolumn{1}{c}{95,774}   \\

%% file: tables/height_education_vacc1to6_pgisplit_tablefragment.tex
          &\multicolumn{3}{c}{Years of education}&\multicolumn{3}{c}{Height in cm}      \\\cmidrule(lr){2-4}\cmidrule(lr){5-7}
          &\multicolumn{1}{c}{(1)}   &\multicolumn{1}{c}{(2)}   &\multicolumn{1}{c}{(3)}   &\multicolumn{1}{c}{(4)}   &\multicolumn{1}{c}{(5)}   &\multicolumn{1}{c}{(6)}   \\
\midrule
\textbf{Panel A - High PGI:}&            &            &            &            &            &            \\
\midrule Post-vaccine share 1 to 6 $\times$ Pre-vacc. measles&    0.009   &    0.103   &   -0.073   &    0.432   &    0.684** &    0.712** \\
          &  (0.130)   &  (0.134)   &  (0.142)   &  (0.361)   &  (0.334)   &  (0.329)   \\
\midrule
N         &\multicolumn{1}{c}{85,180}   &\multicolumn{1}{c}{81,044}   &\multicolumn{1}{c}{85,176}   &\multicolumn{1}{c}{85,321}   &\multicolumn{1}{c}{87,201}   &\multicolumn{1}{c}{85,331}   \\
\midrule
\textbf{Panel B - Low PGI:}&            &            &            &            &            &            \\
\midrule Post-vacc. share 1 to 6 $\times$ Pre-vacc. measles&    0.095   &   -0.033   &    0.004   &   -0.293   &   -0.122   &   -0.377   \\
          &  (0.153)   &  (0.140)   &  (0.134)   &  (0.329)   &  (0.303)   &  (0.315)   \\
\midrule
N         &\multicolumn{1}{c}{84,867}   &\multicolumn{1}{c}{88,892}   &\multicolumn{1}{c}{84,884}   &\multicolumn{1}{c}{85,329}   &\multicolumn{1}{c}{83,327}   &\multicolumn{1}{c}{85,318}   \\
\midrule
\textbf{Controls for:} \\ Gender&    {Yes}   &    {Yes}   &    {Yes}   &    {Yes}   &    {Yes}   &    {Yes}   \\
Month of birth FE&    {Yes}   &    {Yes}   &    {Yes}   &    {Yes}   &    {Yes}   &    {Yes}   \\
District of birth FE&    {Yes}   &    {Yes}   &    {Yes}   &    {Yes}   &    {Yes}   &    {Yes}   \\
School year of birth FE&    {Yes}   &    {Yes}   &    {Yes}   &    {Yes}   &    {Yes}   &    {Yes}   \\
County-specific birthdate trend &    {Yes}   &    {Yes}   &    {Yes}   &    {Yes}   &    {Yes}   &    {Yes}   \\
Comp. schooling 16 $\times$ Pre-vacc. measles &    {Yes}   &    {Yes}   &    {Yes}   &     {No}   &     {No}   &     {No}   \\
\midrule
\textbf{PGI Source:}&\multicolumn{1}{c}{EA3}   &\multicolumn{1}{c}{Repository}   &\multicolumn{1}{c}{UKB}   &\multicolumn{1}{c}{GIANT}   &\multicolumn{1}{c}{Repository}   &\multicolumn{1}{c}{UKB}   \\

%% file: tables/height_education_vacc1to6_GxE_UKB_tablefragment.tex
          &\multicolumn{4}{c}{Years of education}             &\multicolumn{4}{c}{Height in cm}                   \\\cmidrule(lr){2-5}\cmidrule(lr){6-9}
          &\multicolumn{1}{c}{(1)}   &\multicolumn{1}{c}{(2)}   &\multicolumn{1}{c}{(3)}   &\multicolumn{1}{c}{(4)}   &\multicolumn{1}{c}{(5)}   &\multicolumn{1}{c}{(6)}   &\multicolumn{1}{c}{(7)}   &\multicolumn{1}{c}{(8)}   \\
\midrule
Post-vaccine share 1 to 6 $\times$ Pre-vacc. measles&    0.225***&    0.199***&    0.034   &   -0.023   &    2.054***&    2.046***&    0.348   &    0.086   \\
          &  (0.025)   &  (0.022)   &  (0.101)   &  (0.104)   &  (0.061)   &  (0.054)   &  (0.221)   &  (0.222)   \\
PGI       &    0.552***&    0.487***&    0.488***&    0.487***&    3.096***&    3.061***&    3.065***&    3.064***\\
          &  (0.009)   &  (0.007)   &  (0.007)   &  (0.007)   &  (0.015)   &  (0.015)   &  (0.015)   &  (0.015)   \\
Post-vaccine share 1 to 6 $\times$ Pre-vacc. measles $\times$ PGI&   -0.009   &   -0.013   &   -0.014   &   -0.014   &    0.077*  &    0.082*  &    0.073   &    0.087** \\
          &  (0.019)   &  (0.019)   &  (0.019)   &  (0.020)   &  (0.044)   &  (0.044)   &  (0.045)   &  (0.045)   \\
\midrule \textbf{Controls for:} \\ Gender&    {Yes}   &    {Yes}   &    {Yes}   &    {Yes}   &    {Yes}   &    {Yes}   &    {Yes}   &    {Yes}   \\
Month of birth FE&    {Yes}   &    {Yes}   &    {Yes}   &    {Yes}   &    {Yes}   &    {Yes}   &    {Yes}   &    {Yes}   \\
District of birth FE&     {No}   &    {Yes}   &    {Yes}   &    {Yes}   &     {No}   &    {Yes}   &    {Yes}   &    {Yes}   \\
School year of birth FE&     {No}   &     {No}   &    {Yes}   &    {Yes}   &     {No}   &     {No}   &    {Yes}   &    {Yes}   \\
County-specific birthdate trend &     {No}   &     {No}   &     {No}   &    {Yes}   &     {No}   &     {No}   &     {No}   &    {Yes}   \\
Compulsory schooling 16 &    {Yes}   &    {Yes}   &     {No}   &     {No}   &     {No}   &     {No}   &     {No}   &     {No}   \\
Compulsory schooling 16 $\times$ Pre-vacc. measles &    {Yes}   &    {Yes}   &    {Yes}   &    {Yes}   &     {No}   &     {No}   &     {No}   &     {No}   \\
Comp. schooling 16 $\times$ Pre-vacc. measles $\times$ PGI &    {Yes}   &    {Yes}   &    {Yes}   &    {Yes}   &     {No}   &     {No}   &     {No}   &     {No}   \\
Principal components 1-20 &    {Yes}   &    {Yes}   &    {Yes}   &    {Yes}   &    {Yes}   &    {Yes}   &    {Yes}   &    {Yes}   \\
\midrule
N         &\multicolumn{1}{c}{170,158}   &\multicolumn{1}{c}{170,134}   &\multicolumn{1}{c}{170,134}   &\multicolumn{1}{c}{170,134}   &\multicolumn{1}{c}{170,750}   &\multicolumn{1}{c}{170,725}   &\multicolumn{1}{c}{170,725}   &\multicolumn{1}{c}{170,725}   \\

%% file: tables/height_education_vacc1to6_sibsample_GxE_GIANT_EA3_tablefragment.tex
          &\multicolumn{4}{c}{Years of education}             &\multicolumn{4}{c}{Height in cm}                   \\\cmidrule(lr){2-5}\cmidrule(lr){6-9}
          &\multicolumn{1}{c}{(1)}   &\multicolumn{1}{c}{(2)}   &\multicolumn{1}{c}{(3)}   &\multicolumn{1}{c}{(4)}   &\multicolumn{1}{c}{(5)}   &\multicolumn{1}{c}{(6)}   &\multicolumn{1}{c}{(7)}   &\multicolumn{1}{c}{(8)}   \\
\midrule
Post-vaccine share 1 to 6 $\times$ Pre-vacc. measles&    0.228***&    0.190** &    0.190   &    0.124   &    1.891***&    2.176***&    2.500** &    2.203*  \\
          &  (0.085)   &  (0.085)   &  (0.415)   &  (0.433)   &  (0.224)   &  (0.220)   &  (1.085)   &  (1.265)   \\
PGI       &    0.651***&    0.592***&    0.591***&    0.591***&    3.097***&    3.100***&    3.102***&    3.092***\\
          &  (0.030)   &  (0.033)   &  (0.032)   &  (0.033)   &  (0.073)   &  (0.077)   &  (0.077)   &  (0.078)   \\
Post-vacc. share 1 to 6 $\times$ Pre-vacc. measles $\times$ PGI&    0.000   &   -0.018   &   -0.013   &   -0.013   &    0.114   &    0.130   &    0.113   &    0.155   \\
          &  (0.086)   &  (0.089)   &  (0.090)   &  (0.093)   &  (0.231)   &  (0.251)   &  (0.252)   &  (0.258)   \\
\midrule \textbf{Controls for:} \\ Gender&    {Yes}   &    {Yes}   &    {Yes}   &    {Yes}   &    {Yes}   &    {Yes}   &    {Yes}   &    {Yes}   \\
Month of birth FE&    {Yes}   &    {Yes}   &    {Yes}   &    {Yes}   &    {Yes}   &    {Yes}   &    {Yes}   &    {Yes}   \\
District of birth FE&     {No}   &    {Yes}   &    {Yes}   &    {Yes}   &     {No}   &    {Yes}   &    {Yes}   &    {Yes}   \\
School year of birth FE&     {No}   &     {No}   &    {Yes}   &    {Yes}   &     {No}   &     {No}   &    {Yes}   &    {Yes}   \\
County-specific birthdate trend &     {No}   &     {No}   &     {No}   &    {Yes}   &     {No}   &     {No}   &     {No}   &    {Yes}   \\
Compulsory schooling 16 &    {Yes}   &    {Yes}   &     {No}   &     {No}   &     {No}   &     {No}   &     {No}   &     {No}   \\
Compulsory schooling 16 $\times$ Pre-vacc. measles &    {Yes}   &    {Yes}   &    {Yes}   &    {Yes}   &     {No}   &     {No}   &     {No}   &     {No}   \\
Comp. schooling 16 $\times$ Pre-vacc. measles $\times$ PGI &    {Yes}   &    {Yes}   &    {Yes}   &    {Yes}   &     {No}   &     {No}   &     {No}   &     {No}   \\
Principal components 1-20 &    {Yes}   &    {Yes}   &    {Yes}   &    {Yes}   &    {Yes}   &    {Yes}   &    {Yes}   &    {Yes}   \\
\midrule
N         &\multicolumn{1}{c}{10,763}   &\multicolumn{1}{c}{10,553}   &\multicolumn{1}{c}{10,553}   &\multicolumn{1}{c}{10,553}   &\multicolumn{1}{c}{10,803}   &\multicolumn{1}{c}{10,592}   &\multicolumn{1}{c}{10,592}   &\multicolumn{1}{c}{10,592}   \\

%% file: tables/height_education_vacc1to6_gendersplit_tablefragment.tex
          &\multicolumn{2}{c}{Years of education}&\multicolumn{2}{c}{Height in cm}\\\cmidrule(lr){2-3}\cmidrule(lr){4-5}
          &\multicolumn{1}{c}{(1)}   &\multicolumn{1}{c}{(2)}   &\multicolumn{1}{c}{(3)}   &\multicolumn{1}{c}{(4)}   \\
\midrule
\textbf{Panel A - Women:}&            &            &            &            \\
\midrule Post-vaccine share 1 to 6 $\times$ Pre-vacc. measles&   -0.023   &    0.072   &   -0.086   &   -0.161   \\
          &  (0.143)   &  (0.132)   &  (0.336)   &  (0.296)   \\
PGI       &            &    0.616***&            &    2.930***\\
          &            &  (0.010)   &            &  (0.019)   \\
Post-vacc. share 1 to 6 $\times$ Pre-vacc. measles $\times$ PGI&            &    0.008   &            &    0.113*  \\
          &            &  (0.025)   &            &  (0.066)   \\
\midrule
N         &\multicolumn{1}{c}{94,346}   &\multicolumn{1}{c}{94,032}   &\multicolumn{1}{c}{94,682}   &\multicolumn{1}{c}{94,366}   \\
\midrule
\textbf{Panel B - Men:}&            &            &            &            \\
\midrule Post-vaccine share 1 to 6 $\times$ Pre-vacc. measles&   -0.048   &    0.003   &    0.165   &    0.135   \\
          &  (0.138)   &  (0.128)   &  (0.391)   &  (0.330)   \\
PGI       &            &    0.611***&            &    3.191***\\
          &            &  (0.012)   &            &  (0.024)   \\
Post-vacc. share 1 to 6 $\times$ Pre-vacc. measles $\times$ PGI&            &   -0.039   &            &    0.040   \\
          &            &  (0.027)   &            &  (0.070)   \\
\midrule
N         &\multicolumn{1}{c}{76,348}   &\multicolumn{1}{c}{76,018}   &\multicolumn{1}{c}{76,606}   &\multicolumn{1}{c}{76,277}   \\
\midrule
\textbf{Controls for:} \\ Month of birth FE&    {Yes}   &    {Yes}   &    {Yes}   &    {Yes}   \\
District of birth FE&    {Yes}   &    {Yes}   &    {Yes}   &    {Yes}   \\
School year of birth FE&    {Yes}   &    {Yes}   &    {Yes}   &    {Yes}   \\
County-specific birthdate trend &    {Yes}   &    {Yes}   &    {Yes}   &    {Yes}   \\
Compulsory schooling 16 $\times$ Pre-vacc. measles &    {Yes}   &    {Yes}   &     {No}   &     {No}   \\
Comp. schooling 16 $\times$ Pre-vacc. measles $\times$ PGI &     {No}   &    {Yes}   &     {No}   &     {No}   \\
Principal components 1-20 &     {No}   &    {Yes}   &     {No}   &    {Yes}   \\

%% file: tables/height_education_measlesvacc_difftrends_tablefragment.tex
          &\multicolumn{4}{c}{Years of education}             &\multicolumn{4}{c}{Height in cm}                   \\\cmidrule(lr){2-5}\cmidrule(lr){6-9}
          &\multicolumn{1}{c}{(1)}   &\multicolumn{1}{c}{(2)}   &\multicolumn{1}{c}{(3)}   &\multicolumn{1}{c}{(4)}   &\multicolumn{1}{c}{(5)}   &\multicolumn{1}{c}{(6)}   &\multicolumn{1}{c}{(7)}   &\multicolumn{1}{c}{(8)}   \\
\midrule
Post-vaccine share 1 to 6 $\times$ Pre-vacc. measles&   -0.027   &   -0.047   &   -0.035   &   -0.050   &    0.090   &   -0.102   &   -0.015   &   -0.200   \\
          &  (0.105)   &  (0.152)   &  (0.111)   &  (0.154)   &  (0.256)   &  (0.345)   &  (0.281)   &  (0.357)   \\
\midrule \textbf{Controls for:} \\ Gender&    {Yes}   &    {Yes}   &    {Yes}   &    {Yes}   &    {Yes}   &    {Yes}   &    {Yes}   &    {Yes}   \\
Month of birth FE&    {Yes}   &    {Yes}   &    {Yes}   &    {Yes}   &    {Yes}   &    {Yes}   &    {Yes}   &    {Yes}   \\
District of birth FE&    {Yes}   &    {Yes}   &    {Yes}   &    {Yes}   &    {Yes}   &    {Yes}   &    {Yes}   &    {Yes}   \\
School year of birth FE&    {Yes}   &    {Yes}   &    {Yes}   &    {Yes}   &    {Yes}   &    {Yes}   &    {Yes}   &    {Yes}   \\
Compulsory schooling 16 $\times$ Pre-vacc. measles &    {Yes}   &    {Yes}   &    {Yes}   &    {Yes}   &     {No}   &     {No}   &     {No}   &     {No}   \\
County-specific birthdate trend &    {Yes}   &     {No}   &     {No}   &     {No}   &    {Yes}   &     {No}   &     {No}   &     {No}   \\
Birthdate trend $\times$ Pre-vacc. measles &     {No}   &    {Yes}   &     {No}   &     {No}   &     {No}   &    {Yes}   &     {No}   &     {No}   \\
Admin. county-specific birthdate trend &     {No}   &     {No}   &    {Yes}   &     {No}   &     {No}   &     {No}   &    {Yes}   &     {No}   \\
District-specific birthdate trend &     {No}   &     {No}   &     {No}   &    {Yes}   &     {No}   &     {No}   &     {No}   &    {Yes}   \\
\midrule
N         &\multicolumn{1}{c}{170,778}   &\multicolumn{1}{c}{170,778}   &\multicolumn{1}{c}{170,778}   &\multicolumn{1}{c}{170,778}   &\multicolumn{1}{c}{171,370}   &\multicolumn{1}{c}{171,370}   &\multicolumn{1}{c}{171,370}   &\multicolumn{1}{c}{171,370}   \\

%% file: tables/height_education_vacc1to6_GxE_difftrends_tablefragment.tex
          &\multicolumn{4}{c}{Years of education}             &\multicolumn{4}{c}{Height in cm}                   \\\cmidrule(lr){2-5}\cmidrule(lr){6-9}
          &\multicolumn{1}{c}{(1)}   &\multicolumn{1}{c}{(2)}   &\multicolumn{1}{c}{(3)}   &\multicolumn{1}{c}{(4)}   &\multicolumn{1}{c}{(5)}   &\multicolumn{1}{c}{(6)}   &\multicolumn{1}{c}{(7)}   &\multicolumn{1}{c}{(8)}   \\
\midrule
Post-vaccine share 1 to 6 $\times$ Pre-vacc. measles&    0.041   &    0.025   &    0.026   &    0.015   &    0.024   &   -0.076   &   -0.084   &   -0.205   \\
          &  (0.096)   &  (0.144)   &  (0.102)   &  (0.146)   &  (0.223)   &  (0.305)   &  (0.250)   &  (0.315)   \\
PGI       &    0.614***&    0.615***&    0.614***&    0.613***&    3.046***&    3.047***&    3.046***&    3.046***\\
          &  (0.009)   &  (0.009)   &  (0.009)   &  (0.009)   &  (0.015)   &  (0.015)   &  (0.015)   &  (0.016)   \\
Post-vacc. share 1 to 6 $\times$ Pre-vacc. measles $\times$ PGI&   -0.012   &   -0.013   &   -0.012   &   -0.013   &    0.084*  &    0.078   &    0.082*  &    0.087*  \\
          &  (0.017)   &  (0.017)   &  (0.017)   &  (0.018)   &  (0.049)   &  (0.049)   &  (0.049)   &  (0.050)   \\
\midrule \textbf{Controls for:} \\ Gender&    {Yes}   &    {Yes}   &    {Yes}   &    {Yes}   &    {Yes}   &    {Yes}   &    {Yes}   &    {Yes}   \\
Month of birth FE&    {Yes}   &    {Yes}   &    {Yes}   &    {Yes}   &    {Yes}   &    {Yes}   &    {Yes}   &    {Yes}   \\
District of birth FE&    {Yes}   &    {Yes}   &    {Yes}   &    {Yes}   &    {Yes}   &    {Yes}   &    {Yes}   &    {Yes}   \\
School year of birth FE&    {Yes}   &    {Yes}   &    {Yes}   &    {Yes}   &    {Yes}   &    {Yes}   &    {Yes}   &    {Yes}   \\
Compulsory schooling 16 $\times$ Pre-vacc. measles &    {Yes}   &    {Yes}   &    {Yes}   &    {Yes}   &     {No}   &     {No}   &     {No}   &     {No}   \\
Compulsory schooling 16 $\times$ Pre-vacc. measles $\times$ PGI &    {Yes}   &    {Yes}   &    {Yes}   &    {Yes}   &     {No}   &     {No}   &     {No}   &     {No}   \\
County-specific birthdate trend &    {Yes}   &     {No}   &     {No}   &     {No}   &    {Yes}   &     {No}   &     {No}   &     {No}   \\
Birthdate trend $\times$ Pre-vacc. measles &     {No}   &    {Yes}   &     {No}   &     {No}   &     {No}   &    {Yes}   &     {No}   &     {No}   \\
Admin.-county-specific birthdate trend &     {No}   &     {No}   &    {Yes}   &     {No}   &     {No}   &     {No}   &    {Yes}   &     {No}   \\
District-specific birthdate trend &     {No}   &     {No}   &     {No}   &    {Yes}   &     {No}   &     {No}   &     {No}   &    {Yes}   \\
\midrule
N         &\multicolumn{1}{c}{170,134}   &\multicolumn{1}{c}{170,134}   &\multicolumn{1}{c}{170,134}   &\multicolumn{1}{c}{170,134}   &\multicolumn{1}{c}{170,725}   &\multicolumn{1}{c}{170,725}   &\multicolumn{1}{c}{170,725}   &\multicolumn{1}{c}{170,725}   \\

%% file: tables/height_education_measlesvacc_county_cid_yob_cluster_tablefragment.tex
          &\multicolumn{4}{c}{Years of education}             &\multicolumn{4}{c}{Height in cm}                   \\\cmidrule(lr){2-5}\cmidrule(lr){6-9}
          &\multicolumn{1}{c}{(1)}   &\multicolumn{1}{c}{(2)}   &\multicolumn{1}{c}{(3)}   &\multicolumn{1}{c}{(4)}   &\multicolumn{1}{c}{(5)}   &\multicolumn{1}{c}{(6)}   &\multicolumn{1}{c}{(7)}   &\multicolumn{1}{c}{(8)}   \\
\midrule
\textbf{Panel A:}&            &            &            &            &            &            &            &            \\
\midrule Post-vaccine share 1 to 6 $\times$ Pre-vacc. measles&    0.203***&    0.197***&    0.380*  &   -0.114   &    2.093***&    2.086***&    1.265** &   -0.069   \\
          &  (0.075)   &  (0.028)   &  (0.217)   &  (0.239)   &  (0.157)   &  (0.090)   &  (0.633)   &  (0.754)   \\
\midrule
\textbf{Panel B:}&            &            &            &            &            &            &            &            \\
\midrule Post-vaccine share 1 to 2 $\times$ Pre-vacc. measles&    0.023   &   -0.002   &   -0.081   &   -0.187   &    0.391*  &    0.316** &    0.344   &    0.031   \\
          &  (0.116)   &  (0.054)   &  (0.140)   &  (0.138)   &  (0.207)   &  (0.151)   &  (0.612)   &  (0.614)   \\
Post-vaccine share 3 to 4 $\times$ Pre-vacc. measles&    0.100   &    0.094** &    0.139   &   -0.008   &   -0.042   &   -0.040   &    0.191   &   -0.208   \\
          &  (0.135)   &  (0.047)   &  (0.150)   &  (0.154)   &  (0.252)   &  (0.125)   &  (0.387)   &  (0.390)   \\
Post-vaccine share 5 to 6 $\times$ Pre-vacc. measles&    0.062   &    0.077***&    0.248*  &    0.052   &    1.482***&    1.511***&    0.677*  &    0.121   \\
          &  (0.099)   &  (0.028)   &  (0.127)   &  (0.126)   &  (0.202)   &  (0.080)   &  (0.402)   &  (0.429)   \\
\midrule
\textbf{Controls for:} \\ Gender&    {Yes}   &    {Yes}   &    {Yes}   &    {Yes}   &    {Yes}   &    {Yes}   &    {Yes}   &    {Yes}   \\
Month of birth FE&    {Yes}   &    {Yes}   &    {Yes}   &    {Yes}   &    {Yes}   &    {Yes}   &    {Yes}   &    {Yes}   \\
County of birth FE&     {No}   &    {Yes}   &    {Yes}   &    {Yes}   &     {No}   &    {Yes}   &    {Yes}   &    {Yes}   \\
School year of birth FE&     {No}   &     {No}   &    {Yes}   &    {Yes}   &     {No}   &     {No}   &    {Yes}   &    {Yes}   \\
County-specific birthdate trend &     {No}   &     {No}   &     {No}   &    {Yes}   &     {No}   &     {No}   &     {No}   &    {Yes}   \\
Compulsory schooling 16 &    {Yes}   &    {Yes}   &     {No}   &     {No}   &     {No}   &     {No}   &     {No}   &     {No}   \\
Compulsory schooling 16 $\times$ Pre-vacc. measles &    {Yes}   &    {Yes}   &    {Yes}   &    {Yes}   &     {No}   &     {No}   &     {No}   &     {No}   \\
\midrule
N         &\multicolumn{1}{c}{147,919}   &\multicolumn{1}{c}{147,919}   &\multicolumn{1}{c}{147,919}   &\multicolumn{1}{c}{147,919}   &\multicolumn{1}{c}{148,422}   &\multicolumn{1}{c}{148,422}   &\multicolumn{1}{c}{148,422}   &\multicolumn{1}{c}{148,422}   \\

%% file: tables/height_education_measlesvacc_binary_tablefragment.tex
          &\multicolumn{3}{c}{Years of education}&\multicolumn{3}{c}{Height in cm}      \\\cmidrule(lr){2-4}\cmidrule(lr){5-7}
          &\multicolumn{1}{c}{(1)}   &\multicolumn{1}{c}{(2)}   &\multicolumn{1}{c}{(3)}   &\multicolumn{1}{c}{(4)}   &\multicolumn{1}{c}{(5)}   &\multicolumn{1}{c}{(6)}   \\
\midrule
\midrule Post-vacc. share 1 to 6 $\times$ High pre-vacc. measles&    0.024   &            &            &    0.115   &            &            \\
          &  (0.046)   &            &            &  (0.123)   &            &            \\
Any post-vacc. exposure 1 to 6 $\times$ Pre-vacc. measles&            &    0.029   &            &            &    0.004   &            \\
          &            &  (0.057)   &            &            &  (0.155)   &            \\
Any post-vacc. exposure 1 to 6 $\times$ High pre-vacc. measles&            &            &    0.034   &            &            &    0.069   \\
          &            &            &  (0.029)   &            &            &  (0.079)   \\
\midrule \textbf{Controls for:} \\ Gender&    {Yes}   &    {Yes}   &    {Yes}   &    {Yes}   &    {Yes}   &    {Yes}   \\
Month of birth FE&    {Yes}   &    {Yes}   &    {Yes}   &    {Yes}   &    {Yes}   &    {Yes}   \\
District of birth FE&    {Yes}   &    {Yes}   &    {Yes}   &    {Yes}   &    {Yes}   &    {Yes}   \\
School year of birth FE&    {Yes}   &    {Yes}   &    {Yes}   &    {Yes}   &    {Yes}   &    {Yes}   \\
County-specific birthdate trend &    {Yes}   &    {Yes}   &    {Yes}   &    {Yes}   &    {Yes}   &    {Yes}   \\
Comp. schooling 16 $\times$ High pre-vacc. measles &    {Yes}   &     {No}   &    {Yes}   &     {No}   &     {No}   &     {No}   \\
Comp. schooling 16 $\times$ Pre-vacc. measles &     {No}   &    {Yes}   &     {No}   &     {No}   &     {No}   &     {No}   \\
\midrule
N         &\multicolumn{1}{c}{170,778}   &\multicolumn{1}{c}{170,778}   &\multicolumn{1}{c}{170,778}   &\multicolumn{1}{c}{171,370}   &\multicolumn{1}{c}{171,370}   &\multicolumn{1}{c}{171,370}   \\

%% file: tables/height_education_measlesvacc_GxE_binary_tablefragment.tex
          &\multicolumn{3}{c}{Years of education}&\multicolumn{3}{c}{Height in cm}      \\\cmidrule(lr){2-4}\cmidrule(lr){5-7}
          &\multicolumn{1}{c}{(1)}   &\multicolumn{1}{c}{(2)}   &\multicolumn{1}{c}{(3)}   &\multicolumn{1}{c}{(4)}   &\multicolumn{1}{c}{(5)}   &\multicolumn{1}{c}{(6)}   \\
\midrule
PGI       &    0.606***&    0.614***&    0.606***&    3.051***&    3.047***&    3.052***\\
          &  (0.009)   &  (0.009)   &  (0.009)   &  (0.015)   &  (0.016)   &  (0.015)   \\
Post-vacc. share 1 to 6 $\times$ High pre-vacc. measles&    0.039   &            &            &   -0.001   &            &            \\
          &  (0.043)   &            &            &  (0.104)   &            &            \\
Post-vacc. share 1 to 6 $\times$ High pre-vacc. measles $\times$ PGI&   -0.005   &            &            &    0.086   &            &            \\
          &  (0.022)   &            &            &  (0.062)   &            &            \\
Any post-vacc. exposure 1 to 6 $\times$ Pre-vacc. measles&            &    0.077   &            &            &    0.057   &            \\
          &            &  (0.053)   &            &            &  (0.136)   &            \\
Any post-vacc. exposure 1 to 6 $\times$ Pre-vacc. measles $\times$ PGI&            &   -0.003   &            &            &    0.042   &            \\
          &            &  (0.015)   &            &            &  (0.032)   &            \\
Any post-vacc. exposure 1 to 6 $\times$ High pre-vacc. measles&            &            &    0.055*  &            &            &    0.021   \\
          &            &            &  (0.028)   &            &            &  (0.068)   \\
Any post-vacc. exposure 1 to 6 $\times$ High pre-vacc. measles $\times$ PGI&            &            &    0.020   &            &            &    0.042   \\
          &            &            &  (0.019)   &            &            &  (0.040)   \\
\midrule \textbf{Controls for:} \\ Gender&    {Yes}   &    {Yes}   &    {Yes}   &    {Yes}   &    {Yes}   &    {Yes}   \\
Month of birth FE&    {Yes}   &    {Yes}   &    {Yes}   &    {Yes}   &    {Yes}   &    {Yes}   \\
District of birth FE&    {Yes}   &    {Yes}   &    {Yes}   &    {Yes}   &    {Yes}   &    {Yes}   \\
School year of birth FE&    {Yes}   &    {Yes}   &    {Yes}   &    {Yes}   &    {Yes}   &    {Yes}   \\
County-specific birthdate trend &    {Yes}   &    {Yes}   &    {Yes}   &    {Yes}   &    {Yes}   &    {Yes}   \\
Comp. schooling 16 $\times$ High pre-vacc. measles &    {Yes}   &     {No}   &    {Yes}   &     {No}   &     {No}   &     {No}   \\
Compulsory schooling 16 $\times$ High pre-vacc. measles $\times$ PGI &    {Yes}   &     {No}   &    {Yes}   &     {No}   &     {No}   &     {No}   \\
Comp. schooling 16 $\times$ Pre-vacc. measles &     {No}   &    {Yes}   &     {No}   &     {No}   &     {No}   &     {No}   \\
Compulsory schooling 16 $\times$ Pre-vacc. measles $\times$ PGI &     {No}   &    {Yes}   &     {No}   &     {No}   &     {No}   &     {No}   \\
\midrule
N         &\multicolumn{1}{c}{170,134}   &\multicolumn{1}{c}{170,134}   &\multicolumn{1}{c}{170,134}   &\multicolumn{1}{c}{170,725}   &\multicolumn{1}{c}{170,725}   &\multicolumn{1}{c}{170,725}   \\

%% file: tables/ed_combined_measles_1to6_norosla_tablefragment.tex
          &\multicolumn{1}{c}{(1)}   &\multicolumn{1}{c}{(2)}   &\multicolumn{1}{c}{(3)}   &\multicolumn{1}{c}{(4)}   \\
\midrule
\textbf{Panel A - Years of education:}&            &            &            &            \\
\midrule Post-vaccine share 1 to 6 $\times$ Pre-vacc. measles&    0.319***&    0.322***&    0.100   &    0.000   \\
          &  (0.029)   &  (0.021)   &  (0.084)   &  (0.089)   \\
\midrule
\textbf{Panel B - Any qualification:}&            &            &            &            \\
\midrule Post-vaccine share 1 to 6 $\times$ Pre-vacc. measles&    0.097***&    0.096***&    0.031***&    0.000   \\
          &  (0.003)   &  (0.003)   &  (0.012)   &  (0.010)   \\
\midrule
\textbf{Panel C - Upper secondary qualification:}&            &            &            &            \\
\midrule Post-vaccine share 1 to 6 $\times$ Pre-vacc. measles&    0.057***&    0.056***&    0.013   &   -0.011   \\
          &  (0.005)   &  (0.004)   &  (0.016)   &  (0.018)   \\
\midrule
\textbf{Panel D - Degree qualification:}&            &            &            &            \\
\midrule Post-vaccine share 1 to 6 $\times$ Pre-vacc. measles&    0.043***&    0.045***&    0.015   &    0.008   \\
          &  (0.006)   &  (0.005)   &  (0.020)   &  (0.020)   \\
\midrule
\textbf{Controls for:} \\ Gender&    {Yes}   &    {Yes}   &    {Yes}   &    {Yes}   \\
Month of birth FE&    {Yes}   &    {Yes}   &    {Yes}   &    {Yes}   \\
District of birth FE&     {No}   &    {Yes}   &    {Yes}   &    {Yes}   \\
School year of birth FE&     {No}   &     {No}   &    {Yes}   &    {Yes}   \\
County-specific birthdate trend &     {No}   &     {No}   &     {No}   &    {Yes}   \\
\midrule
N         &\multicolumn{1}{c}{170,802}   &\multicolumn{1}{c}{170,778}   &\multicolumn{1}{c}{170,778}   &\multicolumn{1}{c}{170,778}   \\